\documentclass[twocolumn,amsmath,amssymb,aps,prd,reprint,longbibliography,floatfix,nobibnotes,12pt]{revtex4-1}

\usepackage{cancel}
\usepackage{enumitem}
\usepackage[utf8x]{inputenc}
\usepackage{graphicx}
\usepackage{dcolumn}
\usepackage{bm}
\usepackage{hyperref}
\usepackage[switch]{lineno}
\usepackage{float}             
\usepackage{subfigure}
\usepackage{tikz}
\usetikzlibrary{arrows,decorations.pathmorphing,backgrounds,positioning,fit,petri}
\tikzstyle{spring}=[thick,decorate,decoration={zigzag,pre length=0.1cm,post
  length=0.1cm,segment length=6}]
\usepackage{pgfplots}
 
\pgfplotsset{compat = newest}  
  
\newlength\tindent
\begin{document}

\title{Spin-Phonon Relaxation in Magnetic Molecules: Theory, Predictions and Insights.}

\author{Alessandro Lunghi}
\email{lunghia@tcd.ie}
\affiliation{School of Physics, AMBER and CRANN Institute, Trinity College, Dublin 2, Ireland}

\begin{abstract}
{\bf\bf Magnetic molecules have played a central role in the development of magnetism and coordination chemistry and their study keeps leading innovation in cutting-edge scientific fields such as magnetic resonance, magnetism, spintronics, and quantum technologies. Crucially, a long spin lifetime well above cryogenic temperature is a stringent requirement for all these applications. In this chapter we review the foundations of spin relaxation theory and provide a detailed overview of first-principles strategies applied to the problem of spin-phonon relaxation in magnetic molecules. Firstly, we present a rigorous formalism of spin-phonon relaxation based on open quantum systems theory. These results are then used to derive classical phenomenological relations based on the Debye model. Finally, we provide a prescription of how to map the relaxation formalism onto existing electronic structure methods to obtain a quantitative picture of spin-phonon relaxation. Examples from the literature, including both transition metals and lanthanides compounds, will be discussed in order to illustrate how Direct, Orbach and Raman relaxation mechanisms can affect spin dynamics for this class of compounds.}
\end{abstract}

\maketitle
\onecolumngrid

\clearpage
\tableofcontents

\clearpage
\section*{Introduction}\label{intro}

Localized unpaired electrons posses a magnetic moment in virtue of their angular and spin momentum. Such magnetic moment is sensitive to external magnetic fields, which can be used to probe and manipulate it both statically and dynamically. Molecules supporting unpaired electrons are routinely realized with organic radicals, coordination compounds of the first row of transition metals (TMs), and rare-earth ions. On the one hand, the study of molecular unpaired electrons' dynamics opens a windows on electronic structure and magnetism. Indeed, EPR studies of transition metals compounds have long provided a fundamental mean to address the relation between electronic structure, coordination geometry and magnetism\cite{bencini2012epr}. On the other hand, the interaction of spin with magnetic fields provides a rich opportunity to control and drive magnetic moment for specific tasks. Magnetic molecules have been proposed for a wide range of applications. For instance, they are routinely used for sensing, such as the use of spin probes to measure distances and dynamics in biological systems\cite{berliner2006distance}, or as contrast agents in Magnetic Resonance Imaging (MRI)\cite{zhang2001novel}. Magnetism is also another natural area of application for this class of systems. The design of magnetic molecules able to retain the direction of their magnetic moment for long times\cite{sessoli1993magnetic}, commonly known as single-molecule magnets, has been one of the central goals for a large community of chemists, physicists and material scientists for a long time\cite{zabala2021single}. More recently, the advent of disciplines such as spintronics\cite{vzutic2004spintronics,bogani2010molecular,sanvito2011molecular} and quantum science\cite{aromi2012design,gaita2019molecular,atzori2019second,wasielewski2020exploiting,heinrich2021quantum} have also attracted the attention of the molecular magnetism community. For instance, it has been experimentally demonstrated that molecules can be embedded in solid-state spintronics\cite{mannini2009magnetic} devices to influence the flow of currents\cite{urdampilleta2011supramolecular}. Moreover, single molecules with long coherence times and addressable with single-qubit gates have already found application as building blocks of quantum architectures\cite{bader2014room,zadrozny2015millisecond,atzori2016quantum,ariciu2019engineering}. The implementation of two-qubit gates\cite{atzori2018two} and quantum algorithms using molecular spins have already been demonstrated\cite{godfrin2017operating}, and the areas of quantum initialization and read-out\cite{bayliss2020optically}, scaling of the number of qubits\cite{moreno2018molecular,carretta2021perspective,biard2021increasing}, and the interface of molecules with other quantum systems\cite{gimeno2020enhanced,urtizberea2020vanadyl,atzori2021radiofrequency} are also rapidly progressing.  \\

All the applications just listed hinge on the possibility to coherently control the state of a molecule's magnetic moment. Unfortunately, this is only possible on very short time-scales, generally referred to as $T_{2}$. Indeed, the state of spin behaves coherently only in the absence of external perturbations. In this context, external refers to any degrees of freedom other then the one we are interested in. Clearly, such a requirement can never be completely fulfilled in a molecular system. Even for extremely diluted gas-phase molecules, the molecular spin's dynamics is inevitably entangled with the one of the intra-molecular vibrational degrees of freedom or the nuclear spins' one. It is then evident that the study of how $T_2$ is affected by the spin environment becomes a crucial aspect of molecular magnetism, both in terms of how to control it and how to extract information on the environment itself from its measurement. \\

In this chapter we are particularly interested in the study of a specific source of decoherence, namely spin-phonon relaxation. To differentiate it from the total coherence time $T_{2}$, spin-phonon relaxation time is generally called $T_{1}$. The atoms of a molecule, or more generally of a lattice, are always in motion, even at zero Kelvin. This motion perturbs the electronic structure of the system, and in the presence of interactions that mix spin and orbital electronic degrees of freedom, such as spin-orbit or spin-spin coupling, this perturbation transfers to the magnetic moment. This interaction is generally referred to as spin-phonon interaction and it is at the origin of spin thermalization in semiconductors at non-cryogenic temperatures or when strong magnetic dilutions are employed. If the specific heat of the lattice is much higher than the spin one, the former is able to exchange energy with the spin system without changing its state significantly and therefore acts as a thermal reservoir that the spin can use to equilibrate at the same temperature. \\

When a molecular spin relaxes to its thermal equilibrium, after it has been perturbed by external stimuli or a change in thermodynamical conditions, we are witnessing a prototypical dissipative phenomenon. Dissipation and thermalization are not hard-wired processes in quantum mechanics but are emergent ones. Indeed, the Schroedinger equation is a linear differential equation and invariably leads to an oscillatory dynamics of a system variables. However, when the number of such variables is large enough, thermalization arises as the result of the destructive interference among the many non-synchronous oscillations corresponding to the system's eigenvalues. Let us assume we had perfect knowledge of the eigenstates ,$|i\rangle$, and eigenvalues, $E_i$, of a many-body Hamiltonian, $\hat{H}$. In this case, $\hat{H}$ could either represents a system of many spins or one spin interacting with many phonons. Any state prepared in $|\psi\rangle$ would evolve in time according to its decomposition in terms of $\hat{H}$'s eigenvectors
\begin{equation}
    |\psi(t)\rangle=\sum_i \langle i|\psi(t_{0})\rangle |i\rangle e^{-\frac{i}{\hbar}E_i t} = \sum_i C_i |i\rangle e^{-\frac{i}{\hbar}E_i t}\:.
\end{equation}
According to this expression, the operator associated to a local single-particle property $\hat{O}$, such as the single-spin moment $\vec{\mathbf{S}}$, will evolve in time as 
\begin{equation}
    \langle\hat{O}(t)\rangle=\sum_{ij}O_{ij}C_{i}^{*}C_{j}e^{-\frac{i}{\hbar}(E_{j}-E_{i})t} \xrightarrow{t\rightarrow \infty} \sum_{i} O_{ii} |C_{i}|^{2}\:,
    \label{interf}
\end{equation}
Eq. \ref{interf} shows that at long times, the out-of-diagonal terms give destructive interference and only the diagonal time-independent ones survive. Unless particular symmetries are present in the spectrum this is generally fulfilled, and the thermodynamical average of the property $O$ is only given by the initial condition of the system. Interestingly, according to the Eigenstate Thermalization Hypothesis (ETH)\cite{deutsch2018eigenstate}, if this initial condition fulfills some specific conditions, the final average will be consistent with statistical mechanics. This is not always the case and systems exhibiting many-body localization are subject of intense research\cite{nandkishore2015many}. In this work we will always assume that thermalization is possible and we will instead focus on determining the rate at which this thermalized state is reached. \\

Although solving the Schroedinger equation for the entire number of degrees of freedom composing a physical system would always lead to exact predictions of its behaviour, this is far from practical or even possible. The actual implementation of Eq. \ref{interf} would require the knowledge of the spectrum and eigenvectors of the many-body Hamiltonian, whose size grows exponentially with the number of degrees of freedom. This very unfavourable scaling law of the Hamiltonian's size makes it possible to tackle only very small system sizes, in the order of 20 particles with only two-level states each, and at massive computational costs. Fortunately, very commonly we are only interested in monitoring the behaviour of a small fraction of the total number of degrees of freedom. In a such a case it is possible to simplify the theoretical treatment of relaxation by employing the theory of open-quantum systems\cite{breuer2002theory}. In this framework one tries to map the many-body problem into a few-body problem (the degrees of freedom we are interested in) where time evolution is driven by an effective equation of motion that incorporates the action of hidden degrees of freedom, commonly referred to as the bath. \\

In more quantitative terms, let us describe the total system with a density operator $\hat{\rho}(t)$, which evolves in time according to the unitary propagator $\hat{U}(t,t_{0})$ as $\hat{\rho}(t)=\hat{U}(t,t_{0})\hat{\rho}(t_{0})$. Let us also assume that the system is composed by many degrees of freedom that can be separated into two classes, the system $S$ and the bath, $B$. As the name suggests the sub-portion $S$ of the total system is the one we are interested in monitoring in detail, while the sub-system $B$ is only relevant in terms of its effect on $S$. The theory of open quantum systems seeks to find a mathematical way to map the evolution of the entire density operator $\hat{\rho}(t)$ to the evolution of the reduced density operator 
\begin{equation}
    \hat{\rho}_{S}(t)=\mathrm{Tr}_{B}\{\hat{\rho}(t)\}=\hat{V}(t,t_{0})\hat{\rho}_{S}(t_{0}) \:,
    \label{dynamicalmap}
\end{equation}
where the trace operation is applied to all the bath degrees of freedom, thus reducing the dimensionality of the problem to the number of degrees of freedom of $S$. $\hat{V}(t,t_{0})$ is an effective propagator that acts on the sole $S$ sub-space. It turns out that an expression for $\hat{V}(t,t_{0})$ that verifies Eq. \ref{dynamicalmap} exists and that it can be derived with the formalism of projection operators. This approach can be used to derive either the Nakajima-Zwanzig or time-convolutionless equations\cite{breuer2002theory}. The latter provides an exact time-local equation of motion for $\hat{\rho}_{S}(t)$ in the form
\begin{equation}
    \frac{d\hat{\rho}_{S}(t)}{dt}=\hat{K}(t,t_{0})\hat{\rho}_{S}(t)\:.
    \label{tcl}
\end{equation}
It should be noted that solving Eq. \ref{tcl} is not necessarily easier than solving the Schroedinger equation for the total system and in practice we have only formally re-written the problem in terms of the reduced density matrix $\hat{\rho}_{S}$. The great strength of this formalism in fact comes from the possibility to expand the operator $\hat{K}(t,t_{0})$ in a perturbative series\cite{breuer2001time,timm2011time}. For those systems where only a few terms of this expansion are needed, the advantage becomes immense. The theory of open quantum systems just briefly sketched perfectly applies to the problem of spin-phonon relaxation. Indeed, spin-lattice relaxation is a prototypical example of open-quantum system dynamics, where the (few) spin degrees of freedom are the ones we are interested in monitoring and that undergo a weak interaction with a complex environment (the phonons).  \\

Attempts to the formulation of a spin-phonon relaxation theory started already in 30' and 40' with the seminal contributions of Waller\cite{waller1932magnetisierung} and Van Vleck\cite{van1940paramagnetic}. In these early efforts the role of lattice waves in modulating inter-molecular dipolar spin-spin interactions and the magnetic ions' crystal field were considered. The latter mechanism is commonly used to explain spin-phonon relaxation and in the contributions of Orbach\cite{orbach1961spin} it was adapted to the formalism of the effective spin Hamiltonian, which still today represents the language of magnetic resonance. Many others have contributed to the adaptation of these models to specific systems, for instance including the effect of localized vibrational states due to defects and impurities\cite{klemens1962localized} or optical phonons\cite{singh1979optical}, and in general accounting for the large phenomenology of both spin and vibrational spectra, which can span orders of magnitude of splittings and possess many different symmetries\cite{gill1975establishment,stevens1967theory,shrivastava1983theory}. \\

As a result of this prolific theoretical work, we now posses a plethora of models that describe spin-phonon relaxation time as function of external thermodynamical conditions, such as temperature ($T$) and magnetic field ($B$), and that can fit experimental observations in many different scenarios\cite{eaton2002relaxation}. These phenomenological models of spin-phonon relaxation represent the backbone of our understanding of this process and have served the scientific community for almost a century. However, some open questions remain beyond the reach of these models. The measurement of spin relaxation is only an indirect way to probe spin-lattice interactions and, even for relatively simple systems, many different alternative explanations of the observed phenomenology are often possible. Indeed, when many possible relaxation mechanisms overlap with one another, assigning the correct one is not an easy task and a comprehensive experimental characterization becomes necessary\cite{sato2007impact,ding2018field}.
Moreover, some important details on spin-phonon relaxation and interaction remain hidden. For instance, the nature of the vibrations involved in spin relaxation and the exact mechanism by which they couple to spin can hardly be extracted from experiments without any theoretical or computational aid\cite{moseley2018spin,chiesa2020understanding}. Last but not least, if we set ourselves into a materials science frame of mind, where the goalpost shifts from explaining observed phenomena to being able to engineer new chemical systems for target properties, the impossibility to perform quantitative and blind predictions of relaxations rate represents a crucial downside of a phenomenological approach to spin-phonon relaxation theory. \\

All these open questions naturally lead to looking at electronic structure theory and materials modelling as alternative to a phenomenological descriptions of spin systems. In recent decades these fields have made gigantic strides in terms of accuracy, efficiency and range of systems that can be studied\cite{neese2019chemistry}. The description of magnetism and heavy elements is among the most challenging aspects of electronic structure theory. Nonetheless, methods such as density functional theory and correlated wave-function theory can nowadays be used routinely to obtain close-to-quantitative predictions of molecular magnetic properties\cite{atanasov2015first,ungur2017ab}. Importantly, many efforts have been devoted to the mapping of electronic structure theory to the formalism of spin Hamiltonian\cite{maurice2009universal,neese2009spin}, therefore providing a transparent way for theory and experiments to communicate. Similarly, in the case of lattice dynamics, density functional theory has reached a level of sophistication and efficiency that makes it possible to treat thousands of atoms and perform accurate predictions of phonon spectra\cite{garlatti2020unveiling}. All these methods can be used as building blocks of an \textit{ab initio} theory of spin-phonon relaxation, which stands out as the ultimate computational and theoretical framework for the description of spin dynamics in realistic and complex systems.\\

In this chapter we provide an overview of spin-phonon relaxation theory and how it can be applied to the case of crystals of magnetic molecules in a fully \textit{ab initio} fashion. We will begin with a brief section on lattice dynamics in order to introduce the mathematical formalism used in the rest of the chapter and to recap some fundamental concepts. The second section is centered around the theory of open quantum systems and its application to spin-phonon relaxation. Contributions at different orders of perturbation theory and spin-phonon coupling strength will be detailed and analyzed providing all the tools to derive the classical results of spin-phonon relaxation based on the Debye model and used throughout literature until now. The latter will be presented in section three. Section four will instead provide a practical way to the \textit{ab initio} simulation of all the parameters entering the spin relaxation equations and will thus outline a prescription to predict spin-phonon lifetime in realistic solid-state systems. In section five we will review recent literature, with emphasis on the author's work, and show how \textit{ab initio} simulations can provide new insights on spin-phonon coupling and relaxation in magnetic molecules. A conclusive outlook section will summarize the main points of the chapter and propose a way forward for this field. 

\clearpage
\section{Fundamentals of Lattice Dynamics}\label{solidstate}

Let us consider a mono-dimensional system composed of a single particle of mass $m$. The particle position is described by the Cartesian coordinate $X$ and its momentum by $P$. Assuming that the particle experiences an harmonic potential with equilibrium position at $X^{0}=0$, the system's Hamiltonian can be written as
\begin{equation}
\hat{H}_{1D}=\frac{\hat{P}^{2}}{2m}+\frac{1}{2}m\omega^{2} \hat{X}^{2}\:,
\label{Ham}
\end{equation}
where $\omega$ is the angular frequency of the harmonic oscillator. This Hamiltonian is often transformed in an equivalent form by mean of a transformation of coordinates that renders the operators $\hat{X}$ and $\hat{P}$ unit-less. The new position and momentum $\hat{\bar{X}}$ and $\hat{\bar{P}}$ are defined as
\begin{equation}
\hat{\bar{P}}=\sqrt{\frac{1}{m\hbar\omega}}\hat{P} \:,\quad \hat{\bar{X}}=\sqrt{\frac{m\omega}{\hbar}}\hat{X}  \:.
\end{equation}
The Hamiltonian \ref{Ham} in this new basis becomes 
\begin{equation}
\hat{H}_{1D}=\frac{\hbar\omega}{2}\left( \hat{\bar{P}}^{2} + \hat{\bar{X}}^{2} \right)   \:.
\label{Ham2}
\end{equation}
In order to find the eigenstates of Eq. \ref{Ham2} we introduce the basis of occupation numbers, where $|n\rangle$ represents a state with $n$-phonon excitations and $|0\rangle$ is the  vacuum state, i.e. the harmonic oscillator is in its ground state. It is also convenient to introduce creation/annihilation and particle number operators 
\begin{equation}
\hat{a}^{\dag}=\frac{1}{\sqrt{2}}\left(\hat{\bar{X}}-i\hat{\bar{P}}\right) \:, \quad  \hat{a}=\frac{1}{\sqrt{2}}\left(\hat{\bar{X}}+i\hat{\bar{P}}\right) \:\quad \text{and} \quad \hat{n}=\hat{a}^{\dag}\hat{a} \:,
\label{ladder}
\end{equation}
with commuting rule $[\hat{a},\hat{a}^{\dag}]=1$. Starting from Eq. \ref{ladder}, it is possible to demonstrate the following relations
\begin{gather}
    \hat{a}^{\dag} |n\rangle = \sqrt{n+1} |n+1\rangle \\
    \hat{a} |n\rangle = \sqrt{n} |n-1\rangle \\
    \hat{n} |n\rangle = n |n\rangle 
\end{gather}

Finally, according to the transformations of Eq. \ref{ladder} the Hamiltonian takes the form
\begin{equation}
\hat{H}_{1D}=\hbar\omega(\hat{n}+\frac{1}{2}) \:,
\label{Ham3}
\end{equation}
where it is now transparent that every phonon excitation in the system leads to an increase in the vibrational energy of a quantum $\hbar\omega$ with respect to the zero-point energy of the vacuum state $\hbar\omega/2$. \\

Although the Hamiltonian of Eq. \ref{Ham3} only represents a single harmonic oscillator, it forms the basis for a generalization to more complex systems. A molecular system made by $N$ interacting particles can be described in the harmonic approximation assuming the potential energy surface $U$ to be well described by its Taylor expansion around the minimum energy position
\begin{equation}
U(\vec{\Delta\mathbf{X}})=\frac{1}{2}\sum_{ij}^{N}\sum_{st}^{3}\left(\frac{\partial E_{el}}{\partial X_{is}\partial X_{jt}}\right)_{0} \Delta X_{is} \Delta X_{jt} = \frac{1}{2}\sum_{ij}^{3N}\left(\frac{\partial E_{el}}{\partial x_{i}\partial x_{j}}\right)_{0} x_{i} x_{j}\:,
\label{HarmUN}
\end{equation} 
where $E_{el}$ is the electronic energy of the adiabatic ground state. In \ref{HarmUN} we have introduced the notation $\Delta X_{is}=X_{is}-X^{0}_{is}$ to identify the displacement of atom $i$ along the Cartesian coordinate $s$ with respect to the equilibrium position, $X^{0}_{is}$. Alternatively, we can compress the atomic and Cartesian indexes into a single vector of dimension $3N$, $x_{i}$, where $x_{1}=\Delta X_{11}$, $x_{2}=\Delta X_{12}$, $x_{3}=\Delta X_{13}$ , $x_{4}=\Delta X_{21}$, etc. 
This system can still be mapped on a set of $3N$ decoupled 1D harmonic oscillators by introducing the basis of the normal modes of vibration. We first start defining mass-weighted Cartesian coordinates $u_{i}=\sqrt{m_{i}}x_{i}$ and by diagonalizing the force-constant matrix of the energy second-order derivatives $\mathbf{\Phi}$
\begin{equation}
\frac{1}{2}\sum_{ij}^{3N}\left(\frac{\partial E_{el}}{\partial x_{i}\partial x_{j}}\right)x_{i}x_{j}=\frac{1}{2}\sum_{ij}^{3N}\left(\frac{\partial E_{el}}{\partial u_{i}\partial u_{j}}\right) u_{i}u_{j}=\frac{1}{2}\sum_{ij}^{3N}\Phi_{ij}u_{i}u_{j} \rightarrow \frac{1}{2}\sum_{i}^{3N}\textit{diag}(\mathbf{\Phi})_{ii}Q^{2}_{i}
\label{diagH}
\end{equation}
The system Hamiltonian can now be written in a similar fashion to Eq. \ref{Ham} by mapping the $Q_{i}$ coordinates, expressed by the $\mathbf{\Phi}$ eigenvectors $L_{ij}$, on the 1D mass-weighted Cartesian coordinates $\sqrt{m}X$. Doing so, the frequencies of normal vibrations are defined as $\omega_{i}=\sqrt{\textit{diag}(\mathbf{\Phi})_{ii}}$ while $\bar{Q}_{i}$ coordinates are defined as function of Cartesian coordinates $x_{j}$ through $\mathbf{\Phi}$ eigenvectors $L_{ij}$
\begin{equation}
\bar{Q}_{j}=\sum_{i}^{3N}L_{ij}\sqrt{m_{i}}x_{i}
\end{equation}
As done in the previous section we here define a set of unit-less normal modes $Q_{i}$
\begin{equation}
Q_{j}=\sqrt{\frac{\omega_{j}}{\hbar}}\sum_{i}^{3N}L_{ij}\sqrt{m_{i}}x_{i}
\label{normalmodeN}
\end{equation}
and the inverse transformation that defines the Cartesian displacement associated to a unit-less normal mode $Q_{j}$ amount of displacement
\begin{equation}
x_{i}=\sqrt{\frac{\hbar}{m_{i}\omega_{j}}}L_{ij}Q_{j}\:.
\end{equation}
Finally, we can also write the $i$-th unit-less normal mode operator as function of creation and annihilation operators as
\begin{equation}
\hat{Q}_{i}=\frac{1}{\sqrt{2}}\left(\hat{a}^{\dag}_{i}+\hat{a}_{i}\right).
\end{equation}
with the commutation rule $[\hat{a}_{i},\hat{a}_{j}^{\dag}]=\delta_{ij}$ and total Hamiltonian
\begin{equation}
\hat{H}_{3N}=\sum_{i}^{3N}\hbar\omega(\hat{n}_{i}+\frac{1}{2}) \:,
\label{HarmHN}
\end{equation}
where $\hat{n}_{i}=\hat{a}_{i}^{\dag}\hat{a}_{i}$ is the number operator for the $i$-normal mode. \\

The definitions of Eqs. \ref{normalmodeN} to \ref{HarmHN} are suitable to describe any molecular system, whose potential energy surface can be written as in Eq. \ref{HarmUN}. However, when dealing with extended systems like solids, the number of degrees of freedom, $3N$, diverges to infinity, making these definitions unpractical for numerical calculations. However, if the solid has a crystalline structure, it is possible to exploit the translational symmetry of the lattice to simplify the problem. \\

Let us define a lattice of $L$ unit cells, each at position $\vec{R}_{l}$ with respect to an arbitrary cell chosen as the origin of the lattice. Each one of these cells contains a replica of the same group of atoms, with equilibrium coordinates $X^0_{lis}$, where the index $i$ refers to the atom index and $s$ to the Cartesian component and $l$ identify the cell. By virtue of the translational invariance of the lattice, the following expression is fulfilled by all the atoms in the crystal
\begin{equation}
  X^{0}_{lis}=X^{0}_{0is}+R_{ls}\:.
 \label{translation}    
\end{equation}
Let us define the Fourier transform of mass-weighed Cartesian coordinates 
\begin{equation}
 x_{qis}=L^{-1/2}\sum_{l}^{L}x_{lis}e^{i \mathbf{q}\cdot \mathbf{R}_{l}}
\end{equation}

Accordingly, the Hessian matrix $\mathbf{\Phi}$ in this new set of variables reads 
\begin{equation}
    L^{-1}\sum_l\sum_v \Phi^{l,v-l}_{ij} e^{i\mathbf{q}\cdot \mathbf{R}_l}e^{i\mathbf{q'}\cdot (\mathbf{R}_v-\mathbf{R}_l)}=L^{-1}\sum_l e^{i(\mathbf{q}-\mathbf{q'})\cdot \mathbf{R}_l} \sum_v \Phi^{0,v}_{ij} e^{i\mathbf{q'}\cdot\mathbf{R}_v}=\delta_{\mathbf{q}\mathbf{q'}}\sum_v \Phi^{0,v}_{ij} e^{i\mathbf{q'}\cdot\mathbf{R}_v}\:,
    \label{trasl}
\end{equation}
where we have used the fact that $\Phi^{l,v}_{ij}=\Phi^{l-t,v-t}_{ij}$ due to the translational symmetry of the lattice. Eq. \ref{trasl} thus suggests the introduction of the $3N \times 3N$ dynamical matrix
\begin{equation}
 D_{ij}(\mathbf{q}) = \sum_v \Phi^{0,v}_{ij} e^{i\mathbf{q}\cdot\mathbf{R}_v}\:.
\end{equation}
As a consequence of Eq. \ref{trasl}, we can decouple the infinite number of degrees of freedom into an infinite sum (each depending of $\mathbf{q}$) of $3N$ decoupled normal modes. In fact, it is possible to compute a dynamical matrix for every value of $\mathbf{q}$ and compute its eigenvalues as in the case of a single molecule with $3N$ degrees of freedom. The eigenvalues of $\mathbf{D}(\mathbf{q})$ are $\omega^{2}(\mathbf{q})$, while the eigenvectors $\mathbf{L}(\mathbf{q})$ defines the normal modes of vibration. The normal modes $Q_{\alpha\mathbf{q}}$ are now characterized by both a reciprocal lattice vectors, $\mathbf{q}$, and the angular frequency $\omega_{\alpha\mathbf{q}}$.  The periodic atomic displacement wave reads
\begin{equation}
   x_{li}=\frac{1}{\sqrt{N_q}}\sum_{i}^{3N}\sqrt{\frac{\hbar}{\omega_{\alpha\mathbf{q}}m_{i}}}e^{i\mathbf{q}\cdot \mathbf{R}_{l}} L^{\mathbf{q}}_{i\alpha}   Q_{\alpha\mathbf{q}}\:,
    \label{Dispq}
\end{equation}
where the eigenvectors of $\mathbf{D}(\mathbf{q})$, $L^{\mathbf{q}}_{i\alpha}$, describe the composition of the normal mode in terms of atomic displacements.
The Hamiltonian corresponding to periodic system now becomes
\begin{equation}
\hat{H}_\mathrm{ph}=\sum_{\alpha\mathbf{q}}\hbar\omega_{\alpha\mathbf{q}}(\hat{n}_{\alpha\mathbf{q}}+\frac{1}{2})\:,
\label{HPH}
\end{equation}
where $\hat{n}_{\alpha\mathbf{q}}=\hat{a}^{\dag}_{\alpha \mathbf{q}}\hat{a}_{\alpha \mathbf{q}}$ is the phonon's number operator, and $\hat{a}^{\dag}_{\alpha \mathbf{q}}$ ($\hat{a}_{\alpha \mathbf{q}}$) is the creation (annihilation) operator. The phonon operator $\hat{Q}_{\alpha\mathbf{q}}$ can also be written in terms of these operators as
\begin{equation}
\hat{Q}_{\alpha\mathbf{q}}=\frac{1}{\sqrt{2}} (a^{\dag}_{\alpha \mathbf{q}} + a_{\alpha \mathbf{-q}} )
\end{equation}

\clearpage
\section{Spin-Phonon Relaxation Theory}\label{theory}

In this section we are going to address the theory of spin-lattice relaxation in solid-state materials following the formalism open-quantum system dynamics\cite{breuer2002theory}. We will start outlining the nature of spin-phonon coupling at a phenomenological level, and then we will illustrate how to model the time evolution of the spin density matrix under the influence of spin-phonon interaction. \\

Let us start writing a general spin Hamiltonian
\begin{equation}
\hat{H}_\mathrm{s}=\sum_{i}^{N_\mathrm{s}}\beta_{i}\vec{\mathbf{B}}\cdot\mathbf{g}_{i}\cdot\vec{\mathbf{S}}_{i}+ \sum_{i}^{N_\mathrm{s}}\sum_{ l=2}^{2S_{i}}\sum_{m=-l}^{l} B_{m}^{l}\hat{O}^{l}_{m}(\vec{\mathbf{S}}_{i}) +\frac{1}{2}\sum_{i\ne j}^{N_{s}}\vec{\mathbf{S}}_{i}\cdot\mathbf{J}_{ij}\cdot\vec{\mathbf{S}}_{j}\:.
\label{SH}
\end{equation}
The first term in Eq. \ref{SH} describes the Zeeman interaction between the $i$-th spin, $\mathbf{S}_{i}$, and the external magnetic field $\vec{\mathbf{B}}$, as mediated by the electron/nuclear Bohr magneton, $\beta_{i}$, and the Landè tensor, $\mathbf{g}_{i}$. The second term represents a generalized version of the single-spin zero-field-splitting Hamiltonian written in terms of Tesseral operators of order $l$ and component $m$, $\hat{O}^{l}_{m}$\cite{tennant2000rotation}. The five components of order $l=2$ are equivalent to the canonical trace-less Cartesian tensor $\mathbf{D}$ used to model single-spin zero-field splitting in transition metals with the Hamiltonian $\hat{H}_{s}=\vec{\mathbf{S}}\cdot\mathbf{D}\cdot\vec{\mathbf{S}}$. The Cartesian tensors $\mathbf{J}_{ij}$ automatically includes both scalar Heisenberg exchange $J_{ij}=\mathrm{Tr}(\mathbf{J}_{ij})/3$, the Dzialoszynski-Moria interaction, $\vec{\mathbf{d}}_{ij}=(\mathbf{J}_{ij}-\mathbf{J}_{ij}^{\dag})/2$, and the trace-less anisotropic exchange interaction $\mathbf{D}_{ij}=(\mathbf{J}_{ij}+\mathbf{J}_{ij}^{\dag})/2-J_{ij}\mathbf{E}$. At this level, Eq. \ref{SH} can account for both electronic and nuclear spins. For instance, the spin-spin interaction tensor, $\mathbf{J}$, may coincide with the point-dipole interaction, 
$\mathbf{D}^\mathrm{dip}$, or the hyperfine tensor, $\mathbf{A}$, depending on whether the interaction is among electronic and/or nuclear spins. \\

The coefficients appearing in Eq. \ref{SH} are generally fitted to reproduce either the experimental or \textit{ab initio} data available for a specific compound and provide a compact representation of a magnetic system's spectral properties\cite{neese2009spin,bencini2012epr,ungur2017ab}. The values of the spin Hamiltonian's coefficients are strongly dependent on the spin environment and they can be used as an analytical instrument to investigate the relation between chemical structure and magnetic/electronic properties. The strong dependence of the coefficients of Eq. \ref{SH} with respect to molecular structure is not limited to radical changes such as the replacement of ligands or the change of an ion's coordination number. In fact, the sensitivity of magnetic properties with respect of structural distortions is so strong that even small molecular distortions due to thermal strain are able to modulate the coefficients of the spin Hamiltonian\cite{qian2015does}. Therefore, molecular dynamics introduces a time dependence into the spin Hamiltonian  $\hat{H}_{s}(t)$. Such time dependency acts as a perturbation of the equilibrium degrees of freedom and it is at the origin of spin-lattice relaxation itself.\\

Each phonon $\hat{Q}_{\alpha\mathbf{q}}$ contributes to distort the molecular geometry to some degree and as a consequence they can all influence the magnitude of the coefficients in Eq. \ref{SH}. However, molecular structure is only slightly affected during its dynamics and it is possible to model its effect on magnetic properties as perturbation. In a more rigorous way, we assume a weak-coupling regime between phonons and spins and perform a Taylor expansion of the spin Hamiltonian with respect to the phonon variables
\begin{equation}   
\hat{H}_\mathrm{s}(t)=(\hat{H}_{s})_{0}+\sum_{\alpha\mathbf{q}}\Big(\frac{\partial \hat{H}_{s}}{\partial Q_{\alpha\mathbf{q}}}\Big)_{0}\hat{Q}_{\alpha\mathbf{q}}(t) +\sum_{\alpha\mathbf{q}}\sum_{\beta\mathbf{q'}}\Big(\frac{\partial^{2} \hat{H}_{s}}{\partial Q_{\alpha\mathbf{q}}\partial Q_{\beta\mathbf{q'}}}\Big)_{0}\hat{Q}_{\alpha\mathbf{q}}(t)\hat{Q}_{\beta\mathbf{q'}}(t)+... \:,
\label{SPH}
\end{equation}
where the naught symbol points to those quantities evaluated for the crystal geometry at zero-temperature, i.e. for the structure in its energy minimum. The terms of Eq. \ref{SPH} that explicitly depends on time define the spin-phonon coupling Hamiltonian, while the derivatives of the spin Hamiltonian themselves are called spin-phonon coupling coefficients. It is also convenient to introduce a norm definition to compare spin-phonon coupling coefficients among them and easily visualize them. For instance, assuming that our spin Hamiltonian to be dominated by the Zeeman term, a simple and effective choice of norm is
\begin{equation}   
V_{\mathrm{sph}}^{2}(\omega_{\alpha})=\frac{1}{N_{q}}\sum_{\mathbf{q}}\sum_{\alpha\beta} \Big(\frac{\partial g_{\alpha\beta}}{\partial Q_{\alpha\mathbf{q}}}\Big)^{2}\:, \quad
V_{\mathrm{sph}}^{2}(i)=\sum_{s}\sum_{\alpha\beta} \Big(\frac{\partial g_{\alpha\beta}}{\partial X_{is}}\Big)^{2}\:,
\label{norm_def}
\end{equation}
for different phonons and Cartesian degrees of freedom, respectively. The generalization of Eq. \ref{norm_def} to other spin Hamiltonian terms and higher orders of derivatives is straightforward. In conclusion the total Hamiltonian that describe the entangled dynamics of spin and harmonic phonons include the terms
\begin{equation}
(\hat{H}_{s})_{0}\:, \quad \hat{H}_{ph}=\sum_{\alpha \mathbf{q}} \hbar\omega_{\alpha \mathbf{q}} (\hat{n}_{\alpha \mathbf{q}} + \frac{1}{2} )\:, 
\label{hamiltonians0}
\end{equation}
and
\begin{equation}
\hat{H}_{sph}=\sum_{\alpha \mathbf{q}}\Big(\frac{\partial \hat{H}_{s}}{\partial Q_{\alpha\mathbf{q}}}\Big)_{0}\hat{Q}_{\alpha\mathbf{q}}(t)
+\sum_{\alpha\mathbf{q} \ge \beta\mathbf{q'}}\Big(\frac{\partial^{2} \hat{H}_{s}}{\partial Q_{\alpha\mathbf{q}}\partial Q_{\beta\mathbf{q'}}}\Big)_{0}\hat{Q}_{\alpha\mathbf{q}}(t)\hat{Q}_{\beta\mathbf{q'}}(t)+...
\:.
\label{hamiltonians}
\end{equation}
Now that we have formalized the nature of the total Hamiltonian driving the dynamics of spin and phonons, we can proceed to analyze the equations describing the time evolution of our system. In order to do this we introduce the total density operator $\hat{\varrho}$, which spans the total Hilbert space of spin and phonons and evolves in time according to the Liouville equation
\begin{equation}
\frac{d\hat{\varrho}}{dt}=-\frac{i}{\hbar}[\hat{H},\hat{\varrho}]\:.
\label{UN}
\end{equation} 

Expressing $\hat{\varrho}$ in the interaction picture, $\hat{\rho}$, the explicit dynamics is driven by the sole spin-phonon coupling Hamiltonian
\begin{equation}
 \frac{d\hat{\rho}(t)}{dt}=-\frac{i}{\hbar}[\hat{H}_{sph}(t),\hat{\rho}(t)]
\label{eqmot}
\end{equation} 
Here we are interested in solving Eq. \ref{eqmot} perturbatively. To do so we start by writing a formal integration of Eq. \ref{eqmot}
\begin{equation}
 \hat{\rho}(t)=\hat{\rho}(0)-\frac{i}{\hbar}\int_{0}^{t}ds[\hat{H}_{sph}(s),\hat{\rho}(s)]
\label{eqmot2}
\end{equation}
Eq. \ref{eqmot2} can now be inserted back into Eq. \ref{eqmot} as many times as necessary to obtain a series of terms with increasing negative powers of $\hbar$, which will denote the perturbative order. The first two terms of such an expansion read
\begin{equation}
 \frac{d\hat{\rho}(t)}{dt}=-\frac{i}{\hbar}[\hat{H}_{sph}(t),\hat{\rho}(0)]-\frac{1}{\hbar^{2}}\int_{0}^{t}ds[\hat{H}_{sph}(t),[\hat{H}_{sph}(s),\hat{\rho}(s)]]+ O(\hbar^{-2}) \:.
 \label{eqmot3}
\end{equation} 

As the interest is on dynamics of only spin degrees of freedom, it is convenient to operate the Born approximation, which assumes the absence of quantum correlation between spins and phonons at $t=0$, and that the bath is in its thermal equilibrium state. The latter condition is generally fulfilled when the spin system is very dilute in its solid-state host or at high temperature, \textit{i.e.} when the lattice specific heat is much larger than the spin system one and phonon bottle-neck effects are negligible. The Born approximation translates into assuming that $\hat{\rho}(t)=\hat{\rho}^{s}(t)\otimes\hat{\rho}_{B}^{eq}$, where $\hat{\rho}_{B}^{eq}$ is the equilibrium phonons bath density matrix. Under such conditions it is possible to take the trace over bath degrees of freedom and obtain an effective equation for the reduced spin density operator under the influence of a thermalized phonon bath, $\hat{\rho}^{s}(t)=\mathrm{Tr_{B}}(\hat{\rho}(t))$. Eq. \ref{eqmot3} then becomes
\begin{equation}
 \frac{d\hat{\rho}^{s}(t)}{dt}=-\frac{1}{\hbar^{2}}\int_{0}^{t}ds\quad \mathrm{Tr_{B}}[\hat{H}_{sph}(t),[\hat{H}_{sph}(s),\hat{\rho}(s)\otimes\hat{\rho}_{B}]]\:,
 \label{eqmot4}
\end{equation} 
where we have assumed that $\mathrm{Tr_{B}}[\hat{H}_{sph}(t),\hat{\rho}(0)]=0$. Although this condition is not generally fulfilled, it is always possibly to impose it by rescaling the zero-order spin Hamiltonian $(\hat{H}_{s})_0$ as $(\hat{H}_{s})_0 \rightarrow (H_{s})_0+\langle \hat{H}_{sph} \rangle$ and subtract the same quantity from the spin-phonon one, $\hat{H}_{sph} \rightarrow \hat{H}_{sph}-\langle \hat{H}_{sph} \rangle$. With this new definition, $(\hat{H}_{s})_0$ is now corrected for the average thermal effects of phonons and more accurately corresponds to the spin Hamiltonian measured in EPR experiments. Eq. \ref{eqmot4} now only contains the time-dependence of the spin density matrix but it is still an integro-differential equation. A further simplification of this equation is obtained with the \textit{Markov} approximation, which corresponds to assuming that vibrational degrees of freedom relax much faster than the spin system. To do so, the substitution $t'=t-s$ should be done and the t' superior integration boundary can be brought to $+\infty$:
\begin{equation}
 \frac{d\hat{\rho}^{s}(t)}{dt}=-\frac{1}{\hbar^{2}}\int_{0}^{\infty}dt' \mathrm{Tr_{B}}[\hat{H}_{sph}(t),[\hat{H}_{sph}(t-t'),\hat{\rho}^{s}(t)\otimes\hat{\rho}_{B}]]
 \label{eqmot5}
\end{equation} 

\subsection*{Second-order master equation and linear coupling}

In order to further simply Eq. \ref{eqmot2} into something numerically appealing, it is now necessary to make some assumptions on the specific form of $\hat{H}_{sph}$. Following the steps of ref. \cite{lunghi2019phonons}, let us start by considering the sole linear term of the spin-phonon coupling Hamiltonian appearing in Eq. \ref{hamiltonians}.  Introducing the simplified notation $\hat{V}^{\alpha\mathbf{q}}=\Big(\partial \hat{H}_{s} / \partial Q_{\alpha\mathbf{q}}\Big)$ and substituting the definition of $\hat{H}_{sph}$ of Eq. \ref{hamiltonians} into Eq. \ref{eqmot2}, we obtain
\begin{align}
 \frac{d\hat{\rho}^{s}(t)}{dt}=&-\frac{1}{\hbar^{2}}\int_{0}^{\infty}dt'\sum_{\alpha\mathbf{q}}\Big\{ \\
& \Big[ \hat{V}^{\alpha\mathbf{q}}(t)\hat{V}^{\alpha\mathbf{-q}}(t-t')\hat{\rho}^{s}(t) - \hat{V}^{\alpha\mathbf{q}}(t-t')\hat{\rho}^{s}(t)\hat{V}^{\alpha\mathbf{-q}}(t) \Big]  \\ 
& \mathrm{Tr_{B}}\Big(\hat{Q}_{\alpha\mathbf{q}}(t)\hat{Q}_{\alpha\mathbf{-q}}(t-t')\hat{\rho}_{B}^{eq}\Big)- \\
& \Big[ \hat{V}^{\alpha\mathbf{q}}(t)\hat{\rho}^{s}(t)\hat{V}^{\alpha\mathbf{-q}}(t-t') -\hat{\rho}^{s}(t)\hat{V}^{\alpha\mathbf{q}}(t-t')\hat{V}^{\alpha\mathbf{-q}}(t)  \Big]  \\ 
& \mathrm{Tr_{B}}\Big(\hat{Q}_{\alpha\mathbf{q}}(t-t')\hat{Q}_{\alpha\mathbf{-q}}(t)\hat{\rho}_{B}^{eq}\Big)\Big\}
\label{eqmot6}
\end{align}
where the terms $\int_{0}^{\infty}dt'e^{-i\omega_{ij}t'}\mathrm{Tr_{B}}\Big(\hat{Q}_{\alpha\mathbf{q}}(t)\hat{Q}_{\alpha\mathbf{-q}}(t-t')\hat{\rho}_{B}^{eq}\Big)$ are the Fourier transforms of the single phonon bath equilibrium correlation functions and they contain all the information on the bath dynamics, temperature dependence included. Eq. \ref{eqmot3} is derived by taking into account that only $\hat{Q}_{\alpha\mathbf{q}}\hat{Q}_{\alpha\mathbf{-q}}$ products survive the thermal average (\textit{vide infra}). 

Taking the matrix elements of $\hat{\rho}^{s}(t)$ in the eigenket basis of $\hat{H}_{s}|a\rangle=E_{a}|a\rangle$, it is possible to obtain:
\begin{align}
 \frac{d\rho^{s}_{ab}(t)}{dt}=&-\frac{1}{\hbar^{2}}\int_{0}^{\infty}dt'\sum_{\alpha\mathbf{q}}\sum_{cd}\Big\{ \\
&\Big[ V^{\alpha\mathbf{q}}_{ac}(t)V^{\alpha\mathbf{-q}}_{cd}(t-t')\rho^{s}_{db}(t)- V^{\alpha\mathbf{q}}_{ac}(t-t')\rho^{s}_{cd}(t)V^{\alpha\mathbf{-q}}_{db}(t) \Big] \\
& \mathrm{Tr_{B}}\Big(\hat{Q}_{\alpha\mathbf{q}}(t)\hat{Q}_{\alpha\mathbf{-q}}(t-t')\hat{\rho}_{B}^{eq}\Big)- \\
&\Big[ 
V^{\alpha\mathbf{q}}_{ac}(t)\rho^{s}_{cd}(t)V^{\alpha\mathbf{-q}}_{db}(t-t') -\rho^{s}_{ac}(t)V^{\alpha\mathbf{q}}_{cd}(t-t')V^{\alpha\mathbf{-q}}_{db}(t)  \Big] \\
& \mathrm{Tr_{B}}\Big(\hat{Q}_{\alpha\mathbf{q}}(t-t')\hat{Q}_{\alpha\mathbf{-q}}(t)\hat{\rho}_{B}^{eq}\Big)\Big\}
\label{eqmot7}
\end{align}

Making the time dependency of spin degrees of freedom explicit, $V_{ab}(t)=\langle a |e^{iH_{0}t}Ve^{-iH_{0}t}| b \rangle=e^{i\omega_{ab}t}V_{ab}$, and doing some algebra, it is possible to arrive at the one-phonon non-secular Redfield equations\cite{lunghi2019phonons}
\begin{equation}
\frac{d\rho^{s}_{ab}(t)}{dt}=\sum_{cd}e^{i(\omega_{ac}+\omega_{db})t}R^{1-ph}_{ab,cd}\rho^{s}_{cd}(t)\:,
\label{redfield}
\end{equation}
where 
\begin{align}
R^{1-ph}_{ab,cd} = -\frac{1}{\hbar^{2}}\sum_{\alpha\mathbf{q}}\Big\{ 
& \sum_{j} \delta_{bd}V^{\alpha\mathbf{q}}_{aj}V^{\alpha\mathbf{-q}}_{jc}\int_{0}^{\infty}dt'e^{-i\omega_{jc}t'}\mathrm{Tr_{B}}\Big(\hat{Q}_{\alpha\mathbf{q}}(t)\hat{Q}_{\alpha\mathbf{-q}}(t-t')\hat{\rho}_{B}^{eq}\Big) \\
&-V^{\alpha\mathbf{q}}_{ac}V^{\alpha\mathbf{-q}}_{db}\int_{0}^{\infty}dt' e^{-i\omega_{ac}t'}\mathrm{Tr_{B}}\Big(\hat{Q}_{\alpha\mathbf{q}}(t)\hat{Q}_{\alpha\mathbf{-q}}(t-t')\hat{\rho}_{B}^{eq}\Big) \\
&-V^{\alpha\mathbf{q}}_{ac}V^{\alpha\mathbf{-q}}_{db}\int_{0}^{\infty}dt'e^{-i\omega_{db}t'} \mathrm{Tr_{B}}\Big(\hat{Q}_{\alpha\mathbf{q}}(t-t')\hat{Q}_{\alpha\mathbf{-q}}(t)\hat{\rho}_{B}^{eq}\Big) \\
&+\sum_{j}\delta_{ca}V^{\alpha\mathbf{q}}_{dj}V^{\alpha\mathbf{-q}}_{jb}\int_{0}^{\infty}dt'e^{-i\omega_{dj}t'}\mathrm{Tr_{B}}\Big(\hat{Q}_{\alpha\mathbf{q}}(t-t')\hat{Q}_{\alpha\mathbf{-q}}(t)\hat{\rho}_{B}^{eq}\Big)\Big\}
\end{align}
The secular version of \ref{redfield} is obtained by setting to zero the terms that do not verify the equation $(\omega_{ac}+\omega_{db})=0$, \textit{i.e.} the fast oscillatory terms that would get averaged to zero on the time-scale of spin relaxation. The secular approximation is therefore valid only in the case that spin relaxation timescales are longer than the natural oscillations of the spin system, which is usually the case. \\

Finally, we explicit the form of the Fourier transform of the bath correlation functions, starting from substituting the relation between the normal modes of vibration $\hat{Q}_{\alpha\mathbf{q}}$ and the creation (annihilation) operators, and their time evolution
\begin{equation}
e^{i\hat{H}_{ph}t}\hat{a}^{\dag}_{\alpha\mathbf{q}}e^{-i\hat{H}_{ph}t}=e^{i\omega_{\alpha\mathbf{q}}t}\hat{a}^{\dag}_{\alpha\mathbf{q}}\:, \quad
e^{i\hat{H}_{ph}t}\hat{a}_{\alpha\mathbf{q}}e^{-i\hat{H}_{ph}t}=e^{-i\omega_{\alpha\mathbf{q}}t}\hat{a}_{\alpha\mathbf{q}}  \:,
\label{harmonic}
\end{equation}
leading to
\begin{align}
&\int_{0}^{\infty}dt'e^{-i\omega_{ij}t'}\mathrm{Tr_{B}}\Big(\hat{Q}_{\alpha\mathbf{q}}(t)\hat{Q}_{\alpha\mathbf{-q}}(t-t')\hat{\rho}_{B}^{eq}\Big)= \\
& \frac{1}{2}\int_{0}^{\infty}dt'e^{-i\omega_{ij}t'}\mathrm{Tr_{B}}\Big[ e^{i\omega_{\alpha\mathbf{q}}t'}\hat{a}^{\dag}_{\alpha\mathbf{q}}\hat{a}_{\alpha\mathbf{q}} + e^{-i\omega_{\alpha\mathbf{q}}t'}\hat{a}_{\alpha\mathbf{-q}}\hat{a}^{\dag}_{\alpha\mathbf{-q}} \Big]=\\
&\frac{1}{2}\int_{0}^{\infty}dt'e^{-i(\omega_{ij}-\omega_{\alpha\mathbf{q}})t'}\hat{n}_{\alpha\mathbf{q}} + \int_{0}^{\infty}dt'e^{-i(\omega_{ij}+\omega_{\alpha\mathbf{q}})t'}(\hat{n}_{\alpha\mathbf{q}}+1)\:,
\end{align}
where we have used the property $\omega_{\alpha\mathbf{q}}=\omega_{\alpha\mathbf{-q}}$ and the definition of the average phonon number $\bar{n}_{\alpha\mathbf{q}}=\mathrm{Tr_{B}}\Big[\hat{a}^{\dag}_{\alpha\mathbf{q}}\hat{a}_{\alpha\mathbf{q}}\Big]$. It should now be clear that any other combination of indexes $\alpha\mathbf{q},\beta\mathbf{q'}$ in $\hat{Q}_{\alpha\mathbf{q}}\hat{Q}_{\beta\mathbf{q'}}$ would lead to a vanishing trace over the bath degrees of freedom.

Using the definition $\pi\delta(\omega)=\int_{0}^{\infty}dt'e^{-i\omega t'}$ we can rewrite the Redfiled operator $R^{1-ph}_{ab,cd}$ as
\begin{align}
R^{1-ph}_{ab,cd} =-\frac{\pi}{2\hbar^{2}}\sum_{\alpha\mathbf{q}}\Big\{&\sum_{j} \delta_{bd}V^{\alpha\mathbf{q}}_{aj}V^{\alpha\mathbf{-q}}_{jc} G^{1-ph}(\omega_{jc},\omega_{\alpha\mathbf{q}})-V^{\alpha\mathbf{q}}_{ac}V^{\alpha\mathbf{-q}}_{db}G^{1-ph}(\omega_{ac},\omega_{\alpha\mathbf{q}})\\
&-V^{\alpha\mathbf{q}}_{ac}V^{\alpha\mathbf{-q}}_{db}G^{1-ph}(\omega_{bd},\omega_{\alpha\mathbf{q}})+\sum_{j}\delta_{ca}V^{\alpha\mathbf{q}}_{dj}V^{\alpha\mathbf{-q}}_{jb}G^{1-ph}(\omega_{jd},\omega_{\alpha\mathbf{q}})\Big\}\:,
\label{Red21}
\end{align}

where
\begin{equation}
 G^{1-ph}(\omega_{ij},\omega_{\alpha\mathbf{q}})= \delta(\omega_{ij}-\omega_{\alpha\mathbf{q}})\bar{n}_{\alpha\mathbf{q}} + \delta(\omega_{ij}+\omega_{\alpha\mathbf{q}})(\bar{n}_{\alpha\mathbf{q}}+1)
\end{equation}

Eqs. \ref{hamiltonians} and \ref{harmonic} assume the vibrational dynamics to be perfectly harmonic and undamped. Such assumption is however in conflict with the Markov approximation, which requires the bath correlation function to decay to zero before the spin had time to significantly change its status. In order to reconcile this aspect of the theory it is only necessary to recall that in real systems the harmonic approximation is never perfectly fulfilled due to the presence of anharmonic terms in the real phonon Hamiltonian $\hat{H}_{ph}$. In a perturbative regime the presence of anharmonic terms enable phonon-phonon scattering events which limits the lifetime of phonons  and lead to an exponential decay profile of the phonons correlations functions with rate $\Delta^{-1}$. With this in mind we can modify the function $G^{1-ph}$ as
\begin{equation}
 G^{1-ph}(\omega_{ij},\omega_{\alpha\mathbf{q}})= \frac{1}{\pi} \Big[ \frac{\Delta_{\alpha\mathbf{q}}}{\Delta_{\alpha\mathbf{q}}^{2}+(\omega_{ij}-\omega_{\alpha\mathbf{q}})^{2}}\bar{n}_{\alpha\mathbf{q}} + \frac{\Delta_{\alpha\mathbf{q}}}{\Delta_{\alpha\mathbf{q}}^{2}+(\omega_{ij}+\omega_{\alpha\mathbf{q}})^{2}}(\bar{n}_{\alpha\mathbf{q}}+1)\Big]
\end{equation}

\subsection*{Second-order master equation and quadratic coupling}

Let us now consider the quadratic term of the spin-phonon coupling Hamiltonian appearing in Eq. \ref{hamiltonians} and follow the same steps as for the linear term as presented in ref. \cite{lunghi2020limit}. The matrix elements of the spin-phonon coupling Hamiltonain depends on two sets of phonon indexes, as we are now dealing with a two-phonon operators:  $\hat{V}^{\alpha\mathbf{q}\beta\mathbf{q'}}=\Big(\partial^{2} \hat{H}_{s} / \partial Q_{\alpha\mathbf{q}}\partial Q_{\beta\mathbf{q'}}\Big)$. Similarly to what observed for the linear term of the spin-phonon coupling Hamiltonian, we now obtain the following expression for the time evolution of of the reduced spin density matrix under the Born-Markov approximation
\begin{align}
 \frac{d\hat{\rho}^\mathrm{s}(t)}{dt}=&-\frac{1}{4\hbar^{2}}\int_{0}^{\infty}dt'2\sum_{\alpha\mathbf{q}}\sum_{\beta\mathbf{q'}}\Big\{ \\
& \Big[ \hat{V}^{\alpha\mathbf{q}\beta\mathbf{q'}}(t)\hat{V}^{\alpha\mathbf{-q}\beta\mathbf{-q'}}(t-t')\hat{\rho}^{s}(t) - \hat{V}^{\alpha\mathbf{q}\beta\mathbf{q'}}(t-t')\hat{\rho}^{s}(t)\hat{V}^{\alpha\mathbf{-q}\beta\mathbf{-q'}}(t) \Big] \\
&  \mathrm{Tr_{B}}\Big(\hat{Q}_{\alpha\mathbf{q}}(t)\hat{Q}_{\beta\mathbf{q'}}(t)\hat{Q}_{\alpha\mathbf{-q}}(t-t')\hat{Q}_{\beta\mathbf{-q'}}(t-t')\hat{\rho}^{B}_\mathrm{eq}\Big)- \\
& \Big[ \hat{V}^{\alpha\mathbf{q}\beta\mathbf{q'}}(t)\hat{\rho}^{s}(t)\hat{V}^{\alpha\mathbf{-q}\beta\mathbf{-q'}}(t-t') -\hat{\rho}^{s}(t)\hat{V}^{\alpha\mathbf{q}\beta\mathbf{q'}}(t-t')\hat{V}^{\alpha\mathbf{-q}\beta\mathbf{-q'}}(t)  \Big] \\
& \mathrm{Tr_{B}}\Big(\hat{Q}_{\alpha\mathbf{q}}(t-t')\hat{Q}_{\beta\mathbf{q'}}(t-t')\hat{Q}_{\alpha\mathbf{-q}}(t)\hat{Q}_{\beta\mathbf{-q'}}(t)\hat{\rho}^{B}_\mathrm{eq}\Big)\Big\} \:,
\label{eqmot2_1}
\end{align}
where the terms $\int_{0}^{\infty}dt'e^{-i\omega_{ij}t'}\mathrm{Tr_{B}}\Big(\hat{Q}_{\alpha \mathbf{q}}(t)\hat{Q}_{\beta \mathbf{q'}}(t)\hat{Q}_{\alpha\mathbf{q}}(t-t')\hat{Q}_{\beta \mathbf{q'}}(t-t')\hat{\rho}^{B}_\mathrm{eq}\Big)$ are now the Fourier transforms of the two-phonon bath equilibrium correlation functions. Eq. (\ref{eqmot2_1}) was derived by taking into account that only products of terms such as $\hat{Q}_{\alpha\mathbf{q}}(t)\hat{Q}_{\alpha\mathbf{-q}}(t-t')$ would survive the thermal average and corresponds to transitions at non-zero energy. The factor 2 multiplying the summations on the indexes $\alpha \mathbf{q} \beta \mathbf{q'}$ accounts for the multiplicity of terms coming from developing the full product $\hat{Q}_{\alpha \mathbf{q}}(t)\hat{Q}_{\beta \mathbf{q'}}(t)\hat{Q}_{\gamma \mathbf{q''}}(t-t')\hat{Q}_{\delta \mathbf{q'''}}(t-t')$.

By taking the matrix elements of $\hat{\rho}_\mathrm{s}(t)$ in the eigenket basis of $\hat{H}_\mathrm{s}$ we obtain:

\begin{align}
 \frac{d\rho^\mathrm{s}_{ab}(t)}{dt}=&-\frac{1}{2\hbar^{2}}\int_{0}^{\infty}dt'\sum_{\alpha\mathbf{q}}\sum_{\beta\mathbf{q'}}\sum_{cd}\Big\{ \\
&\Big[ V^{\alpha\mathbf{q}\beta\mathbf{q'}}_{ac}(t)V^{\alpha\mathbf{-q}\beta\mathbf{-q'}}_{cd}(t-t')\rho^\mathrm{s}_{db}(t)-V^{\alpha\mathbf{q}\beta\mathbf{q'}}_{ac}(t-t')\rho^\mathrm{s}_{cd}(t)V^{\alpha\mathbf{-q}\beta\mathbf{-q'}}_{db}(t) \Big] \\
&  \mathrm{Tr_{B}}\Big(\hat{Q}_{\alpha\mathbf{q}}(t)\hat{Q}_{\beta\mathbf{q'}}(t)\hat{Q}_{\alpha\mathbf{-q}}(t-t')\hat{Q}_{\beta\mathbf{-q'}}(t-t')\hat{\rho}^{B}_\mathrm{eq}\Big)- \\
&\Big[ V^{\alpha\mathbf{q}\beta\mathbf{q'}}_{ac}(t)\rho^\mathrm{s}_{cd}(t)V^{\alpha\mathbf{-q}\beta\mathbf{-q'}}_{db}(t-t')  -\rho^\mathrm{s}_{ac}(t)V^{v\alpha\mathbf{q}\beta\mathbf{q'}}_{cd}(t-t')V^{\alpha\mathbf{-q}\beta\mathbf{-q'}}_{db}(t) \Big] \\
&  \mathrm{Tr_{B}}\Big(\hat{Q}_{\alpha\mathbf{q}}(t-t')\hat{Q}_{\beta\mathbf{q'}}(t-t')\hat{Q}_{\alpha\mathbf{-q}}(t)\hat{Q}_{\beta\mathbf{-q'}}(t)\hat{\rho}^{B}_\mathrm{eq}\Big)\Big\}
\label{eqmot2_2}
\end{align}

Making the time dependencies of spin degrees of freedom explicit, $V_{ab}(t)=\langle a |e^{iH_{0}t}Ve^{-iH_{0}t}| b \rangle=e^{i\omega_{ab}t}V_{ab}$, and doing some algebra, it is possible to arrive at the two-phonon contribution of the non-secular Redfield equations for two-phonon processes\cite{lunghi2020limit}
\begin{equation}
\frac{d\rho^\mathrm{s}_{ab}(t)}{dt}=\sum_{cd}e^{i(\omega_{ac}+\omega_{db})t}R^{2-\mathrm{ph}}_{ab,cd}\rho^\mathrm{s}_{cd}(t)\:,
\label{redfield2}
\end{equation}
where 
\begin{align}
& R^{2-\mathrm{ph}}_{ab,cd} = -\frac{1}{2\hbar^{2}}\sum_{\alpha\mathbf{q}}\sum_{\beta\mathbf{q'}} \Big\{ \\
& \sum_{j} \delta_{bd}V^{\alpha\mathbf{q}\beta\mathbf{q'}}_{aj}V^{\alpha\mathbf{-q}\beta\mathbf{-q'}}_{jc} \int_{0}^{\infty}dt'e^{-i\omega_{jc}t'} \mathrm{Tr_{B}}\Big(\hat{Q}_{\alpha\mathbf{q}}(t)\hat{Q}_{\beta\mathbf{q'}}(t)\hat{Q}_{\alpha\mathbf{-q}}(t-t')\hat{Q}_{\beta\mathbf{-q'}}(t-t')\hat{\rho}^{B}_\mathrm{eq}\Big) \\
& - V^{\alpha\mathbf{q}\beta\mathbf{q'}}_{ac}V^{\alpha\mathbf{-q}\beta\mathbf{-q'}}_{db}  \int_{0}^{\infty}dt'e^{-i\omega_{ac}t'}\mathrm{Tr_{B}}\Big(\hat{Q}_{\alpha\mathbf{q}}(t)\hat{Q}_{\beta\mathbf{q'}}(t)\hat{Q}_{\alpha\mathbf{-q}}(t-t')\hat{Q}_{\beta\mathbf{-q'}}(t-t')\hat{\rho}^{B}_\mathrm{eq}\Big) \\
& -V^{\alpha\mathbf{q}\beta\mathbf{q'}}_{ac}V^{\alpha\mathbf{-q}\beta\mathbf{-q'}}_{db} \int_{0}^{\infty}dt'e^{-i\omega_{db}t'} \mathrm{Tr_{B}}\Big(\hat{Q}_{\alpha\mathbf{q}}(t-t')\hat{Q}_{\beta\mathbf{q'}}(t-t')\hat{Q}_{\alpha\mathbf{-q}}(t)\hat{Q}_{\beta\mathbf{-q'}}(t)\hat{\rho}^{B}_\mathrm{eq}\Big) + \\
&\sum_{j}\delta_{ca}V^{\alpha\mathbf{q}\beta\mathbf{q'}}_{dj}V^{\alpha\mathbf{-q}\beta\mathbf{-q'}}_{jb}\int_{0}^{\infty}dt'e^{-i\omega_{dj}t'}\mathrm{Tr_{B}}\Big(\hat{Q}_{\alpha\mathbf{q}}(t-t')\hat{Q}_{\beta\mathbf{q'}}(t-t')\hat{Q}_{\alpha\mathbf{-q}}(t)\hat{Q}_{\beta\mathbf{-q'}}(t)\hat{\rho}^{B}_\mathrm{eq}\Big) \Big\} \:.
\end{align}
In perfect analogy to the linear coupling case, the secular version of \ref{redfield2} is obtained by setting to zero the terms that do not verify the equation $(\omega_{ac}+\omega_{db})=0$. \\

Finally, recalling the definitions of Eq. \ref{harmonic}, we obtain
\begin{align}
&\int_{0}^{\infty}dt'e^{-i\omega_{ij}t'}\mathrm{Tr_{B}}\Big(\hat{Q}_{\alpha\mathbf{q}}(t)\hat{Q}_{\beta\mathbf{q'}}(t)\hat{Q}_{\alpha\mathbf{-q}}(t-t')\hat{Q}_{\beta\mathbf{-q'}}(t-t')\hat{\rho}^{B}_\mathrm{eq}\Big)= \frac{1}{4}\int_{0}^{\infty}dt'e^{-i\omega_{ij}t'} \\ & \mathrm{Tr_{B}}\Big[  e^{i\omega_{\alpha\mathbf{q}}t'}e^{i\omega_{\beta\mathbf{q'}}t'}\hat{a}^{\dag}_{\alpha\mathbf{q}}\hat{a}_{\alpha\mathbf{q}}\hat{a}^{\dag}_{\beta\mathbf{q'}}\hat{a}_{\beta\mathbf{q'}} + e^{-i\omega_{\alpha\mathbf{q}}t'}e^{-i\omega_{\beta\mathbf{q'}}t'}\hat{a}_{\alpha\mathbf{-q}}\hat{a}^{\dag}_{\alpha\mathbf{-q}}\hat{a}_{\beta\mathbf{-q'}}\hat{a}^{\dag}_{\beta\mathbf{-q'}} + \\
&e^{i\omega_{\alpha\mathbf{q}}t'}e^{-i\omega_{\beta\mathbf{q'}}t'}\hat{a}^{\dag}_{\alpha\mathbf{q}}\hat{a}_{\alpha\mathbf{q}}\hat{a}_{\beta\mathbf{-q'}}\hat{a}^{\dag}_{\beta\mathbf{-q'}} + e^{-i\omega_{\alpha\mathbf{q}}t'}e^{i\omega_{\beta\mathbf{q'}}t'}\hat{a}_{\alpha\mathbf{-q}}\hat{a}^{\dag}_{\alpha\mathbf{-q}}\hat{a}^{\dag}_{\beta\mathbf{q'}}\hat{a}_{\beta\mathbf{q'}} \Big]=\\
\frac{1}{4}&\int_{0}^{\infty}dt'e^{-i(\omega_{ij}-\omega_{\beta\mathbf{q'}}-\omega_{\alpha\mathbf{q}})t'}\hat{n}_{\alpha\mathbf{q}}\hat{n}_{\beta\mathbf{q'}} + \frac{1}{4}\int_{0}^{\infty}dt'e^{-i(\omega_{ij}+\omega_{\alpha\mathbf{q}}+\omega_{\beta\mathbf{q'}})t'}(\hat{n}_{\alpha\mathbf{q}}+1)(\hat{n}_{\beta\mathbf{q'}}+1)+ \\
\frac{1}{4} & \int_{0}^{\infty}dt'e^{-i(\omega_{ij}-\omega_{\alpha\mathbf{q}}t'+\omega_{\beta\mathbf{q'}})t'}\hat{n}_{\alpha\mathbf{q}}(\hat{n}_{\beta\mathbf{q'}}+1) +\frac{1}{4}\int_{0}^{\infty}dt'e^{-i(\omega_{ij}+\omega_{\alpha\mathbf{q}}-\omega_{\beta\mathbf{q'}})t'}(\hat{n}_{\alpha\mathbf{q}}+1)\hat{n}_{\beta\mathbf{q'}}\:, 
\end{align}
Using once again the definition $\pi\delta(\omega)=\int_{0}^{\infty}dt'e^{-i\omega t'}$ we can rewrite the Redfiled operator $R^{2-\mathrm{ph}}_{ab,cd}$ as
\begin{align}
R^{2-\mathrm{ph}}_{ab,cd} =-\frac{\pi}{4\hbar^{2}}\sum_{\alpha\mathbf{q} \ge \beta\mathbf{q'}} \Big\{ & \sum_{j} \delta_{bd}V^{\alpha\mathbf{q}\beta\mathbf{q'}}_{aj}V^{\alpha\mathbf{-q}\beta\mathbf{-q'}}_{jc} G^{2-\mathrm{ph}}(\omega_{jc},\omega_{\alpha\mathbf{q}},\omega_{\beta\mathbf{q'}}) \\ 
& - V^{\alpha\mathbf{q}\beta\mathbf{q'}}_{ac}V^{\alpha\mathbf{-q}\beta\mathbf{-q'}}_{db}G^{2-\mathrm{ph}}(\omega_{ac},\omega_{\alpha\mathbf{q}},\omega_{\beta\mathbf{q'}})\\
&- V^{\alpha\mathbf{q}\beta\mathbf{q'}}_{ac}V^{\alpha\mathbf{-q}\beta\mathbf{-q'}}_{db}G(\omega_{bd
},\omega_{\alpha\mathbf{q}},\omega_{\beta\mathbf{q'}}) \\ &+\sum_{j}\delta_{ca}V^{\alpha\mathbf{q}\beta\mathbf{q'}}_{dj}V^{\alpha\mathbf{-q}\beta\mathbf{-q'}}_{jb}G(\omega_{jd},\omega_{\alpha\mathbf{q}},\omega_{\beta\mathbf{q'}})\Big\}\:,
\label{Red22}
\end{align}
where
\begin{align}
 G^{2-\mathrm{ph}}(\omega_{ij},\omega_{\alpha\mathbf{q}},\omega_{\beta\mathbf{q'}})= & 
 \delta(\omega_{ij}-\omega_{\alpha\mathbf{q}}-\omega_{\beta\mathbf{q'}})\bar{n}_{\alpha\mathbf{q}}\bar{n}_{\beta\mathbf{q'}} + \\ & \delta(\omega_{ij}+\omega_{\alpha\mathbf{q}}+\omega_{\beta\mathbf{q'}})(\bar{n}_{\alpha\mathbf{q}}+1)(\bar{n}_{\beta\mathbf{q'}}+1) + \\ 
 & \delta(\omega_{ij}+\omega_{\alpha\mathbf{q}}-\omega_{\beta\mathbf{q'}})(\bar{n}_{\alpha\mathbf{q}}+1)\bar{n}_{\beta\mathbf{q'}} + \\ & \delta(\omega_{ij}-\omega_{\alpha\mathbf{q}}+\omega_{\beta\mathbf{q'}})\bar{n}_{\alpha\mathbf{q}}(\bar{n}_{\beta\mathbf{q'}}+1)\:,
\end{align}
or alternatively in the presence of anharmonic phonon-phonon relaxation,
\begin{align}
 G^{2-\mathrm{ph}}(\omega_{ij},\omega_{\alpha\mathbf{q}},\omega_{\beta\mathbf{q'}})= \frac{1}{\pi} \Big[ & 
 \frac{\Delta_{\alpha\mathbf{q}\beta\mathbf{q'}}}{\Delta_{\alpha\mathbf{q}\beta\mathbf{q'}}^{2}+(\omega_{ij}-\omega_{\alpha\mathbf{q}}-\omega_{\beta\mathbf{q'}})^{2}} \bar{n}_{\alpha\mathbf{q}}\bar{n}_{\beta\mathbf{q'}} + \\ &  \frac{\Delta_{\alpha\mathbf{q}\beta\mathbf{q'}}}{\Delta_{\alpha\mathbf{q}\beta\mathbf{q'}}^{2}+(\omega_{ij}+\omega_{\alpha\mathbf{q}}+\omega_{\beta\mathbf{q'}})^{2}} (\bar{n}_{\alpha\mathbf{q}}+1)(\bar{n}_{\beta\mathbf{q'}}+1) + \\ 
& \frac{\Delta_{\alpha\mathbf{q}\beta\mathbf{q'}}}{\Delta_{\alpha\mathbf{q}\beta\mathbf{q'}}^{2}+(\omega_{ij}-\omega_{\alpha\mathbf{q}}+\omega_{\beta\mathbf{q'}})^{2}} \bar{n}_{\alpha\mathbf{q}}(\bar{n}_{\beta\mathbf{q'}}+1) + \\ & \frac{\Delta_{\alpha\mathbf{q}\beta\mathbf{q'}}}{\Delta_{\alpha\mathbf{q}\beta\mathbf{q'}}^{2}+(\omega_{ij}+\omega_{\alpha\mathbf{q}}-\omega_{\beta\mathbf{q'}})^{2}} (\bar{n}_{\alpha\mathbf{q}}+1)\bar{n}_{\beta\mathbf{q'}} \Big] \:,
\end{align}
where $\Delta_{\alpha\mathbf{q}\beta\mathbf{q'}}=\Delta_{\alpha\mathbf{q}}+\Delta_{\beta\mathbf{q'}}$. \\

\subsection*{Fourth-order master equation and linear coupling}

Eq. \ref{eqmot5} was obtained by truncating the expansion of $\hat{\rho}$ to the second order in $\hbar$, but higher order terms might in principle contribute to the dynamics of the density matrix. In this section we will explore the role of the next perturbative order\cite{lunghi2020multiple,lunghi2021towards}. Iterating the procedure used to derive Eq. \ref{eqmot5}, we obtain the term 
\begin{equation}
\frac{d\hat{\rho}^{s}(t)}{dt}=\frac{1}{\hbar^{4}}\int_{0}^{t}ds\int_{0}^{s}ds'\int_{0}^{s'}ds''\quad \mathrm{Tr_{B}}[\hat{H}_{sph}(t),[\hat{H}_{sph}(s),[\hat{H}_{sph}(s'),[\hat{H}_{sph}(s''),\hat{\rho}(s'')]]]] \:.
\label{master4coupl1}
\end{equation}
Although Eq. \ref{master4coupl1} could in principles be simplified following the same prescription adopted for the second-order master equation, this has yet to be attempted for the case of spin-phonon processes. However, it is possible to derive an expression for the time evolution of the sole diagonal elements of the spin density matrix by following another strategy. Instead of working with the entire density matrix, we will here adopt the strategy commonly used for the derivation of the Fermi golden rule and based on time-dependent perturbation theory.  \\

Firstly, let us briefly review the derivation of the Fermi golden rule and demonstrate that it is equivalent to the secular Redfield equation of Eq. \ref{Red21}. Being $\hat{U}_{I}(t,0)$ the propagator from time $t_0=0$ to time $t$ in the interaction picture, time-dependent perturbation theory leads to an expansion for the transition amplitude between two different eigenstates of the zero-order Hamiltonian (the spin Hamiltonian in our case)
\begin{equation}
    C_{ij}(t)=\langle i | U_I(t,0) | j \rangle =C_{ij}^{(0)}+C_{ij}^{(1)}(t)+C_{ij}^{(2)}(t)+... \:,
\end{equation}
where
\begin{align}
    C_{ij}^{(0)}(t)  = & \quad const. \:, \\
    C_{ij}^{(1)}(t)  = & -\frac{i}{\hbar}\int_{0}^{t} dt' e^{i \omega_{ij} t'} \tilde{V}_{ij}(t') \:, \label{c1def} \\
    C_{ij}^{(2)}(t)  = & \left(-\frac{i}{\hbar}\right)^{2}\sum_{m}\int_{0}^{t} dt'\int_{0}^{t'} dt'' e^{i \omega_{im} t'}e^{i \omega_{mj} t''} \tilde{V}_{im}(t')\tilde{V}_{mj}(t'') \:, \label{c2def}
\end{align}
and
\begin{equation}
    \tilde{V}_{ij}(t)= \langle i | \sum_{\alpha} \left( \frac{\partial \hat{H}_s}{\partial Q_{\alpha\mathbf{q}}} \right) \hat{Q}_{\alpha\mathbf{q}}(t) | j \rangle \:.
    \label{vsphdef}
\end{equation}
In order to simplify the notation without any loss of generality, we will drop the index over the Brillouin vector $\mathbf{q}$ for the rest of this section. \\

Starting from the definition of $C_{ij}(t)$ and assuming that the system is in the state $j$ at the initial time, the probability that the system will be found in the state $|i\rangle$ at time $t$ is simply $|C_{ij}(t)|^{2}$. Here, we are however interested in deriving an equation describing the time variation of such a probability at long times with respect to the natural dynamics of the spin system (see Markov and secular approximations in the previous sections) and therefore we define the transition probability $W$ from state $j$ to state $i$ as 
\begin{equation}
    W_{ij}= \frac{d}{dt} \quad \lim_{ t \to \infty} \quad |C_{ij}(t)|^{2} \:.
\end{equation}

Let us now insert the definition of Eq. \ref{vsphdef} into Eq. \ref{c1def} for $C_{ij}^{(1)}(t)$. Firstly, we need to perform the integration with respect to $t'$, leading to 
\begin{equation}
    C_{ij}^{(1)}(t)  =  -\frac{i}{\sqrt{2}\hbar} \sum_{\alpha} V^{\alpha}_{ij} \left[  \frac{( e^{i (\omega_{ij}+\omega_{\alpha} )t} - 1 )}{i(\omega_{ij}+\omega_{\alpha} ) } a^{\dag}_{\alpha} + \frac{( e^{i (\omega_{ij}-\omega_{\alpha} )t} -1)}{i(\omega_{ij}-\omega_{\alpha} ) } a_{\alpha}  \right] \:.
    \label{c1_eq1}
\end{equation}
Taking the square of Eq. \ref{c1_eq1} and retaining only the terms in $a^{\dag}_{\alpha}$ and $a_{\alpha}$ that conserve the number of phonons, after a little algebra we obtain 
\begin{equation}
    |C_{ij}^{(1)}(t)|^{2}  =  \frac{1}{2\hbar^{2}} \sum_{\alpha} |V^{\alpha}_{ij}|^{2} \left[  \frac{ 4 sin^{2}((\omega_{ij}+\omega_{\alpha}/2 )t )}{(\omega_{ij}+\omega_{\alpha} ) } a_{\alpha}a^{\dag}_{\alpha} + \frac{ 4 sin^{2}((\omega_{ij}-\omega_{\alpha}/2 )t )}{i(\omega_{ij}-\omega_{\alpha} ) } a^{\dag}_{\alpha}a_{\alpha}  \right] \:
\end{equation}
We are now in the position to take the limit for $t \to \infty$, which brings us to
\begin{equation}
    \lim_{t \to \infty}| C^{(1)}_{ij}(t)|^{2}  =  \frac{\pi}{\hbar^{2}} \sum_{\alpha} |V^{\alpha}_{ij}|^{2} \left[  \delta(\omega_{ij}+\omega_{\alpha}) a_{\alpha}a^{\dag}_{\alpha} + \delta(\omega_{ij}-\omega_{\alpha}) a^{\dag}_{\alpha}a_{\alpha}  \right] t \:.
\end{equation}
Finally we can take the derivative with respect to $t$ and obtain an expression equivalent to the canonical Fermi golden rule
\begin{equation}
    W_{ij} =  \frac{\pi}{\hbar^{2}} \sum_{\alpha} |V^{\alpha}_{ij}|^{2} \left[  \delta(\omega_{ij}+\omega_{\alpha}) a_{\alpha}a^{\dag}_{\alpha} + \delta(\omega_{ij}-\omega_{\alpha}) a^{\dag}_{\alpha}a_{\alpha}  \right] \:.
\end{equation}
Since we are here interested in a thermal bath at the equilibrium we can take one further step by averaging $W_{ij}$ with respect to the phonon variables at their thermal equilibrium by multiplying by $\rho_{B}^{eq}$ and tracing over the the phonon states $\mathrm{Tr_{B}}\{\cdot\}$. This leads to the final expression for one-phonon transitions among spin states
\begin{equation}
    W^{1-ph}_{ij} = \frac{\pi}{\hbar^{2}} \sum_{\alpha} |V^{\alpha}_{ij}|^{2} \left[  \delta(\omega_{ij}+\omega_{\alpha}) (\bar{n}_{\alpha}+1) + \delta(\omega_{ij}-\omega_{\alpha}) \bar{n}_{\alpha}  \right] =  \frac{\pi}{\hbar^{2}} \sum_{\alpha} |V^{\alpha}_{ij}|^{2} G^{1-ph}(\omega_{ij},\omega_{\alpha}) \:.
    \label{c1_eq2}
\end{equation}
In order to complete our mapping between the secular Redfield equations and the Fermi golden rule, let us now focus on the dynamics of the population terms in Eq. \ref{redfield}. Taking Eq. \ref{redfield} and setting to zero all the matrix elements except $R^{1-ph}_{aa,bb}$ we are left with a master matrix with transition rates
\begin{align}
R^{1-ph}_{aa,bb} =&-\frac{\pi}{2\hbar^{2}}\sum_{\alpha\mathbf{q}}\Big\{\sum_{j} \delta_{ab}V^{\alpha\mathbf{q}}_{aj}V^{\alpha\mathbf{-q}}_{jb} G^{1-ph}(\omega_{jb},\omega_{\alpha\mathbf{q}})- V^{\alpha\mathbf{q}}_{ab}V^{\alpha\mathbf{-q}}_{ba}G^{1-ph}(\omega_{ab},\omega_{\alpha\mathbf{q}}) \\
&-V^{\alpha\mathbf{q}}_{ab}V^{\alpha\mathbf{-q}}_{ba}G^{1-ph}(\omega_{ab},\omega_{\alpha\mathbf{q}})+\sum_{j}\delta_{ba}V^{\alpha\mathbf{q}}_{bj}V^{\alpha\mathbf{-q}}_{ja}G^{1-ph}(\omega_{jb},\omega_{\alpha\mathbf{q}})\Big\} \\
 = & -\frac{\pi}{\hbar^{2}}\sum_{\alpha\mathbf{q}}\Big\{\sum_{j} \delta_{ab}V^{\alpha\mathbf{q}}_{aj}V^{\alpha\mathbf{-q}}_{jb} G^{1-ph}(\omega_{jb},\omega_{\alpha\mathbf{q}})-V^{\alpha\mathbf{q}}_{ab}V^{\alpha\mathbf{-q}}_{ba}G^{1-ph}(\omega_{ab},\omega_{\alpha\mathbf{q}})\Big\}\:,
 \label{c1_eq3}
\end{align}
The second term of Eq. \ref{c1_eq3} is identical to Eq. \ref{c1_eq2}, as we intended to show. We also note that the first term of Eq. \ref{c1_eq3}  is equivalent to imposing the condition $\sum_{i}W_{ij}=0$ as it is commonly done for the master matrix of a Markov process. \\

Now that we established a relation between the common time-dependent perturbation theory and the Redfield formalism, we can exploit it to derive an expression that describes the time evolution of the diagonal element of $\hat{\rho}^{s}$ valid at the second order of time-dependent perturbation theory and equivalent to a fourth-order perturbation theory for the density matrix. To do this we start by developing the the second-order term of time dependent perturbation theory,
\begin{equation}
 C^{(2)}_{ij}(t) =  \frac{1}{\hbar^2} \sum_{c} \int_0^{t} dt' \int_0^{t'} dt'' e^{i \omega_{ic} t'} \tilde{V}_{ic}(t') e^{i \omega_{cj} t''} \tilde{V}_{cj}(t'')\:.
 \label{pt2_1}
\end{equation}
Making the time dependence of the operators $\tilde{V}$ explicit, as done in Eq. \ref{c1_eq1}, we obtain
\begin{align}
 C^{(2)}_{ij}(t)= \frac{1}{2\hbar^2} \sum_{c} \sum_{\alpha,\beta} & \int_0^{t} dt' \int_0^{t'} dt'' V^{\alpha}_{ic} V^{\beta}_{cj} \{  \\
 & e^{i (\omega_{ic}+\omega_{\alpha}) t'} e^{i (\omega_{cj}+\omega_{\beta}) t''} a^{\dag}_{\alpha}a^{\dag}_{\beta} +
   e^{i (\omega_{ic}+\omega_{\alpha}) t'} e^{i (\omega_{cj}-\omega_{\beta}) t''} a^{\dag}_{\alpha}a_{\beta} + \\
 & e^{i (\omega_{ic}-\omega_{\alpha}) t'} e^{i (\omega_{cj}+\omega_{\beta}) t''} a_{\alpha}a^{\dag}_{\beta} +
  e^{i (\omega_{ic}-\omega_{\alpha}) t'} e^{i (\omega_{cj}-\omega_{\beta}) t''} a_{\alpha}a_{\beta} \} \:,
 \label{pt2_2}
\end{align}
Let us now unravel the algebra for the first of the four terms in parenthesis. Integrating with respect to $dt''$ we obtain
\begin{equation}
    C^{(2)}_{ij}(t)= -\frac{i}{2\hbar^2} \sum_{c} \sum_{\alpha,\beta} \frac{ V^{\alpha}_{ic} V^{\beta}_{cj}}{\omega_{cj}-\omega_{\beta}} a^{\dag}_{\alpha}a^{\dag}_{\beta} \int_0^{t} dt'\left( e^{i (\omega_{ij}+\omega_{\alpha}+\omega_{\beta}) t'} - e^{i (\omega_{ic}+\omega_{\alpha}) t'}  \right)
    \label{pt2_3}
\end{equation}

Retaining only the first exponential term and repeating the same steps for all four terms in Eq. \ref{pt2_2}, we obtain
\begin{align}
 C^{(2)}_{ij}(t)= -\frac{i}{2\hbar^2} \sum_{c} \sum_{\alpha,\beta} & \int_0^{t} dt' \Big[ \\ & \frac{V^{\alpha}_{ic} V^{\beta}_{cj}}{\omega_{cj}+\omega_{\beta}} 
  e^{i (\omega_{ij}+\omega_{\alpha}+\omega_{\beta}) t'} a^{\dag}_{\alpha}a^{\dag}_{\beta} +
  \frac{V^{\alpha}_{ic} V^{\beta}_{cj}}{\omega_{cj}-\omega_{\beta}} 
  e^{i (\omega_{ij}+\omega_{\alpha}-\omega_{\beta}) t'} a^{\dag}_{\alpha}a_{\beta} + \\
  & \frac{V^{\alpha}_{ic} V^{\beta}_{cj}}{\omega_{cj}+\omega_{\beta}} 
  e^{i (\omega_{ij}-\omega_{\alpha}+\omega_{\beta}) t'} a_{\alpha}a^{\dag}_{\beta} +
  \frac{V^{\alpha}_{ic} V^{\beta}_{cj}}{\omega_{cj}-\omega_{\beta}} 
  e^{i (\omega_{ij}-\omega_{\alpha}-\omega_{\beta}) t'} a_{\alpha}a_{\beta} \Big] \:,
 \label{pt2_4}
\end{align}
The next step requires squaring $C^{(2)}_{ij}(t)$. In doing so, one obtains many terms, each with a product of four creation/annihilation operators. One needs to remember that only those that contains an equal number of creation and annihilation operators for each phonon will ultimately survive the thermal average. For instance, taking the square of the first term of Eq. \ref{pt2_4}, one obtains 
\begin{equation}
    \mathrm{Tr_{B}} \left[ a_{\gamma}a_{\delta}a^{\dag}_{\alpha}a^{\dag}_{\beta} \mathrm{\rho^{eq}_{B}} \right]= (\delta_{\gamma \alpha} \delta_{\delta \beta}  + \delta_{\gamma \beta} \delta_{\delta \alpha} ) \mathrm{Tr_{B}} \left[  a_{\gamma}a_{\delta}a^{\dag}_{\alpha}a^{\dag}_{\beta} \mathrm{\rho^{eq}_{B}} \right]= 2(\bar{n}_{\alpha}+1)(\bar{n}_{\beta}+1)\:.
\end{equation}
Following this criterion, the first and last terms of Eq. \ref{pt2_4} will only contribute when multiplied by their complex conjugate. On the other hand, the second and third terms will give rise to all four mix products among them. \\

With some algebra it is possible to demonstrate that by squaring Eq. \ref{pt2_4}, taking the limit for $t \rightarrow \infty$, deriving with respect to $t$, and averaging with respect to the bath degrees of freedom, we arrive at the expression
\begin{align}
& \mathrm{Tr_{B}}\left [ \frac{d}{dt} \lim_{t \to \infty} |C_{ij}|^{2} \rho_{\mathrm{B}}^{\mathrm{eq}}\right] = \\
& \frac{\pi}{2\hbar^{2}} \sum_{\alpha \ge\beta} \left [ A_{\alpha \beta} W^{--}_{ab}(\alpha \beta) + A_{\alpha \beta} W^{++}_{ab}(\alpha \beta) + B_{\alpha \beta} W^{+-}_{ab}(\alpha \beta) + B_{\alpha \beta} W^{-+}_{ab}(\alpha \beta) \right] \:,
\label{Red41}
\end{align}
where
\begin{align}
  &   W^{--}_{ab}(\alpha \beta) =  \Big|\sum_{c} \frac{\langle a |\hat{V}_{\alpha}|c\rangle\langle c|\hat{V}_{\beta} | b \rangle}{E_{c}-E_{b}-\hbar\omega_{\beta}}+
    \frac{\langle a |\hat{V}_{\beta}|c\rangle\langle c|\hat{V}_{\alpha} | b \rangle}{E_{c}-E_{b}-\hbar\omega_{\alpha}} \Big|^{2} \bar{n}_{\alpha} \bar{n}_{\beta} \delta(\omega_{ab}-\omega_{\alpha}-\omega_{\beta})  \\
  &  W^{++}_{ab}(\alpha \beta) =   \Big|\sum_{c}
    \frac{\langle a |\hat{V}_{\alpha}|c\rangle\langle c|\hat{V}_{\beta} | b \rangle}{E_{c}-E_{b}+\hbar\omega_{\beta}}+
    \frac{\langle a |\hat{V}_{\beta}|c\rangle\langle c|\hat{V}_{\alpha} | b \rangle}{E_{c}-E_{b}+\hbar\omega_{\alpha}} \Big|^{2} (\bar{n}_{\alpha}+1) (\bar{n}_{\beta}+1) \delta(\omega_{ab}+\omega_{\alpha}+\omega_{\beta})  \\
  &  W^{+-}_{ab}(\alpha \beta) =   \Big|\sum_{c}
    \frac{\langle a |\hat{V}_{\alpha}|c\rangle\langle c|\hat{V}_{\beta} | b \rangle}{E_{c}-E_{b}+\hbar\omega_{\beta}}+
    \frac{\langle a |\hat{V}_{\beta}|c\rangle\langle c|\hat{V}_{\alpha} | b \rangle}{E_{c}-E_{b}-\hbar\omega_{\alpha}} \Big|^{2} \bar{n}_{\alpha} (\bar{n}_{\beta}+1) \delta(\omega_{ab}-\omega_{\alpha}+\omega_{\beta})  \\
  &   W^{-+}_{ab}(\alpha \beta) =   \Big|\sum_{c}
    \frac{\langle a |\hat{V}_{\alpha}|c\rangle\langle c|\hat{V}_{\beta} | b \rangle}{E_{c}-E_{b}-\hbar\omega_{\beta}}+
    \frac{\langle a |\hat{V}_{\beta}|c\rangle\langle c|\hat{V}_{\alpha} | b \rangle}{E_{c}-E_{b}+\hbar\omega_{\alpha}} \Big|^{2} (\bar{n}_{\alpha}+1) \bar{n}_{\beta} \delta(\omega_{ab}+\omega_{\alpha}-\omega_{\beta})  \:,
    \label{Red41_2}
\end{align}
$A_{\alpha \beta}=(1-3/4\delta_{\alpha\beta})$ and $B_{\alpha \beta}=(1-1/2\delta_{\alpha\beta})$. The expression generalizes to phonons with finite lifetimes as seen previously for what concern the Dirac's delta function and by adding a term $i\Delta$ at the denominators.\\

We note that Eq. \ref{Red41} have been derived in the present form in Ref. \cite{lunghi2021towards} and that it differs from the one previously proposed in ref. \cite{lunghi2020multiple}. In the latter contribution not all the possible contributions arising form the squaring of Eq. \ref{pt2_4} where taken into account. It should also be remarked that Eq. \ref{Red41} is only valid in the absence of degeneracies in the spin spectrum. Such condition arises from the fact that in restricting ourselves to the study of the sole population terms of the density matrix we are neglecting the dynamics of coherence terms. However, in the presence of degenerate states, this separation is no longer justified and the dynamics of the entire density matrix must be studied.\cite{lunghi2021towards}

\newpage
\section{Phenomenological models and the canonical picture of spin-phonon relaxation} \label{oldmodels}

In the previous section we have derived a number of equations that describe the transition between spin states according to different spin-phonon interactions. Although these equations have a very general validity, their format cannot be easily interpreted to make predictions about spin relaxation time. There are two ways around this, i) make use of assumptions that simplify the equations, or ii) solve them numerically after having determined all their coefficients from first principles. In this section we will explore the former approach, which leads to the canonical picture of spin-phonon relaxation in solid-state materials as proposed by Van Vleck\cite{van1940paramagnetic}, Orbach\cite{orbach1961spin} and many others starting from the 40s\cite{shrivastava1983theory}.

\subsection*{The Debye model}

The main theoretical framework we will use to derive the canonical picture of spin-phonon relaxation is based on the dynamical equations of a one-dimensional system of identical atoms. We will show at the end that this approach leads to the same conclusions of the Debye model often referenced in the literature\cite{shrivastava1983theory}.  

\begin{figure}
\begin{center}
\begin{tikzpicture}

\node[] at (-9,6) {\textbf{A)}};
\node[] at (-2,6) {\textbf{B)}};

\draw[spring] (-9,3) -- (-8,3);
\draw[spring] (-8,3) -- (-7,3);
\draw[spring] (-7,3) -- (-6,3);
\draw[spring] (-6,3) -- (-5,3);
\draw[spring] (-5,3) -- (-4,3);
\draw[spring] (-4,3) -- (-3,3);
\filldraw[red] (-8,3) circle (0.15);
\filldraw[red] (-7,3) circle (0.15);
\filldraw[blue] (-6,3) circle (0.15);
\filldraw[red] (-5,3) circle (0.15);
\filldraw[red] (-4,3) circle (0.15);

\node[] at (-9,3) {...};
\node[] at (-3,3) {...};

\node[] at (-5.5,3.7) {$r_{1}$};
\draw [ <->,>=stealth]  (-6,3.5) -- (-5,3.5);
\node[] at (-6.5,3.7) {$r_{2}$};
\draw [ <->,>=stealth]  (-7,3.5) -- (-6,3.5);

\draw (-7,1.55) -- (-7,1.45);
\draw (-8,1.55) -- (-8,1.45);
\draw (-8,1.5) -- (-7,1.5);
\node[] at (-7.5,1) {$a$};

\node[] at (-9,2.4) {$R=$};
\node[] at (-7,2.4) {$-1$};
\node[] at (-6,2.4) {$0$};
\node[] at (-5,2.4) {$1$};

\draw (-4,0.45) -- (-4,0.55);
\draw (-8,0.45) -- (-8,0.55);
\draw (-8,0.5) -- (-4,0.5);
\node[] at (-6,0) {$L$};

\begin{axis}[
    xmin = 0, xmax = 30,
    ymin = 0, ymax = 2.0,
    xlabel = {$\omega$ (cm$^{-1}$)},
    ylabel = {Density of states (Arbitrary units)}]
    \addplot[
        samples = 200,
        smooth,
        line width = 2,
        red,
        domain = 0:20,
    ] { 0.004*x^2 };
    \addplot[
        samples = 200,
        smooth,
        line width = 2,
        red,
        domain = 20:30,
    ] { 0 };
\end{axis}
\end{tikzpicture}
\end{center}
\caption{\textbf{Schematic representation of the Debye model.} Panel A) reports the atomic structure of a one-dimensional system of identical atoms underpinning the Debye model. The central blue atom is colored differently to indicate that it is magnetic, although with the same mass as the other atoms and connected to them with a spring of the same force constant. Panel B) reports the typical form of the Debye density of states with an arbitrary cutoff Debye frequency of $\omega_{D}=20$ cm$^{-1}$.}
\label{debye}
\end{figure}
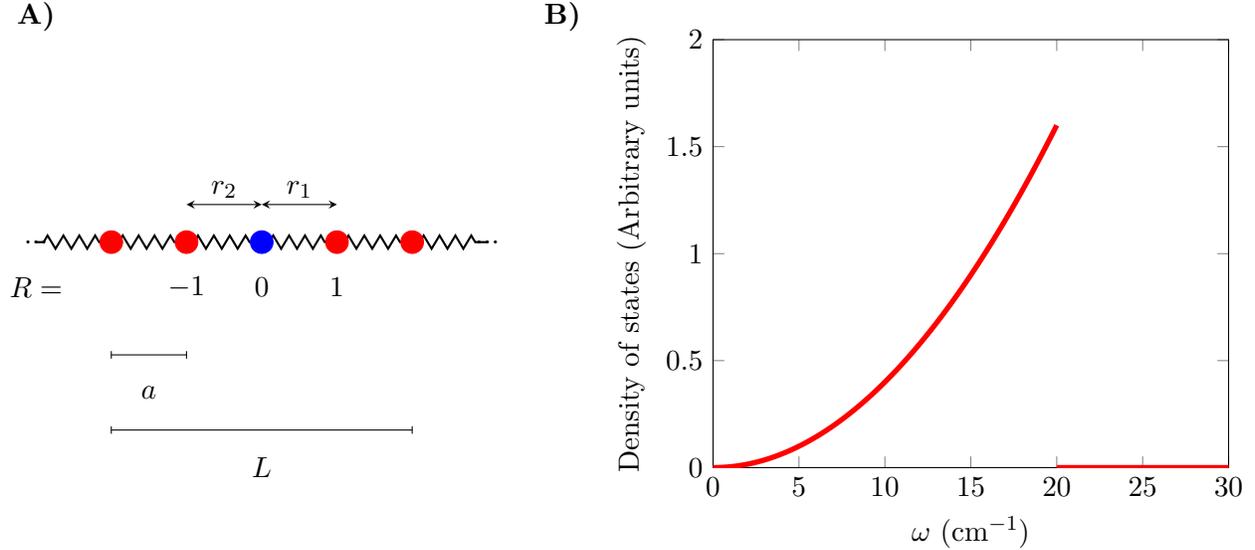

Let us assume that all our atoms are evenly spaced by an amount $a$, posses a mass $m$, and are connected by a harmonic force constant $k_m$. Let us also assume to have a finite number of atoms so that the linear chain has total length $L$. A schematic representation of this system is presented in Fig. \ref{debye}A. 

\vspace{0.5cm}
\vspace{0.5cm}

In this setup, the atomic motion can be described by a single acoustic mode
\begin{equation}
    Q_{q}(t)=A_{q}e^{i(qR-\omega_{q}t)}\:, \quad \text{where } \omega_{q}=\sqrt{\frac{2k_{m}}{m}(1-cos(qa))}
    \label{1Dchain}
\end{equation}
where $q=2\pi\:n/L$ with $ n\in \mathbb{N}_{0}$, and R is index of the lattice site. In the limit of small $q$ points, namely for long wave-lengths, $\omega_q=a\sqrt{(k_{m}/m)}q=v_{s}|q|$, where $v_s$ is the sounds velocity in the crystal medium. At the same time, the full expression of $\omega_q$ shows that the frequencies in lattice has a maximum value at the Brillouin zone boundary $q=\pi/a$. The generalization of Eq. \ref{1Dchain} to the case of a cubic three-dimensional lattice with isotropic sound velocity is performed by simply substituting the scalar $q$ with the vector $\vec{\mathbf{q}}=q_x \mathbf{e_1}+q_y \mathbf{e_2}+q_z\mathbf{e_3}$, where $\mathbf{e_i}$ are the reciprocal space basis vector. The Debye model can be seen as crude version of this classical model, where it is assumed the presence of a linear dispersion relation between angular frequency and reciprocal lattice vector $\omega_q=v_sq$, and that $\omega$ has a maximum value called the Debye cutoff frequency $\omega_D$ (see Fig. \ref{1Dchain}B). 

With these assumptions in mind, let us now simplify the sum over vibrational states that appear in the various equations of the previous section. The Redfield matrix $R_{ab,cd}$ contains a summation over $q$-points and vibrational bands $\alpha$. In our cubic isotropic lattice, the latter index will run over the three pseudo-acoustic Debye modes. However these modes are identical to one another for each value of $\mathbf{q}$ and it is therefore possible to drop the summation over $\alpha$ and multiply by 3 the final result. Regarding the sum over $\mathbf{q}$, this can be simplified by noting that \textbf{q} can only assume the values $2\pi\: n/L$, with $n \in \mathbb{N}_{0}$. The Brillouin zone is therefore discretized with a minimal element volume  $d\mathbf{q}=(2\pi / L)^{3}$. For a macroscopic crystal ($L\rightarrow\infty$), and the volume $d\mathbf{q}$ goes to zero. Therefore by multiplying and dividing by  $d\mathbf{q}$, we obtain 
\begin{align}
    \sum_{\mathbf{q}}\sum_{\alpha} \rightarrow \frac{3V}{(2\pi)^{3}}\int d{\mathbf{q}} \rightarrow & \frac{3V}{(2\pi)^{3}} \int_0^{2\pi} d\phi \int_0^{\pi} sin\theta d\theta \int_0^{\Theta_D} q^2 d{q} = \\
    & \frac{3V}{2\pi^{2}}  \int_0^{\infty} q^2 d{q} = \frac{3V}{2\pi^{2}v^3_{s}}\int_0^{\omega_D} \omega^2 d{\omega}\:,
    \label{DebyeDOS}
\end{align}
where the last equality arise from invoking the two additional assumptions of the Debye model, i.e. linear dispersion relation and $\omega_{max}=\omega_D$. 

Now that we know how to simplify the Brillouin zone integration appearing in the Redfield equation we only need a simple rule for estimating the spin-phonon coupling coefficients. Let us consider once again the one-dimensional chain of atoms. In this simple picture we can assume that one atom at position $R_l=0$ is substituted by a magnetic impurity. Moreover, we will also assume that this substitution does not lead to any change in terms of masses and force constants, so to preserve the discussion of the Debye model obtained so far. The interaction of the central magnetic atom with the two nearest neighbours provides a crude representation of the coordination sphere of a magnetic ion in a solid-state matrix or molecular compound and the size of this interaction is assume to depend on the reciprocal distance of these atoms, i.e. $\hat{H}_{s}\sim\hat{H}_{s}(r_1,r_2)$, where $r_1=|X_{1}-X_{0}|$ and $r_2=|X_{0}-X_{-1}|$. On the base of these assumptions, the spin-phonon coupling will read 
\begin{align}
    \left(\frac{\partial \hat{H}_s}{\partial Q_q}\right)=&\left(\frac{\partial \hat{H}_s}{\partial X_{-1}}\right) \left(\frac{\partial X_{-1}}{\partial Q_q}\right) + \left (\frac{\partial \hat{H}_s}{\partial X_{0}}\right) \left(\frac{\partial X_{0}}{\partial Q_q}\right) + \left (\frac{\partial \hat{H}_s}{\partial X_{1}}\right) \left(\frac{\partial X_{1}}{\partial Q_q}\right)\\ =&
    \sqrt{\frac{\hbar}{m\omega_q}} \left [  \left(\frac{\partial \hat{H}_s}{\partial X_{-1}}\right)e^{-iq} +  \left(\frac{\partial \hat{H}_s}{\partial X_{0}}\right) +  \left(\frac{\partial \hat{H}_s}{\partial X_{1}}\right)e^{iq}\right ]\:,
\end{align}
where the first step is a simple application of the derivation chain-rule and the second one follows from the definition of acoustic displacements in a linear chain. To evaluate the derivative of the spin Hamiltonian, we can apply the derivative chain rule once again
\begin{align}
    & \left(\frac{\partial \hat{H}_s}{\partial X_{0}}\right)= \left(\frac{\partial \hat{H}_s}{\partial r_{1}}\right) \left(\frac{\partial r_1}{\partial X_{0}}\right)+ \left(\frac{\partial \hat{H}_s}{\partial r_{2}}\right) \left(\frac{\partial r_2}{\partial X_{0}}\right) =
    -\left(\frac{\partial \hat{H}_s}{\partial r_{1}}\right) + \left(\frac{\partial \hat{H}_s}{\partial r_{2}}\right) \\
    &  \left(\frac{\partial \hat{H}_s}{\partial X_{1}}\right)=\left(\frac{\partial \hat{H}_s}{\partial r_{1}}\right) \left(\frac{\partial r_1}{\partial X_{1}}\right)=\left(\frac{\partial \hat{H}_s}{\partial r_{1}}\right)\\
    &  \left(\frac{\partial \hat{H}_s}{\partial X_{-1}}\right)= \left(\frac{\partial \hat{H}_s}{\partial r_{2}}\right) \left(\frac{\partial r_2}{\partial X_{-1}}\right)=-\left(\frac{\partial \hat{H}_s}{\partial r_{2}}\right)\:.
\end{align}
Finally, by simple symmetry considerations we have that  
\begin{equation}
    \hat{V}_{sph}=\left(\frac{\partial \hat{H}_s}{\partial r_{1}}\right)=\left(\frac{\partial \hat{H}_s}{\partial r_{2}}\right)\:,
\end{equation}
which makes it possible to write
\begin{equation}
    \left(\frac{\partial \hat{H}_s}{\partial Q_q}\right)=-i2\hat{V}_{sph}sin(q)\sqrt{\frac{\hbar}{m\omega_q}}=-2iv_s\hat{V}_{sph}\sqrt{\frac{\hbar\omega_q}{m}} \:,
    \label{Debyesph}
\end{equation}
where the last equality comes from assuming small values of $q$ and a linear dispersion relation. We are in now in the position to simply the Redfield equation by making use of Eqs. \ref{DebyeDOS} and \ref{Debyesph}.

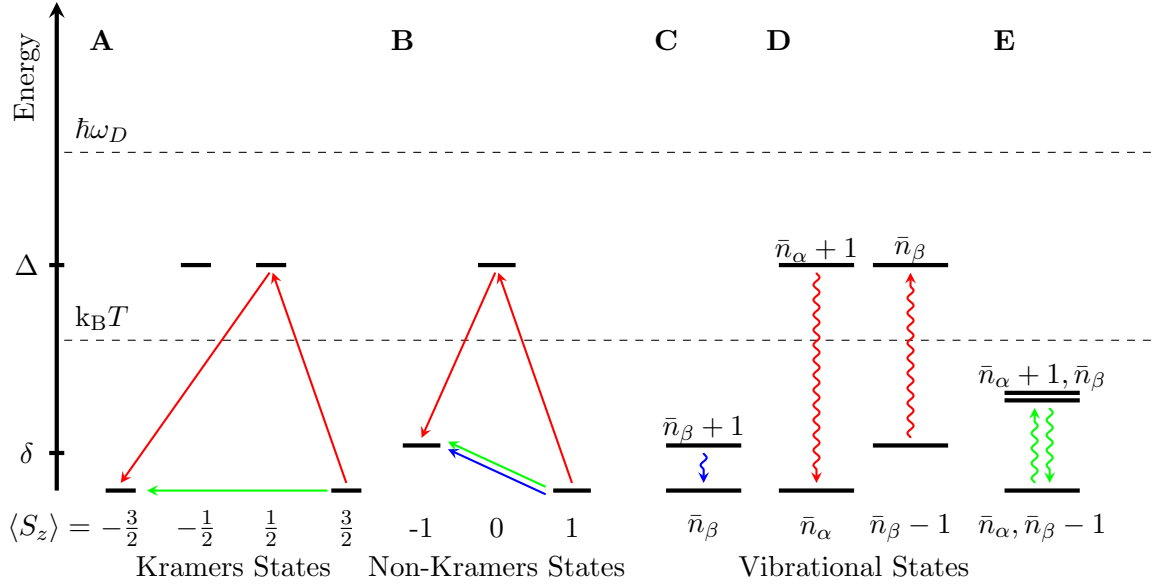
\begin{figure}
\begin{center}
  
\begin{tikzpicture}

\draw [ultra thick, ->,>=stealth] (-8.6,2) -- (-8.6,8.5);
\node [rotate=90] at (-9,7.5) {Energy};

\node at (-8,8) {\textbf{A}};
\node at (-4,8) {\textbf{B}};
\node at (-0.5,8) {\textbf{C}};
\node at (1,8) {\textbf{D}};
\node at (4,8) {\textbf{E}};

\node at (-8,6.8) {$\hbar\omega_{D}$};
\draw [dashed] (-8.5,6.5) -- (6,6.5);
\node at (-8,4.3) {$\mathrm{k_{B}}T$};
\draw [dashed] (-8.5,4) -- (6,4);

\draw [ultra thick] (-8.7,5) -- (-8.5,5);
\draw [ultra thick] (-8.7,2.5) -- (-8.5,2.5);
\node at (-9,5) {$\Delta$};
\node at (-9,2.5) {$\delta$};

\node at (-6.25,1) {Kramers States};

\draw [ultra thick] (-7.55,2) -- (-7.95,2);
\draw [ultra thick] (-6.55,5) -- (-6.95,5);
\draw [ultra thick] (-5.55,5) -- (-5.95,5);
\draw [ultra thick] (-4.55,2) -- (-4.95,2);

\node at (-7.75,1.5) {$-\frac{3}{2}$};
\node at (-6.75,1.5) {$-\frac{1}{2}$};
\node at (-5.75,1.5) {$\frac{1}{2}$};
\node at (-4.75,1.5) {$\frac{3}{2}$};

\node at (-8.7,1.5) {$\langle S_{z} \rangle=$};
\draw [ultra thick] (-4,2.6) -- (-3.5,2.6);
\draw [ultra thick] (-3,5) -- (-2.5,5);
\draw [ultra thick] (-2,2) -- (-1.5,2);

\node at (-3.75,1.5) {-1};
\node at (-2.75,1.5) {0};
\node at (-1.75,1.5) {1};

\node at (-2.75,1) {Non-Kramers States};

\draw [thick, red,->,>=stealth] (-4.75,2.1) -- (-5.73,4.9);
\draw [thick, red,->,>=stealth] (-5.77,4.9) -- (-7.75,2.1);
\draw [thick, green,->,>=stealth] (-5,2) -- (-7.4,2);

\draw [thick, red,->,>=stealth] (-1.75,2.1) -- (-2.73,4.9);
\draw [thick, red,->,>=stealth] (-2.77,4.9) -- (-3.75,2.7);
\draw [thick, blue,->,>=stealth] (-2.1,1.95) -- (-3.4,2.55);
\draw [thick, green,->,>=stealth] (-2.1,2.05) -- (-3.4,2.65);

\node at (2,1) {Vibrational States};


\draw [ultra thick] (-0.5,2) -- (0.5,2);
\draw [ultra thick] (-0.5,2.6) -- (0.5,2.6);

\draw [thick,blue,->,>=stealth,decorate,
decoration={snake,amplitude=.4mm,segment length=2mm,post length=1mm}]
(0,2.5) -- (0,2.1) ;  

\node at (0,1.5) {$\bar{n}_{\beta}$};
\node at (0,2.8) {$\bar{n}_{\beta}+1$};


\draw [ultra thick] (1,5) -- (2,5);
\draw [ultra thick] (1,2) -- (2,2); 

\draw [ultra thick] (2.25,5) -- (3.25,5);
\draw [ultra thick] (2.25,2.6) -- (3.25,2.6);

\draw [thick,red,->,>=stealth,decorate,
decoration={snake,amplitude=.4mm,segment length=2mm,post length=1mm}]
(1.5,4.9) -- (1.5,2.1);
\draw [thick,red,->,>=stealth,decorate,
decoration={snake,amplitude=.4mm,segment length=2mm,post length=1mm}]
(2.75,2.7) -- (2.75,4.9) ;


\node at (1.5,1.5) {$\bar{n}_{\alpha}$};
\node at (1.5,5.2) {$\bar{n}_{\alpha}+1$};

\node at (2.75,1.5) {$\bar{n}_{\beta}-1$};
\node at (2.75,5.2) {$\bar{n}_{\beta}$};


\draw [ultra thick] (4,2) -- (5,2);
\draw [ultra thick] (4,3.3) -- (5,3.3);
\draw [ultra thick] (4,3.2) -- (5,3.2); 


\draw [thick,green,->,>=stealth,decorate,
decoration={snake,amplitude=.4mm,segment length=2mm,post length=1mm}]
(4.4,2.1) -- (4.4,3.1) ;
\draw [thick,green,->,>=stealth,decorate,
decoration={snake,amplitude=.4mm,segment length=2mm,post length=1mm}]
(4.6,3.1) -- (4.6,2.1) ;  

\node at (4.5,1.5) {$\bar{n}_{\alpha},\bar{n}_{\beta}-1$};
\node at (4.5,3.5) {$\bar{n}_{\alpha}+1,\bar{n}_{\beta}$};

\end{tikzpicture}
\end{center}
\caption{\textbf{Schematic representation of spin relaxation mechanisms.} Panel A represents the energy levels of a $S=3/2$ with uni-axial anisotropy in the absence of external. Panel B represents the energy levels of a $S=1$ with uni-axial anisotropy in the presence of a small external field that removes the degeneracy of states with $M_S=\pm1$. Panel C represents the direct relaxation mechanism as the absorption of a phonon resonant with the states $M_S=\pm1$ of the non-Kramers states. Panel D represents an Orbach relaxation mechanism due to the time-sequential absorption and emission of a phonon in resonance with either the $M_S=\pm1 - M_S=0$ or $M_S=\pm3/2 - M_S=\pm 1/2$ transitions for the non-Kramers and Kramers systems, respectively. Panel E represents the Raman relaxation mechanism, where two phonons are simultaneously absorbed/emitted to induce a transition among the states $M_S=\pm S $. In this scheme we assume that the thermal energy ($\mathrm{k_{B}}T$) is much higher that the Zeeman energy $\delta$, but lower that the excited states energy $\Delta$, and that the Debye cutoff energy ($\hbar\omega_{D}$) is much larger than the excited spin states energy $\Delta$.}
\label{RelaxScheme}
\end{figure}

\subsection*{One-phonon relaxation}

So far we have discussed expressions for the transition rate between spin states, but not all those transitions contribute to spin relaxation. In fact, only processes that lead to a spin flip $|S_z\rangle\rightarrow|-S_z\rangle$ should be considered at this stage. When only one phonon at the time can be exchanged between lattice and spin, there are two possible relaxation mechanisms: direct and Orbach. In the first case, as the name suggests, a direct transition between the two states with maximum spin polarization occurs (see Fig. \ref{RelaxScheme}A-C), while in the second case, single-phonon transitions to intermediate states are necessary (see Fig. \ref{RelaxScheme}A,B,D). Generally, we are interested in scenarios where the $|\pm S_z\rangle$ states form a quasi-degenerate ground state, split by an energy $\delta$ and separated by higher excited states by a much larger energy $\Delta$, as depicted in Fig. \ref{RelaxScheme}.\\

Let us start our analysis of spin-phonon relaxation by simplifying the Redfield matrix appearing in Eq. \ref{Red21}. In order to provide a qualitative analysis of what drives the transition rate between two different spin states we can simply focus on the population transfer terms
\begin{equation}
    R_{bb,aa}=\frac{\pi}{\hbar^{2}} \sum_{\alpha\mathbf{q}} \Big|\langle b | \left(  \frac{\partial \hat{H}_s}{\partial Q_{\alpha\mathbf{q}}} \right)| a \rangle \Big|^{2} \hat{n}_{\alpha\mathbf{q}} \delta(\omega_{ba}-\omega_{\alpha\mathbf{q}})\:.
\end{equation}
By applying Eqs. \ref{DebyeDOS} and \ref{Debyesph} we obtain
\begin{align}
R_{bb,aa}= & \frac{\pi}{\hbar^{2}} \int_0^{\omega_D} \omega^{2} \Big|\langle b | \left(  \frac{\partial \hat{H}_s}{\partial Q_{\alpha\mathbf{q}}} \right)| a \rangle \Big|^{2} \left( e^{\beta\hbar\omega}-1\right)^{-1} \delta(\omega_{ba}-\omega) d\omega \\
= & \tilde{V}_{sph} \int_0^{\omega_D} \omega^{3} \left( e^{\beta\hbar\omega}-1\right)^{-1} \delta(\omega_{ba}-\omega) d\omega\:,
\end{align}
where the term $\tilde{V}_{sph}$ is used to absorb all the constants that do not depend on $T$ or $\omega$. Performing the integration with respect to $\omega$ we obtain
\begin{equation}
R_{bb,aa}=\tilde{V}_{sph} \omega_{ba}^{3} \left( e^{\beta\hbar\omega_{ba}}-1\right)^{-1}\:.
\label{Debye1Ph}
\end{equation}
We can now distinguish different scenarios depending on the relaxation process we want to describe. Let us start from the direct relaxation, where $\omega_{ba}=\delta \ll \mathrm{k_B}T$. The last condition comes from the fact that the splitting between $|\pm S_z\rangle$ states arises from either Zeeman or transverse crystal field interactions, which usually do not go above a few cm$^{-1}$, corresponding to a temperature of a few K. In this scenario, we can interpret $R_{bb,aa}$ as the relaxation rate obtaining
\begin{equation}
    \tau^{-1}=\tilde{V}_{sph}\delta^{2}k_BT\:.
    \label{DebyeDir}
\end{equation}
Eq. \ref{DebyeDir} shows that the direct relaxation rate depends linearly on $T$, as commonly derived in literature. Eq. \ref{DebyeDir} also sheds light on the field dependence of spin relaxation. Firstly, we note that $\tau^{-1}$ depends quadratically on $\delta$, which is proportional to the external field through the Zeeman interaction. However, in order to completely unravel the field dependence of $\tau^{-1}$ we must also consider the term $\tilde{V}_{sph}$. $\tilde{V}_{sph}$ contains the square of the matrix element of the spin-phonon coupling operator $\hat{V}_{sph}=(\partial \hat{H}_{s}/\partial r)$, where $r$ is the distance between the central magnetic atom of our one-dimensional chain and its nearest neighbours. In the presence of an external field, the operator $\hat{H}_{s}$ contains the Zeeman interaction and if this interaction is the dominant one, as for the case of $S=1/2$ systems in magnetically diluted conditions and absence of hyperfine fields, it is then  reasonable to expect that the Lande' tensor will lead to spin-phonon coupling due to its dependence on the atomic positions. The latter is due to the presence of spin-orbit coupling in the electronic Hamiltonian, which contributes to the anisotropy of the $\mathbf{g}$ tensor. We also note that this interaction is able to break the time-reversal symmetry in Kramers systems, and it is indeed commonly used to explain the direct relaxation in this class of compounds\cite{van1940paramagnetic,orbach1961spin}. If this spin-phonon coupling mechanism is considered, then the spin-phonon coupling operator reads
\begin{equation}
    \hat{V}_{sph}=\mu_{B}\left[\vec{\mathbf{S}}\cdot\left(\frac{\partial \mathbf{g}}{\partial r}\right)\cdot\vec{\mathbf{B}}\right]\:,
\end{equation}
leading to an additional quadratic field contribution to $\tau^{-1}$. In conclusion we have that direct relaxation follows two possible trends
\begin{equation}
    \text{for k$_{\mathrm{B}}T<\omega$,}\quad\tau^{-1}_{\textrm{Dir}}\sim B^{4}\mathrm{k_B}T\:, \quad \tau^{-1}_{\textrm{Dir}}\sim B^{2}\mathrm{k_B}T\:,
    \label{DebyeDirFinal}
\end{equation}
\begin{equation}
    \text{for k$_{\mathrm{B}}T>\omega$,}\quad \tau^{-1}_{\textrm{Dir}}\sim B^{4}\mathrm{k_B}T^{0}\:, \quad \tau^{-1}_{\textrm{Dir}}\sim B^{2}\mathrm{k_B}T^{0}\:,
    \label{DebyeDirFinal_2}
\end{equation}
depending on whether the modulation of the Zeeman interaction by phonons drives relaxation or not, respectively.\\

Next we can use Eq. \ref{Debye1Ph} to derive an expression similar to \ref{DebyeDirFinal} for the Orbach relaxation mechanism. Let us now assume that we have a system with at least one excited spin states above the ground-state pseudo-doublet, as depicted in Fig. \ref{RelaxScheme}A,B. Let us also assume that the direct relaxation is inhibited. In this circumstances, one-phonon relaxation can only occur in a two-step process with the absorption of a phonon and the subsequent emission of another one. From Eq. \ref{Red21}, we can observe that the relaxation rates of the absorption and emission transitions have the exact same mathematical structure except for the phonon population contribution, which accounts for the spontaneous emission of one phonon through the term $\bar{n}+1$. As a consequence, the emission process will always be faster than the absorption one, leaving the latter to act as the rate determining step of the relaxation process. Therefore we can focus on adapting Eq. \ref{Debye1Ph} to the case of the transition $|a\rangle \rightarrow |c\rangle$ due to the absorption of one phonon, which leads to
\begin{equation}
R_{cc,aa}=\tilde{V}_{sph} \omega_{ca}^{3} \left( e^{\beta\hbar\omega_{ca}}-1\right)^{-1}\:.
\end{equation}
However, differently from the direct relaxation case, we are now in the opposite temperature vs spin splitting limit: $\omega_{ca} \gg \mathrm{k_{B}}T$, which leads to
\begin{equation}
R_{cc,aa}=\tilde{V}_{sph} \omega_{ca}^{3} e^{-\beta\hbar\omega_{ca}}\:.
\end{equation}
We note that this expression has been derived under the additional assumption that the Debye model is still valid and that acoustic modes are driving relaxation, i.e. $\omega_{ca}< \omega_D$. Differently from the direct case, the Orbach necessarily concerns systems with $S>1/2$, where zero-field splitting interactions due to the ion's crystal field are the dominant term of $\hat{H}_s$. Therefore, additional contributions to the field in $\tilde{V}_{sph}$ are not expected and we obtain the final result
\begin{equation}
\tau^{-1}\sim \Delta^{3} e^{-\Delta/\mathrm{k_{B}}T}\:,
    \label{DebyeOrbFinal}
\end{equation}
where we have also assumed that $\Delta$ does not depends on the field as by construction $\Delta \gg \delta \propto |B|$

\subsection*{Two-phonon relaxation}

In section \ref{theory} we have distinguished two different two-phonon spin transition mechanisms: one coming from second-order time-dependent perturbation theory plus second-order spin-phonon coupling and the other one coming from fourth-order time-dependent perturbation theory plus first-order spin-phonon coupling strength. Starting our analysis from the first case and using Eq. \ref{Red22}, we can apply a similar strategy as for the one-phonon case. However, this time we have a double summation over the phonon spectrum and the transition $|a\rangle \rightarrow |b\rangle$ due to the simultaneous absorption and emission of two-phonons reads
\begin{equation}
    R_{bb,aa}=\frac{\pi}{4\hbar^{2}} \sum_{\alpha\mathbf{q} \ne \beta\mathbf{q'}} \Big|\langle b | \left(  \frac{\partial^2 \hat{H}_s}{\partial Q_{\alpha\mathbf{q}} \partial Q_{\beta\mathbf{q'}} } \right)| a \rangle \Big|^{2} \hat{n}_{\alpha\mathbf{q}} (\hat{n}_{\beta\mathbf{q'}}+1) \delta(\omega_{ba}-\omega_{\alpha\mathbf{q}}+\omega_{\beta\mathbf{q'}})\:.
\end{equation}
By substituting the expression of Eq. \ref{DebyeDOS} and accounting that each phonon derivative contributes with a term proportional to $\sqrt{\omega}$ (see Eq. \ref{Debyesph}), we obtain
\begin{equation}
R_{bb,aa}= \tilde{V}_{sph} \int_0^{\omega_D}d\omega\int_0^{\omega_D}d\omega' \omega^{3}\omega'^{3}  \frac{e^{\beta\hbar\omega'}}{(e^{\beta\hbar\omega}-1)(e^{\beta\hbar\omega'}-1)} \delta(\delta-\omega+\omega')\:.
\label{2phDebye0}
\end{equation}
The Dirac delta function in Eq. \ref{2phDebye0} enforces that $\omega=\delta+\omega'$ and, since $\delta \ll \omega$ for ordinary Zeeman splittings and phonons energies, we can assume that $\omega=\omega'$, which leads to
\begin{equation}
R_{bb,aa}= \tilde{V}_{sph} \int_0^{\omega_D}d\omega \omega^{6} \frac{e^{\beta\hbar\omega}}{(e^{\beta\hbar\omega}-1)^2}\:.
\label{2phDebye}
\end{equation}
An important result of Eq. \ref{2phDebye} is that two-phonon relaxation receives contributions from the entire vibrational spectrum without any dependence on the size of the spin splitting $\delta$, nor involving excited spin states. In order to make the $T$ dependence of this expression more transparent, we can make the substitution $x=\beta\omega$ and rewrite it as
\begin{equation}
\tau^{-1}= \tilde{V}_{sph} (\mathrm{k_B} T)^7 \int_0^{\beta\omega_D}dx x^{6} \frac{e^{x}}{(e^{x}-1)^2}\:,
\end{equation}
where the integral does not explicitly depends on temperature anymore for low $T$. \textit{i.e.} when $\beta\omega_{D}\rightarrow\infty$. In the high-$T$ limit, the ratio of exponential functions in Eq. \ref{2phDebye} simplify to $(\beta\omega)^{-2}$, thus leading to an overall $T^{-2}$. \\

Concerning the field dependence of $\tau^{-1}$, we note that it can only arise from the nature of $\tilde{V}_{sph}$, which might be driven by the modulation of the Zeeman interaction, as seen for the direct relaxation. Therefore we can conclude that Raman relaxation follows the trends
\begin{equation}
\text{for k$_{\mathrm{B}}T<\omega$:}\quad \tau_{\textrm{Raman}}^{-1}\sim B^{2}T^7 \:, \quad \tau_{\textrm{Raman}}^{-1}\sim B^{0}T^7 \:,
\end{equation}
or 
\begin{equation}
\text{for k$_{\mathrm{B}}T>\omega$:}\quad \tau_{\textrm{Raman}}^{-1}\sim B^{2}T^2 \:, \quad \tau_{\textrm{Raman}}^{-1}\sim B^{0}T^2 \:,
\end{equation}
depending on whether the modulation of the Zeeman interaction by phonons drives relaxation or not, respectively. \\

Let us now investigate the temperature and field dependence of $\tau$ due to the use of Eq. \ref{Red41} for a non-Kramers system, \textit{e.g.} when only one excited state is available as depicted in Fig. \ref{RelaxScheme}B. By substituting the expression of Eq. \ref{DebyeDOS} into Eq. \ref{Red41} and accounting that each phonon derivative contributes with a term proportional to $\sqrt{\omega}$, we obtain 
\begin{align}
R_{bb,aa}= \tilde{V}_{sph} \int_0^{\omega_D}d\omega\int_0^{\omega_D}d\omega' \omega^{3}\omega'^{3} & \left| \frac{\langle b | V' | c \rangle\langle c | V | a \rangle }{\Delta-\omega} + \frac{\langle b | V | c \rangle\langle c | V' | a \rangle }{\Delta+\omega'}  \right |^{2} \\ & \frac{e^{\beta\hbar\omega'}}{(e^{\beta\hbar\omega}-1)(e^{\beta\hbar\omega'}-1)} \delta(\delta-\omega+\omega')\:.
\label{2phDebye_pt2_0}
\end{align}
As before, the Dirac delta function in Eq. \ref{2phDebye_pt2_0} enforces that $\omega=\delta+\omega'$ and, since $\delta \ll \omega$ for ordinary Zeeman splittings and phonons energies, we can assume that $\omega=\omega'$. Additionally, we are going to assume that only portion of the integral with $\Delta >> \omega$ is going to contribute in virtue of the condition $\mathrm{k_{B}}T<<\Delta$, which leads us to
\begin{equation}
R_{bb,aa}= \tilde{V}_{sph} \Delta^{-2} \int_0^{\omega_D}d\omega \omega^{6} \frac{e^{\beta\hbar\omega}}{(e^{\beta\hbar\omega}-1)^2}\:.
\label{2phDebye_pt2}
\end{equation}
Except for a different pre-factor containing the term $\Delta^{-2}$, this result is identical to Eq. \ref{2phDebye} and have the same $T$ and $B$ dependence. However, while Eq. \ref{Red22} led to identical results for Kramers and non-Kramers systems, the use of Eq. \ref{Red41} does not. Let us call the ground-state Kramers doublet $\pm p$ and the excited one $\pm q$. A transition $p \rightarrow -p$ would then involve a sum over the excited states that reads 
\begin{align}
 \frac{\langle -p | V' | q \rangle\langle q | V | p \rangle }{\Delta-\omega} + &
\frac{\langle -p | V' | -q \rangle\langle -q | V | p \rangle }{\Delta-\omega} + \\ 
&   \frac{\langle -p | V | q \rangle\langle q | V' | p \rangle }{\Delta+\omega'} + 
\frac{\langle -p | V | -q \rangle\langle -q | V' | p \rangle }{\Delta+\omega'} = \\
 \frac{\langle -p | V' | q \rangle\langle q | V | p \rangle }{\Delta-\omega} - &
\frac{\langle p | V' | q \rangle^{*}\langle q | V | -p \rangle^{*} }{\Delta-\omega} + \\ 
&  \frac{\langle -p | V | q \rangle\langle q | V' | p \rangle }{\Delta+\omega'} - \frac{\langle p | V | q \rangle^{*}\langle q | V' | -p \rangle^{*} }{\Delta+\omega'} = \\
 \frac{\langle -p | V' | q \rangle\langle q | V | p \rangle }{\Delta-\omega} - &
\frac{\langle -p | V | q \rangle\langle q | V' | p \rangle }{\Delta-\omega} + \\
&  \frac{\langle -p | V | q \rangle\langle q | V' | p \rangle }{\Delta+\omega'} - \frac{\langle -p | V' | q \rangle\langle q | V | p \rangle }{\Delta+\omega'} \:,
\label{kramers1}
\end{align}
where the first equality follows from the Kramers symmetry and the second one from the Hermitian property of the spin-phonon coupling Hamiltonian. Eq. \ref{kramers1} can then be further simplified by enforcing the condition $\Delta >> \omega=\omega'$
\begin{align}
& \langle -p | V' | q \rangle\langle q | V | p \rangle \left ( \frac{1}{\Delta-\omega} -  \frac{1}{\Delta+\omega'} \right) + 
\langle -p | V | q \rangle\langle q | V' | p \rangle \left ( \frac{1}{\Delta+\omega'} - \frac{1}{\Delta-\omega} \right) = \\ 
& \langle -p | V' | q \rangle\langle q | V | p \rangle \left ( \frac{-2\omega}{\Delta^{2}-\omega^{2}}  \right) +
\langle -p | V | q \rangle\langle q | V' | p \rangle\left ( \frac{2\omega}{\Delta^{2}-\omega^{2}}  \right) = \\
& \frac{2\omega}{\Delta^{2}} \left( \langle -p | V | q \rangle\langle q | V' | p \rangle - \langle -p | V' | q \rangle\langle q | V | p \rangle \right ) \:.
\label{kramers2}
\end{align}
When the result of Eq. \ref{kramers2} is used together to Eq. \ref{Red41}, Eq. \ref{2phDebye_pt2} turns into 
\begin{equation}
R_{bb,aa}= \tilde{V}_{sph} \Delta^{-4} \int_0^{\omega_D}d\omega \omega^{8} \frac{e^{\beta\hbar\omega}}{(e^{\beta\hbar\omega}-1)^2}\:.
\label{2phDebye_pt2_kr}
\end{equation}
In conclusion for Kramers systems that relax according to fourth-order perturbation theory, $\tau$ has the following dependency with respect to field and temperature
\begin{equation}
\text{for k$_{\mathrm{B}}T<\omega$:}\quad \tau_{\textrm{Raman}}^{-1}\sim B^{2}T^9 \:, \quad \tau_{\textrm{Raman}}^{-1}\sim B^{0}T^9 \:,
\end{equation}
or 
\begin{equation}
\text{for k$_{\mathrm{B}}T>\omega$:}\quad \tau_{\textrm{Raman}}^{-1}\sim B^{2}T^2 \:, \quad \tau_{\textrm{Raman}}^{-1}\sim B^{0}T^2 \:,
\end{equation}\\

We have now concluded the review of the most fundamental results presented by Van Vleck \cite{van1940paramagnetic} and Orbach \cite{orbach1961spin} in their seminal works. It must be noted that these derivation were carried out assuming the very specific condition summarized in Fig. \ref{RelaxScheme}. When different conditions are probed, such as $\hbar\omega_{D}<\Delta$ or extending the Debye model to optical phonons, the phenomenology can drastically change. Indeed, it is notable that Shrivastava, in his 1987 review\cite{shrivastava1983theory}, concludes that depending on the sample's properties and external conditions, relaxation time can exhibit a dependence on temperature as exponential or as a power law $T^{-n}$ with any exponent in the range $1\le n \le 9$. The impressive range of different regimes that spin relaxation can exhibit is at the heart of the challenge of interpreting experiments in a unequivocal way. 

\newpage
\section{\textit{Ab initio} spin dynamics simulations}\label{abinitio-methods}

In the previous section we have illustrated how the theory of open quantum systems can be used to derive simple relations between relaxation time and external conditions such as temperature and magnetic field. However, spin relaxation can receive contributions from several different fundamental processes, as expressed by the different perturbation orders involved at both coupling strength and time-scale levels. Moreover, the relation demonstrated in the previous section also depends on assumptions about the reciprocal positioning of the Debye frequency with respect to the spin energy levels and the thermal energy. As a consequence, a plethora of different experimental behaviours can be expected for different materials in different conditions, making the interpretation of experimental results hard at best. Moreover, the phenomenological models derived in the previous section depend on the validity of the Debye model itself, which makes strong assumptions on the nature of phonons and their dispersion relation. Although this model qualitatively accounts for the acoustic dispersions of simple elements, it clearly fails in accounting for the complexity of vibrations in molecular crystals. A way around all these limitations requires to exploit electronic structure theory to quantitatively predict all the terms populating Eqs. \ref{Red21}, \ref{Red22} and \ref{Red41}, and extracts the relaxation time from their numerical solution. In this section we will explore this strategy and how accurately and effectively determine phonons and spin-phonon coupling coefficients in molecular crystals. 

\subsection*{\textit{Ab initio} simulation of phonons}

The first fundamental ingredient we need for the computation of Eqs. \ref{Red21}, \ref{Red22} and \ref{Red41}, is the notion of phonons in a molecular crystal. As anticipated in the section about solid-state vibrations, the Hessian matrix $\Phi$ is all we need to estimate harmonic frequencies and related displacement waves (i.e. the phonons) at any $\mathbf{q}$-point of the Brillouin zone. Let us recall its definition
\begin{equation}
\Phi^{0l}_{ij}=\left ( \frac{\partial^{2} E_{el}}{\partial x_{li} \partial x_{0j}} \right ) \:,
\label{Hessian}
\end{equation}
where $E_{el}$ can be computed with standard electronic structure methods\cite{kuhne2020cp2k,neese2020orca}. It is important to note that the summation over the cells of the molecular lattice (index $l$ in Eq. \ref{Hessian}) is in principles extended to an infinite number of them. However, the forces among two atoms decay with their distance and the elements of $\Phi$ will decays to zero for distant enough lattice cells. It is then possible to estimate all the elements of $\Phi$ by performing electronic structure simulations on a small supercell, i.e. the replication in space of the crystal unit-cell\cite{kresse1995ab}. The size of the supercell actually needed to converge the elements of $\Phi$ should determined case-by-case, depending on the size of the unit-cell and the nature of atomic interactions. For instance, large unit-cells with only short-range interactions will necessitate of only small supercells, such as a $2\times 2\times2$ repetition of the unit-cell along the three different crystallographic direction. Polar crystals usually do not fall in this category due to the long range nature of dipolar electrostatic interactions, but dedicated expressions to account for this effect fortunately exist\cite{wang2010mixed}.\\

Electronic structure methods often also make it possible to compute the forces acting on atoms by mean of the Hellman-Feynman theorem,
\begin{equation}
 \vec{\mathbf{F}}_{i} = -\langle \psi | \vec{\nabla}_{i} E_{el} | \psi \rangle \:,
 \label{HFtheo}
\end{equation}
where $\psi$ is the ground-state wave function for a given molecular geometry. Using atomic forces, the numerical estimation of $\Phi$ only requires a first-order numerical differentiation,
\begin{equation}
    \Phi^{0l}_{ij}= -\left ( \frac{\partial F_{lj}}{\partial x_{0i}} \right )_{0} \:,
    \label{Hess_force}
\end{equation}
where the right-hand term should be interpreted as the force acting on atom $j$ of cell $l$ due to the displacement of atom $i$ in the reference cell with $R_{l}=0$. The estimation of the integral of Eq. \ref{HFtheo} is only slightly more computationally expensive than the determination of the electronic ground state $|\psi\rangle$, making it possible to significantly speed up the numerical evaluation of $\Phi$ compared to using Eq. \ref{Hessian}. It is also important to remark that the use of Eq. \ref{Hess_force} only requires the displacement of the atoms in one unit-cell of the entire super-cell, in virtue of the translational symmetry of the problem. Numerical differentiation of Eq. \ref{Hess_force} is often carried out with a single-step two-sided finite difference method. In a nutshell, this method involves i) optimization of the supercell's lattice parameters and atomic positions, ii) displacement of every atom of the unit-cell by $\pm \delta$ (\AA) along every Cartesian direction and calculation of $F_{lj}(x^{0}_{0i}\pm \delta)$, iii) estimation of $\Phi$ according to the expression
\begin{equation}
    \left ( \frac{\partial F_{lj}}{\partial x_{0i}} \right )_{0}= \frac{F_{lj}(x^{0}_{0i}+\delta)-F^{l}_{j}(x^{0}_{0i}-\delta)}{2\delta}\:,
    \label{finitediff}
\end{equation}
where the size of $\delta$ should be chosen in a way that the $F$ vs $x$ profile is mostly linear and not significantly affected by numerical noise. Values around 0.01 \AA$\:$ are often a good choice for molecular crystals. Once all the elements of $\Phi$ have been computed, the definition of the dynamical matrix in Eq. \ref{trasl} can be used to obtain phonon frequencies and atomic displacements at any $\mathbf{q}$-point. \\

The calculation of $\Phi$ by numerical differentiation invariably carries over some numerical noise. One of the main sources of numerical noise in the simulation of force constants is the partially converged nature of simulations, both in terms of self consistent cycles and geometry optimization. Unless very tight convergence criteria are implemented, forces will result affected by a small amount of error that will then results into inconsistent force constants. Due to their small size, numerical errors in the determination of forces mainly affect the low-energy part of the vibrational spectrum ( from a few to tens of cm$^{-1}$) and often manifest themselves through the appearance of imaginary frequencies (generally plotted as negative values of $\omega_{\alpha}$) and breaking down of translational symmetry. The latter in turns reflects in the break down of the acoustic sum rule, i.e. the condition that force constants must obey as consequence of the translational symmetry of the crystal. This condition reads
\begin{equation}
    \sum_{l}\sum_{j} \Phi^{0l}_{ij} = 0 \:, \quad \text{for any }i\:.
    \label{ASR}
\end{equation}
The relation also imply that $D(\mathbf{q}=0)$ has three eigenvalues equal to zero, which correspond to the frequencies of the three acoustic modes at the $\Gamma$-point. These modes corresponds to a rigid translations of the entire crystals in the three different crystallographic directions, and since they do not create any lattice strain they don't have any energy change associated. In real calculations, Eq. \ref{ASR}, often referred to as acoustic sum rule, breaks down and the three acoustic phonons' frequencies acquire a finite value (either positive or negative) at $\mathbf{q}=0$. The size of this deviations provides indications on the reliability of phonons and can be used to tune simulations' convergence criteria. The maximum value of the acceptable numerical noise in the simulation of phonons' frequencies depends on the specific application but the energy of the first optical modes gives a reference value. More stringent requirements are in general needed if the acoustic modes at small $\mathbf{q}$ are of interest.\\

Besides providing a measure of numerical noise, Eq. \ref{ASR} also suggests a practical route towards correcting it. Generally speaking, the symmetries of the problem can be enforced by computing $\Phi$ as the best compromise between reproducing the values of $(\partial F_{li}/\partial X_{0j})$ and the condition expressed by Eq. \ref{ASR} or similar expressions for other symmetries. This approach is however numerically expensive, as it requires the solution of a linear system $A\mathbf{x}=b$ with $A$ being a matrix of size $(N\times M)$, where $N$ is the number of force constants and $M$ is the number of DFT forces used to infer them. Even for a crystals with a small unit-cell containing 60 atoms and using a $2\times2\times2$ supercell, $N\sim260,000$ and $M\sim500,000$, requiring $\sim 1$ TB of memory to only store $A$. Additionally, the size of $A$ scales to the power of 4 with the number of atoms in the unit-cell, rapidly leading to an intractable problem for the common sizes of molecular crystals (often in the range of a few hundred atoms). Although sparsity arguments can help reducing the complexity of this problem, this is not an efficient route for large molecular crystals. However, if the numerical noise is little, Eq. \ref{ASR} can be used to correct the values of $\Phi$ by rescaling its value in order to enforce the acoustic sum rule. This simple method helps enforcing the correct behaviour of the acoustic phonons at the $\Gamma$-point with no additional costs.  \\

So far we have discussed the practicality of phonons calculations assuming that electronic structure theory can provide an accurate representation of $E_{el}$. This is in fact not always the case and many different approaches to the problem of computing energy and forces for a molecular or solid-state system exist\cite{neese2009prediction,cramer2009density}. When dealing with crystals of magnetic molecules there are three main subtleties that must be carefully considered: i) the treatment of dispersion interactions, fundamental to describe the crystal packing, ii) the treatment of correlation of magnetic states, and iii) temperature effects. \\

Ideally, quantum chemistry methods such as multi-reference coupled cluster should be preferred for the task of computing atomic forces as they automatically include electronic correlation effects leading to dispersion forces as well as to correctly treating the magnetic degrees of freedom. However, their computational cost virtually limits their application to small gas-phase molecules. When it comes to periodic systems, Density Functional Theory (DFT) is the only option. Moreover, for large systems comprising several hundreds of atoms, as often is the case for molecular crystals, DFT in its Generalized Gradient Approximation (GGA) is often the best, if not unique, option. GGA functionals, like PBE\cite{perdew1996generalized}, are generally recognized to provide a good estimation of bond lengths and vibrational force-constants. Alternatively, hybrid functionals also provide a good choice for the simulation of vibrational and thermodynamic properties of molecular systems\cite{perdew1996rationale}, but the computational overheads often hinder their use for solid-state systems. \\

Relatively to point i), common GGA or hybrid functionals fail in capturing vdW interactions, but several correction schemes are available at very little additional computational cost. Some of the most widely employed methods include a parametrization of dispersion interactions by adding a potential term $V(\mathbf{r}_{ij})=-C_{6}/r^{6}_{ij}$ among atoms. The parametrization of the coefficients $C_{6}$ is then carried out according to different schemes, such as the Tkatchenko's or Grimme's methods\cite{tkatchenko2009accurate,grimme2010consistent}. These methods account for the screening effect of covalent bonds on local atomic polarizability, which reflects on a reduction of the coefficients $C_{6}$ with respect to the reference gas-phase atomic values. Concerning point ii) of correctly describing the strongly correlated states of $d$- or $f$-elements, DFT is known to perform poorly, and methods such as CASSCF\cite{neese2019chemistry} or DMRG\cite{chan2011density} are often used to fix the problem. These methods are however not suitable for large periodic systems and the use of GGA DFT is once again a forced choice for current algorithms and computational power. Fortunately, however, the shortcoming of GGA DFT in representing the correct magnetic states of open shell metal compounds does not seem to dramatically spread to the prediction of forces, and phonons calculations of magnetic molecules' crystals based on DFT are routinely carried out with good success. Finally, in relation to point iii), it should be noted that a structural optimization leads to a \textit{zero-temperature} geometry. The latter does not correctly describe the compound structure in the same thermodynamics condition of the experiments. The inclusion of finite-temperature effects can be accomplished by accounting for anharmonic contributions to the crystal potential energy surface. Anharmonic effects in crystals can be modelled in different ways, including quasi-harmonic approaches, perturbative treatments, or by running molecular dynamics simulation in the presence of a thermostat\cite{ruggiero2020invited}. The importance of including anharmonic effects into the description of phonons for spin-phonon coupling calculations has only recently be pointed out\cite{lunghi2017intra,albino2021temperature,ullah2021insights} and a systematic investigation is yet to be presented, but it represents an important area of investigation. \\

\subsection*{\textit{Ab initio} simulation of spin Hamiltonian's parameters}

The next fundamental step in simulating spin-phonon relaxation time is the calculation spin Hamiltonian coefficients. Let us assume that we can solve the Schroedinger equation for the electronic Hamiltonian $\hat{H}_{el}$, inclusive of relativistic interactions such as spin-orbit and spin-spin coupling. From the knowledge of the first $N_{k}$ eigenvalues, $E_{k}$, and eigenvectors, $| k \rangle$, of $\hat{H}_{el}$ we can write 
\begin{equation}
\hat{H}_{el}=\sum_{k=1}^{N_{k}}| k \rangle E_{k}\langle k | \:.
\label{Hel}
\end{equation}
Very often the eigenstates $| k \rangle$ will also be eigenstates of the operator $S^{2}$ or $J^{2}$, for vanishing and non-vanishing orbital angular momentum, respectively. If one of these two cases applies we can perform a mapping of $\hat{H}_{el}$ onto an effective spin Hamiltonian by defining a projector operator
\begin{equation}
\hat{P}=\sum_{M_{S}}^{2S+1} | S,M_{S}\rangle \langle S, M_{S}| \:.
\end{equation}
$\hat{P}$ verifies the relation $\hat{P}^{2}=\hat{P}=\hat{P}^{-1}$ in virtue of the fact the the eigenstates of the operator $S_{z}$, $|S,M_{S}\rangle$, form a complete $(2S+1)$-fold basis set. Similar expressions are build for the case of non-vanishing angular momentum. The operator $\hat{P}$ can now be used to project the first $(2S+1)$ states of the electronic Hamiltonian onto the basis set $|S,M_{S}\rangle$, thus eliminating the explicit dependence over all the electronic degrees of freedom. The resulting operator corresponds to an effective Hamiltonian function of the sole total spin operator, which exactly corresponds to the definition of the spin Hamiltonian\cite{maurice2009universal,ungur2017ab,jung2019derivation}
\begin{equation}
\hat{H}_{s}=\hat{P}\hat{H}_{el}\hat{P}^{-1}=\sum_{M_{S'}}^{2S+1}\sum_{M_{S}}^{2S+1} | S,M_{S}\rangle \langle S, M_{S}| \hat{H}_{el} | S,M_{S'}\rangle \langle S, M_{S'}| \:.
\label{HsDef}
\end{equation}
Eq. \ref{HsDef} can be recast in the form 
\begin{equation}
\langle SM_{S}| \hat{H}_{s} | SM_{S'} \rangle = \langle SM_{S}| \hat{H}_{el} | SM_{S'} \rangle = \sum_{k}^{2S+1}\langle SM_{S}| k \rangle E_{k}\langle k | SM_{S'} \rangle \:,
\label{Hsfit}
\end{equation}
which leads to a linear system of equations to solve once a specific form of $\hat{H}_{s}$ is chosen. For instance, assuming that we are dealing with a transition metal single-ion complex, the spin Hamiltonian can be chosen of the form $\hat{H}_{s}=\sum_{s,t} D_{st} \hat{S}_{s}\hat{S}_{t}$, where $(s,t)=x, y, z$. Under this assumption, Eq. \ref{Hsfit} reads
\begin{equation}
\sum_{s,t} D_{st}\langle SM_{S}| \hat{S}_{s}\hat{S}_{t} | SM_{S'} \rangle =  \sum_{k}^{2S+1}\langle SM_{S}| k \rangle E_{k}\langle k | SM_{S'} \rangle \:.
\label{fitHs2}
\end{equation}
Eq. \ref{fitHs2} has the from of a linear system $\mathbf{Ax}=\mathbf{B}$, where the vector $\mathbf{x}=(D_{11},D_{12},...,D_{33})$ spans the nine elements of $\mathbf{D}$ that we want to determine, and the vector $\mathbf{B}$ spans the $(2S+1)^{2}$ values of the right-hand side of Eq. \ref{fitHs2}. When the linear system is over-determined, it can be solved with common linear algebra  routines for least-square fitting, leading to a set of spin Hamiltonian coefficients that optimally map the electronic Hamiltonian's low-lying eigenstates. This approach is valid for any spin Hamiltonian term that has its origin in the sole electronic degrees of freedom, such as the Heisenberg Hamiltonian and the Crystal Field effective Hamiltonian commonly used for Lanthanide complexes\cite{ungur2017ab}. A similar approach can be used to include the effect of the electronic Zeeman interaction, which leads to the mapping of the effective Zeeman spin Hamiltonian\cite{chibotaru2012ab,singh2018challenges}. Other mapping procedures exist for interactions involving nuclear spins, such as the Hyperfine one\cite{neese2009spin}. It should also be noted that the possibility to find a good solution to Eq. \ref{Hsfit} lies on the assumption that we have knowledge of the function $\hat{H}_{s}$ that fits the spectrum of $\hat{H}_{el}$. For common classes of compounds this is usually known, but some care in this choice should be applied. \\

Now that we have a procedure to map $\hat{H_{s}}$ from the eigenstates of $\hat{H}_{el}$, it remains to address the question of how to determine the latter. Electronic structure methods have evolved during the years in order to address this point and have now reached a very high degree of sophistication and accuracy. Wave function based methods, such as complete active space, are among the most popular methods to implement Eq. \ref{Hsfit}, as they directly provide $E_{k}$ and $| k\rangle$\cite{ungur2017ab,neese2019chemistry}. Density functional theory can also be used to determine spin Hamiltonian coefficients but within a different mapping formalism from the one presented here. DFT's solutions cannot in fact be easily interpreted as electronic excited states with a well defined $S^{2}$ value and a different expression from Eq. \ref{Hsfit} is thus necessary. An extensive review of electronic structure and spin Hamiltonian mapping is beyond the scope of this work and we redirect the interested reader to the rich literature on this field.\cite{neese2009spin,ungur2017ab,jung2019derivation,neese2019chemistry} \\

Regardless the electronic structure method used to perform the mapping, calculated spin Hamiltonian parameters are very sensitive to slight structural changes of the molecule under investigation. A common choice is to use DFT to optimize the molecular structure in gas-phase starting from the atomic positions obtained from crystallography experiments. In case of rigid coordination spheres and ligands, this approach provides a good estimation of the intra-molecular distances, thus leading to an accurate prediction of the molecular magnetic properties. This is however not always the case and accounting for the effect of inter-molecular interactions due to the crystal packing is often of crucial importance. As detailed in the previous section, the most appropriate computational strategy for dealing with molecular structure's optimization in condensed-phase involves the use of periodic DFT, which enables a full unit-cell optimization. It is important to note that the effect of crystal packing on molecular's magnetic properties is mostly indirect and operates through the steric effects of molecule-molecule interactions. This is a very convenient situation, as it makes it possible to assume that the electronic structure of a molecule is the same in- or outside the crystal for the same set of intra-molecular distances. It is then possible to perform the simulation of spin Hamiltonian using CASSCF methods, only available for gas-phase calculations, while steric effects of crystal packing are introduced by using the coordinates of a single molecule as obtained at the end of the periodic crystal's optimization. Early attempts to also include long-range electrostatic effects in the simulation of magnetic properties showed that the gas-phase approximation to the simulation of spin Hamiltonian parameters might not always be appropriate\cite{briganti2019covalency}, but more systematic investigations are needed. 

\subsection*{Linear spin-phonon coupling strength}

Now that we have established a procedure to predict the coefficients of $\hat{H}_{s}$ as function of molecular geometry, we can use it to estimate the spin-phonon coupling coefficients as numerical derivatives of the former. One possible way to approach this problem is to generate a set of distorted molecular structures according to the atomic displacements associated with each phonon $Q_{\alpha \mathbf{q}}$ by mean of the inverse relation of Eq. \ref{Dispq}. Once the spin Hamiltonian coefficients have been determined for each one of these configurations, a numerical first-order derivative can be estimated with finite-difference methods. This method was first used by in Refs. \cite{escalera2017determining,lunghi2017role,goodwin2017molecular} but it can be modified to reduce its computational cost\cite{lunghi2017intra}. \\

We start noting that there is an imbalance between the number of molecular degrees of freedom ($3N_{at}$) and the number of the unit-cell degrees of freedom ($3N$). The two in fact coincide only when the unit cell contains a single molecular unit and no counter-ions or solvent molecules. Since this is rarely the case, performing the numerical differentiation with respect to phonons, which are proportional to the unit-cell degrees of freedom, implies a redundant number of calculations. The situation becomes even more striking when a number $N_{q}$ of $\mathbf{q}$-points are considered to sample the Brillouin zone. In this case the total number of phonons used to describe the crystal's vibrational properties becomes $3N N_{q} \gg 3N_{at}$.\\

\begin{figure}[h!]
 \begin{center}
  \includegraphics[scale=1]{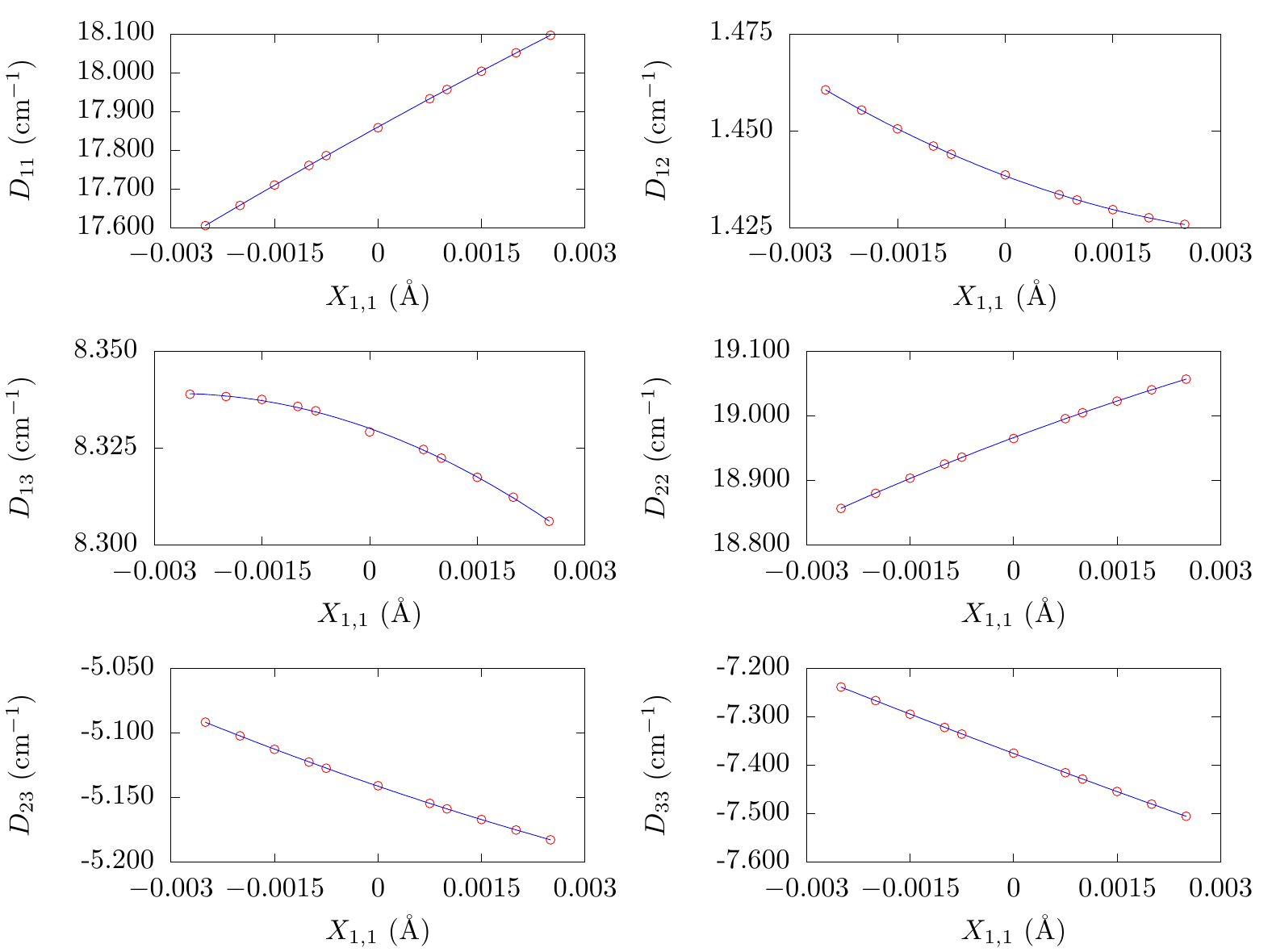}
 \end{center}
 \caption{\textbf{Spin-phonon coupling coefficients.} The derivatives of the anisotropy tensor for the molecule [(tpaPh)Fe]$^{-}$\cite{harman2010slow} are computed by displacing the Cartesian components of the DFT optimized structure.
 The value of each independent component of the tensor $\mathbf{D}$ is plotted as function of the displacement of the Cartesian coordinate $x$ of the central Iron atom. This figure is adapted from ref. \cite{lunghi2017intra}.}
 \label{sph_deriv_fesim}
\end{figure}
\begin{figure}[h!]
 \begin{center}
  \includegraphics[scale=1]{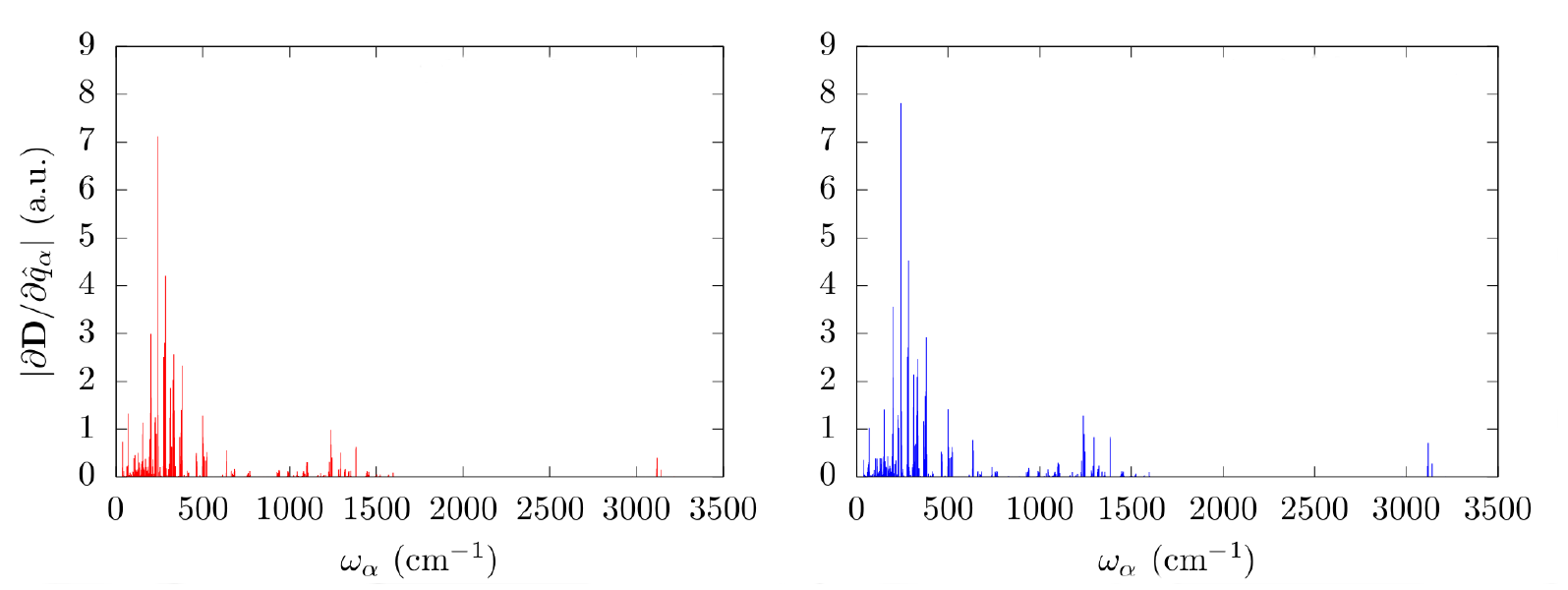}
 \end{center}
 \caption{\textbf{Comparison of spin-phonon coupling simulation methods.} The spin-phonon coupling intensity as function of vibrational frequency for [(tpaPh)Fe]$^{-}$\cite{harman2010slow} is computed starting from coefficients computed by differentiating with respect of Cartesian and phonon displacements, respectively\cite{lunghi2017intra}. This figure is adapted from ref. \cite{lunghi2017intra}.}
 \label{sph_deriv_method}
\end{figure}

A more convenient strategy requires i) the calculation of the derivatives of the spin Hamiltonian parameters with respect to molecular Cartesian coordinates, $(\partial \hat{H}_\mathrm{s}/\partial x_{i})$, and ii) their transformation into the basis set of the crystal coordinates by means of the expression
\begin{equation}   
\Big(\frac{\partial \hat{H}_\mathrm{s}}{\partial Q_{\alpha\mathbf{q}}}\Big)=\sum_{l}^{N_{cells}}\sum_{i}^{N}\sqrt{\frac{\hbar}{N_{q}\omega_{\alpha\mathbf{q}}m_{i}}}e^{i\mathbf{q}\cdot \mathbf{R}_{l}} L^{\mathbf{q}}_{\alpha i} \Big(\frac{\partial \hat{H}_\mathrm{s}}{\partial x_{li}}\Big)\:,
\label{sphcart}
\end{equation}
where $N_{q}$ is the number of $\mathbf{q}$-points used. It is important to note that the sums over $l$ and $i$ involve all the atoms of the crystal, but, according to our assumptions, only the derivatives with respect to the atoms of the single molecule we selected for the spin Hamiltonian calculations will be non-zero, therefore reducing the sum to only $3N_{at}$ degrees of freedom. Nonetheless, the effect of all the phonons of the Brillouin zone can be accounted for by this method and it does not represent an approximation \textit{per se}, but simply the most efficient way to perform the estimation of spin-phonon coupling coefficients for molecular crystals.\\

To summarize, the strategy to estimate spin-phonon coupling coefficients requires distorting the molecular structure along every Cartesian degrees of freedom (one at the time) by a value $\pm\delta$ or multiples. The spin Hamiltonian coefficients of each distorted molecule are computed and then fitted with a polynomial expression
\begin{equation}
    f(X_{is})=C_{0}+C_{1}X_{is}+C_{2}X_{is}^{2}+C_{3}X_{is}^{3}+ ...\:,
\end{equation}
where $C_{1}=(\partial \hat{H}_\mathrm{s}/\partial X_{is})$ and $C_{0}=H_{s}$ evaluated at the equilibrium geometry. The order of the polynomial function should be judged based on the data and can be changed to check the reliability of the fit. Fig. \ref{sph_deriv_fesim} shows the results presented in Ref. \cite{lunghi2017intra} concerning the calculation of the spin-phonon coupling of the Fe$^{2+}$ single-ion magnet [(tpaPh)Fe]$^{−}$\cite{harman2010slow}, where the ligand H$_{3}$tpaPh is tris((5-phenyl-1H-pyrrol-2-yl)methyl)amine.\\

Finally, the excellent agreement between results obtained by differentiating with respect to Cartesian coordinates or phonon displacements\cite{lunghi2017intra} is depicted in Fig. \ref{sph_deriv_method}. The agreement between the two sets of spin-phonon coupling predictions is excellent and the very small differences come from the use of slightly different derivation protocol.

\subsection*{Quadratic spin-phonon coupling strength}

\begin{figure}[t]
 \begin{center}
  \includegraphics[scale=1]{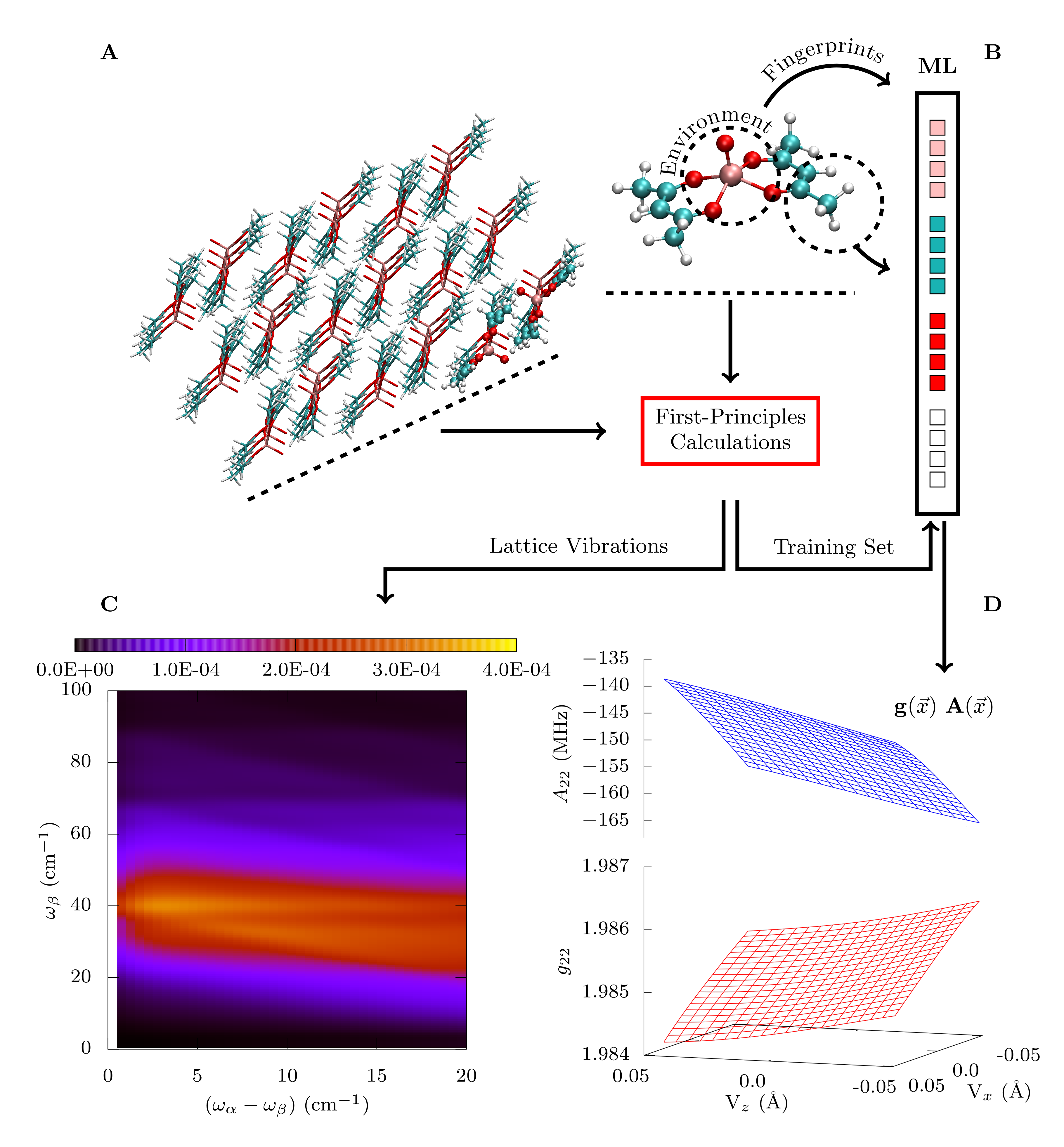}
 \end{center}
 \caption{\textbf{First-principles and ML approach to lattice and spin dynamics.} (A) The $3\times3\times3$ replica of the VO(acac)2 primitive cell used for the simulation of the crystal’s vibrational properties and the structure of the isolated molecular unit used to generate the training set for the ML algorithm. (B) The schematic structure of the ML algorithm used to predict the magnetic properties as a function of the general atomic
displacements. Each atomic environment is converted into a vector of fingerprints that determine the atomic contributions to the $\mathbf{A}$ and $\mathbf{g}$ tensors. (C) The Fourier transform of the two-phonon correlation function integrated over the Brillouin zone. (D) Examples of ML predictions for the hyperfine and Lande' tensors as function of the Vanadium displacements along $x$ and $z$. Reprinted with permission from ref. \cite{lunghi2020limit}. Copyright 2020 American Chemical Society.}
 \label{MLVOacac}
\end{figure}

A similar strategy can be used to estimate second-order derivatives\cite{lunghi2020limit}. However, in this case one needs to scan the spin Hamiltonian parameters with respect to two molecular degrees of freedom at the time and interpolate the resulting profile with a two-variable polynomial functions. The coefficients in front of the second order mix-variables term can then used to compute the second-order spin-phonon coupling coefficients with the relation
\begin{equation}   
\Big(\frac{\partial^{2} \hat{H}_\mathrm{s}}{\partial Q_{\alpha\mathbf{q}}Q_{\beta\mathbf{q'}}}\Big)=\sum_{l}^{N_{cells}}\sum_{is}^{N,3}\sum_{l'}^{N_{cells}}\sum_{i's'}^{N,3}\sqrt{\frac{\hbar}{N_{q}\omega_{\alpha\mathbf{q}}m_{i}}}e^{i\mathbf{q}\cdot \mathbf{R}_{l}} L^{\alpha\mathbf{q}}_{is} \sqrt{\frac{\hbar}{N_{q'}\omega_{\beta\mathbf{q'}}m_{i'}}}e^{i\mathbf{q'}\cdot \mathbf{R}_{l'}} L^{\beta\mathbf{q'}}_{i's'} \Big(\frac{\partial^{2} \hat{H}_\mathrm{s}}{\partial X^{l}_{is}\partial X^{l'}_{i's'}}\Big)\:.
\label{SPH2D}
\end{equation}

The simulation of the second-order derivatives poses serious challenges from a computational point of view, as the number of required simulations scales quadratically with the number of atoms and the mesh used for the polynomial interpolation. In Ref. \cite{lunghi2020limit}, a solution to this problem based on machine learning was brought forward. The idea is to first learn a general function $f_{ML}(X_{is},\alpha_{i})$ that outputs the spin Hamiltonian parameters as function of molecular coordinates, $X_{is}$, and then use it to predict all the data necessary to numerically estimate the second-order derivatives of Eq. \ref{SPH2D}. The learning of such a function is done by optimizing the built-in coefficients $\alpha_{i}$ in order to reproduce the values of spin Hamiltonian of a set of reference calculations\cite{lunghi2019unified,lunghi2020surfing}. It should be noted that the method provides an advantage only if the number of calculations needed to compute the training set of $f_{ML}(X_{is},\alpha_{i})$ is smaller than the number of calculations required for a brute-force finite difference calculation. There is in fact no conceptual difference from using a polynomial function to perform a regression of values of spin Hamiltonian as function of molecular distortions or the use of machine learning. The crucial difference lies in the way the two approaches accounts for non-linearity and their reciprocal learning rate. In the case of polynomial functions this is handled by increasing the polynomial order, which rapidly leads to an increase of coefficients that needs to be determined. Machine learning offers a few advantages: i) symmetries are automatically included in the formalisms, ii) molecular structure can be described in terms of very compact descriptors, such as bispectrum components\cite{bartok2013representing}, which reduces the number of independent variables that needs to be determined, and iii) the function is represented by a (shallow) neural network able to optimally fit any function. Two flavours of machine learning methods were recently adopted to predict the spin-phonon coupling coefficients of magnetic molecules, both in the context of solid-state perturbation theory\cite{lunghi2020limit} and molecular dynamics of molecules in liquid solution\cite{lunghi2020insights}. A schematic representation of the former method, based on Ridge regression and bispectrum components as molecular fingerprints is reported in Fig. \ref{MLVOacac}. In a nutshell, a supercell approach to the calculation of phonons is used in combination with DFT. The optimized molecular structure is then distorted in the range of 1000 times by applying small random perturbations of the atomic positions. The spin Hamiltonian coefficients of each distorted structure are determined with electronic structure methods and form the basis for the generation of a ML model. A part of this set is usually not used for training the ML model but just for testing purposes. The ML model used in Ref. \cite{lunghi2020limit} predicted the spin Hamiltonian coefficients by means of the knowledge of each atom's coordination environment, as graphically sketched in Fig. \ref{MLVOacac}B. After a successful training of the ML, it is possible to very quickly scan the spin Hamiltonian coefficients along any pair of Cartesian directions\cite{lunghi2020surfing}. Fig \ref{MLVOacac}D reports just an example out the many 2D plots needed to computed all the derivatives needed in Eq. \ref{SPH2D}.

\subsection*{Spin-Phonon Coupling Translational Invariance}

Similarly to what noted for the calculation of force constants, also spin-phonon coupling coefficients need to fulfill certain sum rules that arise from the symmetries of the problem. For instance, let us assume that the $\mathbf{D}$ tensor of a molecule depends linearly with respect to atomic positions
\begin{equation}
 D_{\alpha\beta}(X_{is})=D_{\alpha\beta} + \sum_{is}\left(\frac{\partial D_{\alpha\beta}}{\partial X_{is}}\right) (X_{is}-X_{is}^{0}).
\end{equation}
If the entire system is rigidly translated in space by an amount $a$ \AA, then $(X_{is}-X_{is}^{0})=a$, leading to 
\begin{equation}
 D_{\alpha\beta}(X_{is})=D_{\alpha\beta} + a\sum_{is}\left(\frac{\partial D_{\alpha\beta}}{\partial X_{is}}\right).
\end{equation}
Finally, since the value of $\mathbf{D}$ is independent on the particular position of the molecule in space, $D_{\alpha\beta}(X_{is}^{0}+a)=D_{\alpha\beta}(X_{is}^{0})$. Therefore we obtain the condition 
\begin{equation}
 \sum_{i}\left(\frac{\partial D_{\alpha\beta}}{\partial X_{is}}\right)=0 \:, \quad \text{for each } s\:.
\end{equation}

Similarly to the method used to regularize the force constants in phonons calculations, translational invariance conditions can be enforced on the calculated derivatives of the spin Hamiltonian tensors by rescaling the values by a mean deviation of this condition. For instance, for the elements $\alpha\beta$ of the tensor \textbf{D}, the correction for the $s$ Cartesian direction reads
\begin{equation}
\Big( \frac{\partial D_{\alpha\beta}}{\partial X_{is}} \Big)=\Big( \frac{\partial D_{\alpha\beta}}{\partial X_{is}} \Big)-Dev_{s}\:, \quad Dev_{s}=\frac{1}{N_{at}}\sum_{i}^{N_{at}} \Big( \frac{\partial D_{\alpha\beta}}{\partial X_{is}} \Big)\:,
\end{equation}
where $N_{at}$ is the number of atoms in the molecule. An equivalent expression can be used for the components of the second-order derivatives that shares the same Cartesian degrees of freedom\cite{lunghi2020limit}.

\subsection*{Spin-Phonon Relaxation Time}

The calculation of the Redfield matrices is now the last step required to compute the spin-phonon relaxation time.\\

The calculation of the elements of $R_{ab,cd}$ is relatively straightforward once phonon frequencies and spin-phonon coupling coefficients are know, except for two important technical details: the integration of the Brillouion zone and dealing with the presence of a Dirac delta\cite{lunghi2019phonons,lunghi2020multiple}. These two aspects are strongly connected and need to be discussed simultaneously. Let us start by noting that the Dirac delta in not strictly speaking a function, but a distribution in the functional analysis sense, and that it only makes sense when appearing under the integral sign. This does not lead to any issue as all the expressions of $R_{ab,cd}$ contain a sum over $\mathbf{q}$ points, which for a solid-state periodic system extend to an infinite summation, namely an integral. We thus have
\begin{equation}
    R_{ab,cd} \propto \sum_{\mathbf{q}} \delta(\omega-\omega_{\mathbf{q}})\rightarrow \int_{\mathrm{BZ}} \delta(\omega-\omega_{\mathbf{q}}) d\mathbf{q}\:.
    \label{deltasum}
\end{equation}
In fact, differently from the sparse spectra of an isolated cluster of atoms, the vibrational density of states $P(\omega)$ of a solid-state system is a continuous function of $\omega$ in virtue of an infinite number of degrees of system. Let us now discuss how to numerically implement an object such as the one in Eq. \ref{deltasum}. The integration with respect to the q-points can be simply dealt with by discretizing the first Brillouin zone in a finite mesh of uniformly sampled q-points. The value of $R_{ab,cd}$ is then computed for each q-point, which are then summed up. By increasing the finesse of this mesh one eventually reaches convergence to the desired accuracy and the Redfield matrix will be numerically indistinguishable from the one obtained by integration. However, as already discussed, the Dirac delta loses sense when the sign of integral is removed and we need a numerical way to address this issue. The common way to deal with the Dirac delta is to replace it with a real function that mimics its properties. A common choice is to use a Gaussian function with width $\sigma$ in the limit of $\sigma \rightarrow 0$. One can then substitute the Dirac delta functions with a Gaussian  
\begin{equation}
    \delta(x)\sim\frac{1}{\sigma\sqrt{\pi}}e^{-x^{2}/\sigma^{2}}\:,
\end{equation}
and recompute the Redfield matrix for decreasing values of $\sigma$ until convergence is reached. To summarize one needs to evaluate the Redfield matrix in the limit of infinite q-points and vanishing Gaussian smearing
\begin{equation}
R_{ab,cd} \propto \int_{\mathrm{BZ}} \delta(\omega-\omega_{\mathbf{q}}) d\mathbf{q} = \lim_{\sigma \rightarrow 0}\lim_{q \rightarrow \infty} \sum_{\mathbf{q}} \frac{1}{\sigma\sqrt{\pi}}e^{-(\omega-\omega_{\mathbf{q}})^{2}/\sigma^{2}}\:,
\label{Redlim}
\end{equation}

One final important care in evaluating the limits of Eq. \ref{Redlim} is needed: the limit with respect to $\mathbf{q}$ must always be evaluated first than the one with respect to $\sigma$. This is required because in a simulation we are always dealing with a finite number of phonon states, regardless the number of $N_{q}$, and if $\sigma$ is reduced arbitrarily the elements of $R$ will invariably converge to zero. This is because exact degeneracy between two discrete spectra (the spin and the phonons ones) is never achieved and without any smearing, the Dirac delta will always invariably result in a null value. However, we know that the discrete nature of the phonon spectrum is only due to the use of a finite mesh of q-points and it is not intrinsic of a solid-state system. For this reason, $\sigma$ can never be chosen to be much smaller than the finesse of the vibrational spectrum. \\

To summarize, the correct numerical procedure to evaluate Eq. \ref{Redlim} is the following: i) $R$ is evaluated for a large smearing, e.g. $\sigma=10$ cm$^{-1}$, and one q-point, namely the $\Gamma$ point; ii) Keeping $\sigma$ fixed, $R$ is recalculated for increasing values of $N_q$ until the desired level of convergence is reached; iii) $\sigma$ is reduced and $N_q$ is increased further until new convergence is reached; iv) step iii) is repeated until convergence with respect to both $\sigma$ and $N_{q}$ is obtained. \\

If this procedure is employed, the elements of the Redfield matrix converge to the right value without leading to any divergence of the relaxation time\cite{lunghi2019phonons,lunghi2020multiple}. Until now we have considered a Gaussian smearing to reproduce the Dirac delta function but that is not the only option. Another common choice is the use of a Lorentzian profile, with line-width $\Delta$. Both Gaussian and Lorentzian functions converge to the Dirac delta for their width tending to zero, however, the rate of convergence of the two functions is quite different. The tails of the Gaussian function indeed drop to zero exponentially, while those of a Lorentzian function they do so quadratically. Therefore, in order for a Gaussian and Lorentzian smearing to give results in agreement, the latter must be much smaller than the former. Finally, it is important to remark that in this discussion we have assumed that smearing is only used to represent the Dirac delta, which correspond to a perfectly harmonic phonons bath. In section \ref{theory} we have also contemplated the use of Lorentzian smering, but for a different reason. A Lorentzian line-width is in fact the natural line-shape of the Fourier transform of the phonon correlation function in the presence of phonon-phonon dissipation\cite{lunghi2017role}. The use of Lorentzian smearing in these two circumstances should not be confused. In the former case $\Delta$ must tend to zero in order to converge to a Dirac delta, while in the latter case $\Delta$ is set by the phonon bath properties and will in general have a finite value in the order of cm$^{-1}$ or fractions. \\

Once the Redfield matrix is computed, all is left to do is to solve the equation of motion for the density matrix
\begin{equation}
     \frac{d\rho_{ab}(t)}{dt}=\sum_{cd}R_{ab,cd}\rho_{cd}(t)\:.
\end{equation}
In order to treat a mathematical object with four indexes we introduce the Liuville space, $\mathcal{L}$ which is defined as the tensor product of the original Hilbert space, $\mathcal{L}=\mathcal{H}\otimes\mathcal{H}$. The space $\mathcal{L}$ is therefore spanned by vectors of size $size(\mathcal{H})^{2}$ that lists all the possible pairs of vectors of the Hilbert space. What is expressed as a matrix in the Hilbert space, then becomes a vector in the Liuville space. For instance, the density matrix $\rho_{ab}$ itself would read $\rho_{i}$, where $i=ab$. Accordingly, the Redfield matrix $R_{ab,cd}$ becomes $R_{ij}$, where $i=ab$ and $j=cd$. An operator in matrix form in the Liuville space is called a super-operator and indicated with a double hat, e.g. $\hat{\hat{R}}$, to highlight that it acts on normal operators to give another operator. Finally, the equation of motion of the density matrix in the interaction picture reads
\begin{equation}
     \frac{d\rho_{i}(t)}{dt}=\sum_{j}R_{ij}\rho_{j}(t)\:, \quad \text{or equivalently,} \quad \frac{d\hat{\rho}}{dt}=\hat{\hat{R}}\hat{\rho}(t)
\end{equation}
This differential equation can now be solved with common numerical strategies. Once the Redfield matrix has been computed and diagonalized, one can construct the propagator as 
\begin{equation}
    L_{ij}(t)=\sum_{k}V_{ik}e^{i\lambda_{k}t}V^{-1}_{kj}\:,
\end{equation}
where $V_{ij}$ and $\lambda_{k}$ are the eigenvectors and eigenvalues of $R_{ij}$, respectively, and $V^{-1}_{ij}$ are the elements of the inverse matrix of the eigenvectors of $R_{ij}$. The evolution of the density matrix in the interaction picture therefore is 
\begin{equation}
  \rho_{i}(t)=\sum_{j}L_{ij}(t)\rho_{j}(t=0)\:.
\end{equation}
Importantly, given the properties of $R$, $\lambda_{k}$ should always have one null eigenvalue and all the others of negative sign. Moreover, the eigenvector associated with the null eigenvalue will correspond to the equilibrium distribution of spin states' population. These properties of the propagator enforce that the time evolution of $\hat{\rho}$ in fact corresponds to a relaxation process, where out-of-diagonal elements goes to zero an the diagonal terms tend to their equilibrium value for $t\rightarrow\infty$.\\

Once the evolution of $\hat{\rho}$ has been determined, it is possible to compute the time evolution of the magnetization with the expression $\vec{\mathbf{M}}(t)=\mathrm{Tr}\left(\vec{\mathbf{S}}\hat{\rho}(t)\right)$, where the trace is generally performed in the basis of the spin Hamiltonian eigenvectors. The profile of $\vec{\mathbf{M}}(t)$ can be studied as commonly done in experiments and in general different initial conditions for $\hat{\rho}(t=0)$ and different orientations of the molecule in the external field will lead to different demagnetization profiles. A common choice is to orient the molecule with the easy/hard axis along the external field and initialize $\hat{\rho}$ in such a way that the molecule is fully magnetized along the same direction. In this scenario, for a single-spin system, an exponential decay of $M_{z}(t)$ is generally observed. The latter can be fitted with the common expression
\begin{equation}
    M_{z}(t)=\left[ M_{z}(t=0)-M_{z}^{eq} \right]e^{-t/T_1}+M_{z}^{eq}
\end{equation}
where $M_{z}^{eq}$ is the equilibrium value of the magnetization and $T_1$ (or $\tau$) is the relaxation time. Under this conditions, $1/T_{1}$ coincide with the value of the second-smallest eigenvalue of $R_{ij}$ (the first non-zero one).

\newpage
\section{The origin of spin-phonon coupling in magnetic molecules}\label{abinitio-sph}

Now that we have established a robust computational protocol to implement the theory of spin-phonon relaxation, we are well positioned to discuss applications of this method for different molecular systems. In particular we are interested in understanding how spin-phonon coupling is influenced by different molecular motions and by the chemical nature of the metal ions, ligands, and their bond.\\

\subsection*{Normal modes composition}

Let us start our journey form a series of simple $S=1/2$ mono-nuclear coordination compounds based on a V$^{4+}$ metal center. Vanadium in this oxidation state has a $[$Ar$]d^{1}$ electronic configuration and for the common square pyramidal or distorted octahedral coordination geometries the unpaired electron occupies a non-degenerate $d$ orbital. This leads to a fully quenched angular momentum and a well defined $S=1/2$ ground state. The first electronic excited state in V$^{+4}$ compounds is often found at energies above 5000 cm$^{-1}$\cite{albino2019first}, thus much higher energies than any vibrational frequency of molecular crystals. This electronic configuration makes it possible to use Vanadium complexes to study spin-phonon relaxation in a prototypical two-level systems. \\

The spin Hamiltonian for an isolated V$^{4+}$ molecule reads 
\begin{equation}
\hat{H}_{s}=\mu_{B} \vec{\mathbf{S}} \cdot \mathbf{g} \cdot \vec{\mathbf{B}} +  \vec{\mathbf{S}} \cdot \mathbf{A} \cdot \vec{\mathbf{I}}\:,
\label{acacSH}
\end{equation}
where the Lande' tensor $\mathbf{g}$ and the hyperfine coupling tensor $\mathbf{A}$ determine the spin spectrum of the compound. Additionally, when the molecule is embedded in a crystal, the dynamics of the single spin is influenced by the dipolar interactions with all the surrounding molecules. This interaction is captured by a spin Hamiltonian of the form
\begin{equation}
\hat{H}_{s}= \sum_{ij}  \vec{\mathbf{S}}_{i} \cdot \mathbf{D}_{ij}^{Dip} \cdot \vec{\mathbf{S}}_{j}\:,
\label{acacSH2}
\end{equation}
where $\mathbf{D}^{Dip}$ depends on the molecule-molecule distance as $r^{-3}$ and the magnitude and orientation of the molecular magnetic moments\cite{bencini2012epr}. 
\begin{figure*}[t]
 \begin{center}
  \includegraphics[scale=1]{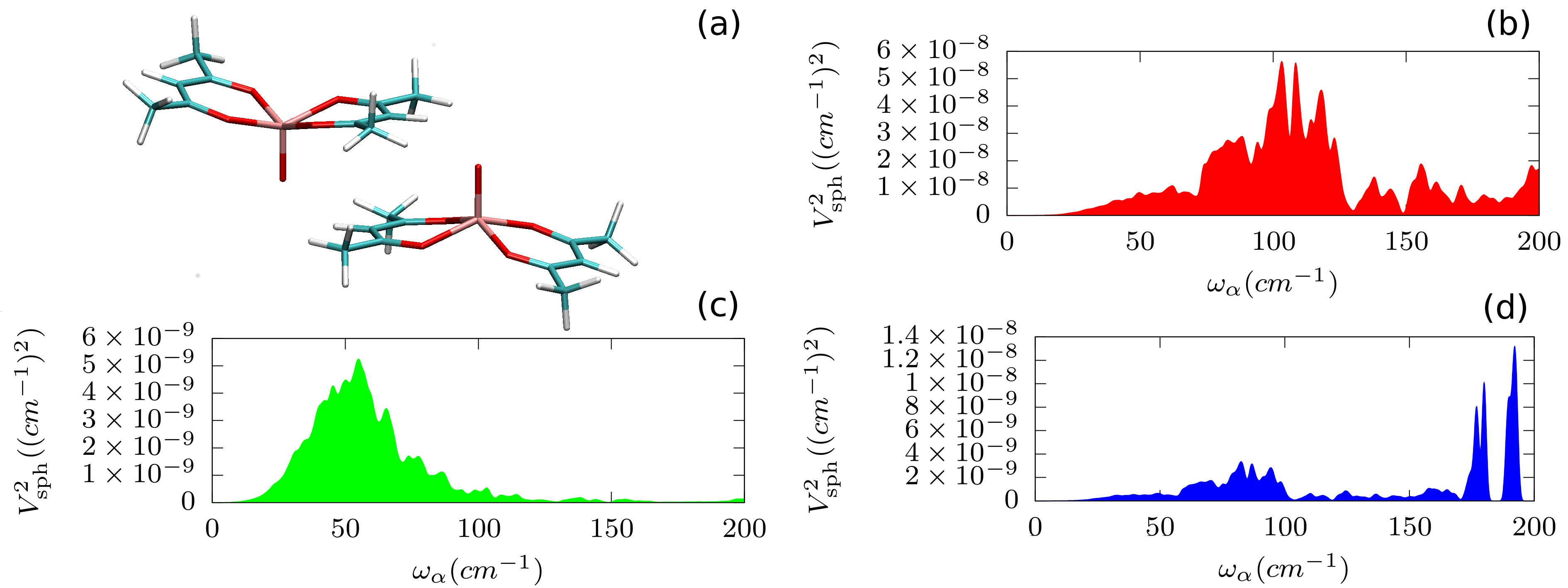}
 \end{center}
 \caption{\textbf{Molecular structure and spin-phonon coupling in VO(acac)$_{2}$.} The unit cell of VO(acac)$_{2}$ is reported in panel a. Vanadium atoms are in pink, Carbon atoms in green, Oxygen atoms in red and Hydrogen atoms in white. The the distribution of spin-phonon coupling intensity as function of phonon frequencies for the g-tensor, the dipolar D-tensor and the the A-tensor are reported in panels b, c, and d, respectively. Reprinted from ref. \cite{lunghi2019phonons}. $\copyright$ The Authors, some rights reserved; exclusive licensee AAAS. Distributed under a CC BY-NC 4.0 License.}
 \label{DOS1VOacac}
\end{figure*}
Besides shaping the static spin spectrum, the tensors $\mathbf{g}$, $\mathbf{A}$ and $\mathbf{D}^{Dip}$ are also the interactions that get modulated by molecular motion and that are at the origin of spin-phonon coupling. In Refs. \cite{lunghi2019phonons,lunghi2020limit,garlatti2020unveiling,albino2021temperature}, the compound VO(acac)$_2$ was studied following the methods detailed in the previous section, shedding light on the nature of molecular vibrations, and spin-phonon coupling in $S=1/2$ molecular crystals. For this purpose a $3\times3\times3$ supercell of the VO(acac)$_2$ crystal was optimized and used to sample the crystal phonons across the entire Brillouin zone\cite{lunghi2019phonons}. The unit cell of the crystal is reported in Fig. \ref{DOS1VOacac}A. Moreover, linear spin-phonon coupling coefficients were computed with Eq. \ref{sphcart} for all the interactions in Eqs. \ref{acacSH} and \ref{acacSH2}. Fig \ref{DOS1VOacac} reports the norm of spin-phonon coupling resolved as function of the phonon's angular frequency and integrated over all the Brillouin zone. This analysis reveals that the modulation of dipolar interactions reaches a maximum around 50 cm$^{-1}$ (Fig. \ref{DOS1VOacac}c), coincidentally very close to the energy of the first optical mode at the $\Gamma$-point. On the other hand, the contribution of hyperfine and Lande' tensors to spin-phonon coupling (Fig. \ref{DOS1VOacac}b,c) maintain a finite value for all energies. This observation can be explained by considering the nature of the tensors $\mathbf{g}$, $\mathbf{A}$ and $\mathbf{D}^{Dip}$. The latter only depends on the inter-molecular distance and is therefore most effectively modulated by rigid molecular translations. This sort of molecular displacements are expected to be associated with acoustic modes, which density of states increases with $\omega$ but up to a finite cut-off value that depends on the inter-molecular force constants and molecular mass. Fig \ref{DOS1VOacac} is compatible with such an interpretation with the exception of the presence of a sharp cutoff frequency. On the other hand, hyperfine and Lande' tensors represent interactions mediated by the electronic structure of the metal ion and it is reasonable to expect them to be influenced by those optical modes that modulate the metal's coordination shell. \\

\begin{figure}[t]
 \begin{center}
  \includegraphics[scale=1]{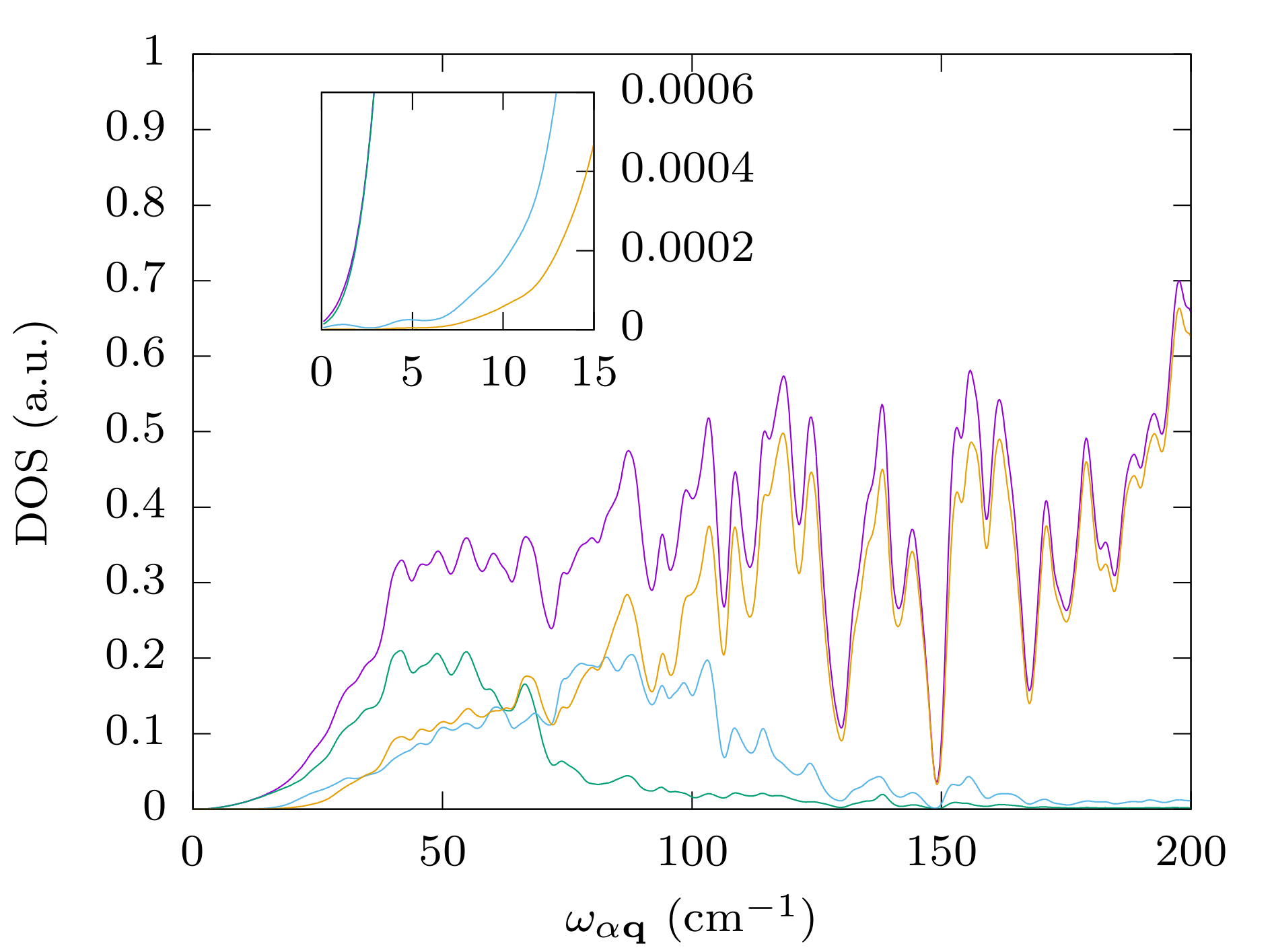}\includegraphics[scale=0.305]{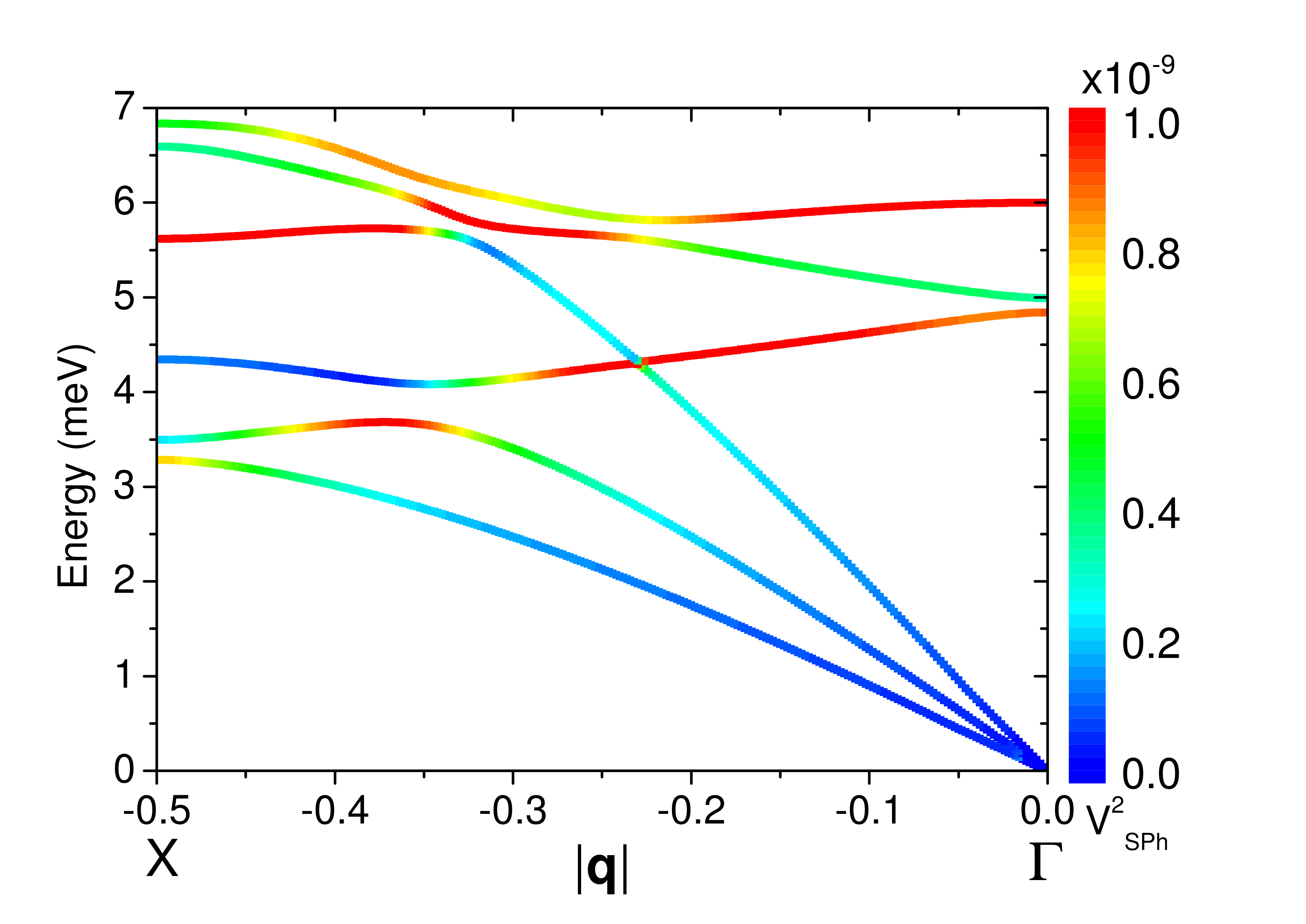}
 \end{center}
 \caption{\textbf{Mode composition and spin-phonon coupling in VO(acac)$_2$.} The left panel reports the total density of states of VO(acac)$_{2}$ (purple) and its decomposition into translational molecular contributions (green), molecular rotations (blue) and intra-molecular distortions (yellow). The right panel reports the spin-phonon coupling intensity due to the modulation of the g-tensor as function of phonon frequency and Brillouin zone vector $\mathbf{q}$ along the path $X-\Gamma$. Left panel is reprinted from ref. \cite{lunghi2019phonons}. $\copyright$ The Authors, some rights reserved; exclusive licensee AAAS. Distributed under a CC BY-NC 4.0 License. Right panel is reprinted from \cite{garlatti2020unveiling}. Copyright 2020, The Author(s).}
 \label{DOS1VOacac_dec}
\end{figure}

A full decomposition of the phonons density of states in terms of molecular motions was also performed in Ref. \cite{lunghi2019phonons} and reported here in Fig. \ref{DOS1VOacac_dec}. The total vibrational density of states is decomposed in terms of rigid molecular translations, rotations and intra-molecular distortions, revealing several important features of molecular crystals' vibrational properties. Low-energy phonons are dominated by rigid translations or rotations up to frequencies of 100 cm$^{-1}$, after which optical modes are fully characterized by intra-molecular displacements. Not surprisingly, the dipolar contribution to spin-phonon coupling closely mimics the translation contribution to phonons which peaks at 50 cm$^{-1}$. Although intra-molecular contributions become dominant only at high-frequencies, it is important to note that they are present at virtually any frequency. As shown by the inset of Fig. \ref{DOS1VOacac_dec}, a tiny amount of intra-molecular contribution to the phonons is always present due to the finite rigidity of molecular structures. Since rigid translations cannot lead to the modulation of local molecular properties such as $\mathbf{g}$ and $\mathbf{A}$, this intra-molecular contribution to acoustic phonons must be at the origin of spin-phonon coupling at frequencies much smaller than optical modes.
The right panel of Fig. \ref{DOS1VOacac_dec} provides an analysis of spin-phonon coupling in VO(acac)$_{2}$ from a slightly different perspective. In this case, the spin-phonon coupling due the modulation of the $g$-tensor was projected on the vibrational modes along a path in the reciprocal space\cite{garlatti2020unveiling}. Fig. \ref{DOS1VOacac_dec} makes it possible to appreciate the three acoustic modes with vanishing frequencies for $q\rightarrow\Gamma$. The same phonons also has a vanishing spin-phonon coupling, as acoustic modes at $q=\Gamma$ correspond to a collective translation of the entire crystal, which cannot lead to any coupling with the spin due to the isotropic nature of space. The spin-phonon coupling norm of acoustic modes reported in Fig. \ref{DOS1VOacac_dec} increases linearly moving away from the $\Gamma$-point up to an energy of $\sim 2$ meV (not shown explicitly), and abruptly changes for higher energies when the admixing with optical phonons becomes important. This is also highlighted by the presence of avoided crossing points in the phonon dispersions. Considering that the norm used in Fig. \ref{DOS1VOacac_dec} is the square of spin-phonon coupling, the latter is then found to vary as $\sqrt{q}$ for small values of $q$, or equivalently as $\sqrt{\omega}$ in virtue of a pseudo-linear dispersion relation. Although this behaviour is in agreement with the predictions of the Debye model, its origin is qualitatively different. The Debye model assumes that the magnetic properties of an atoms are directly influenced by the modulation of its distance with the atoms in the adjacent unit lattice cell as prescribed by acoustic phonons. As already discussed, the rigid translation of molecules in the crystal cannot lead to any sizeable spin-phonon coupling due to the fact that the number of atoms per unit cell is very large and that the magnetic properties of the metal ions are very well screened from inter-molecular interactions. However, since the molecules are not infinitely rigid, when molecules translate in space they also slightly deform, leading to a finite value of spin-phonon coupling. From the results of Ref. \cite{garlatti2020unveiling} we can deduce that the amount of this intra-molecular contribution to acoustic modes depends linearly with $q$ (for small $q$), thus leading to results in qualitative agreement with the Debye model at frequencies much smaller that the optical modes. \\

The study of linear spin-phonon coupling provides important insights on the origin of this interaction in terms of molecular displacements and the same principles can expected to be applicable also to higher order couplings\cite{lunghi2020limit}. \\

\begin{figure}[t]
 \begin{center}
  \includegraphics[scale=1]{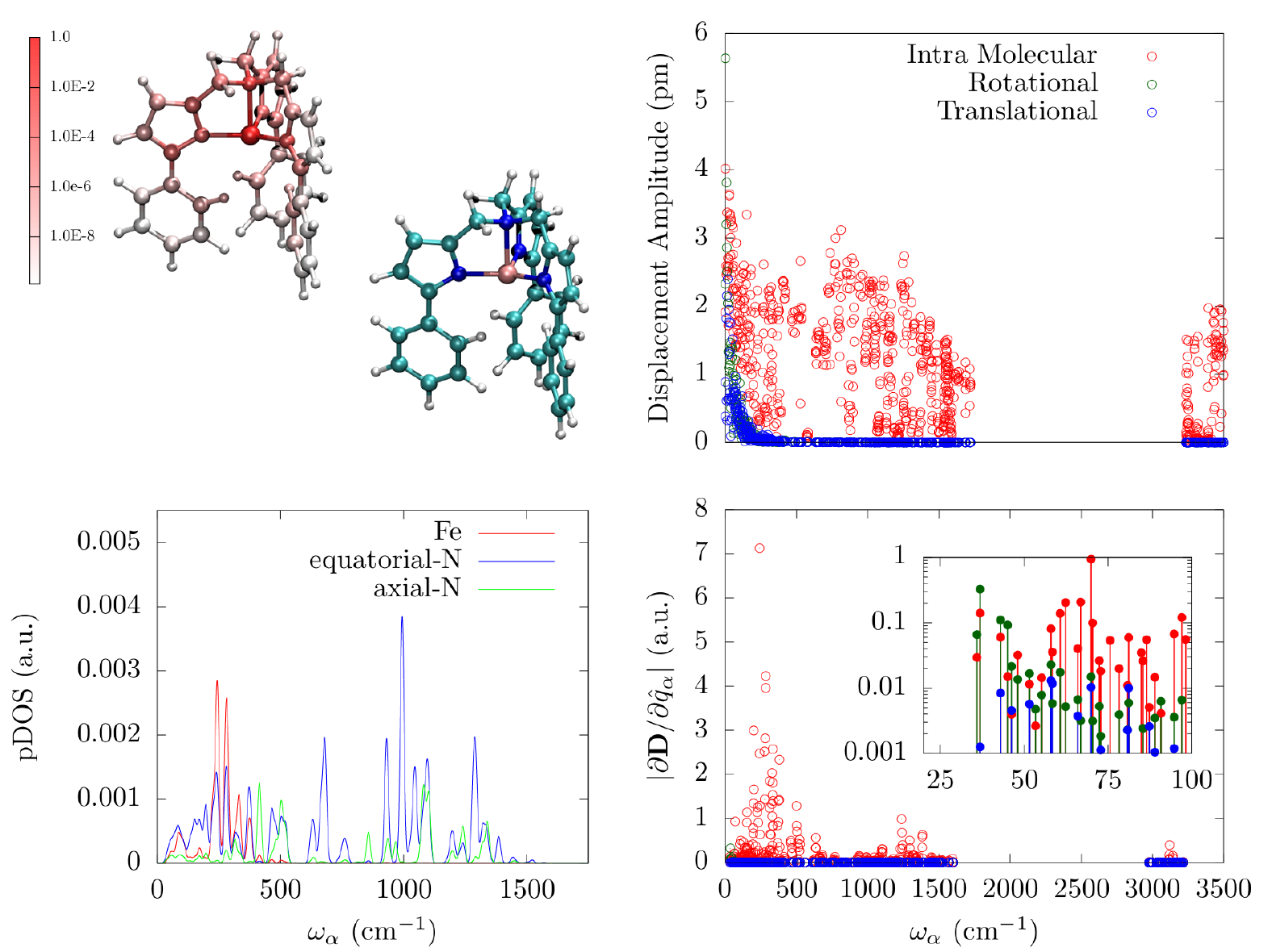}
 \end{center}
 \caption{\textbf{Spin-phonon coupling in [(tpaPh)Fe]$^{-}$.} The top-left panel reports the relative spin-phonon coupling intensity due to the modulation of the $\mathbf{D}$ by the displacement of the Cartesian coordinates of each atom. The molecular structure is reported for reference. The bottom-left panel reports the vibrational density of states due to the motion of the first coordination shell. The top-right panel reports the composition of the normal modes as function of frequency and the bottom-right panel the spin-phonon coupling intensity decomposed across translational, rotational and intra-molecular contributions. This figure is adapted from ref \cite{lunghi2017intra}.}
 \label{Fetpa_2}
\end{figure}

A similar analysis to the one just presented has also been performed for transition metal coordination compounds with S$>$1/2. Ref. \cite{lunghi2017intra} reported a study on mono-nuclear compound based on a Fe$^{2+}$ ion with a nominal $S=1$ ground state, namely [(tpaPh)Fe]$^{−}$ (H$_{3}$tpaPh=tris((5-phenyl-1H-pyrrol-2-yl)methyl)amine). The molecular structure is reported in Fig. \ref{Fetpa_2}.

The vibrations of this system were only studied at the level of the unit-cell, therefore only giving access at the $\Gamma$-point vibrations. Despite this limitation, the study showed a consistent picture of spin-phonon coupling in molecular compounds to the one just presented. For instance, a comparison between the left and right panels of Fig. \ref{Fetpa_2}, shows that local distortions of the molecular units are responsible for the largest spin-phonon coupling and that the spin-phonon coupling due to the motion of atoms outside the first coordination shell rapidly decreases with the distance from the metal atom. Indeed, the intra-molecular distortions that were found to be strongly coupled to spin are those involving the first coordination shell. 

The right panel of Fig. \ref{Fetpa_2} also reports the analysis of the molecular motion associated to each phonon and shows that optical modes are completely characterized by intra-molecular distortions for energies above 100-200 cm$^{-1}$, where the maximum of spin-phonon coupling is achieved.\\

\begin{figure}[t]
 \begin{center}
  \includegraphics[scale=1]{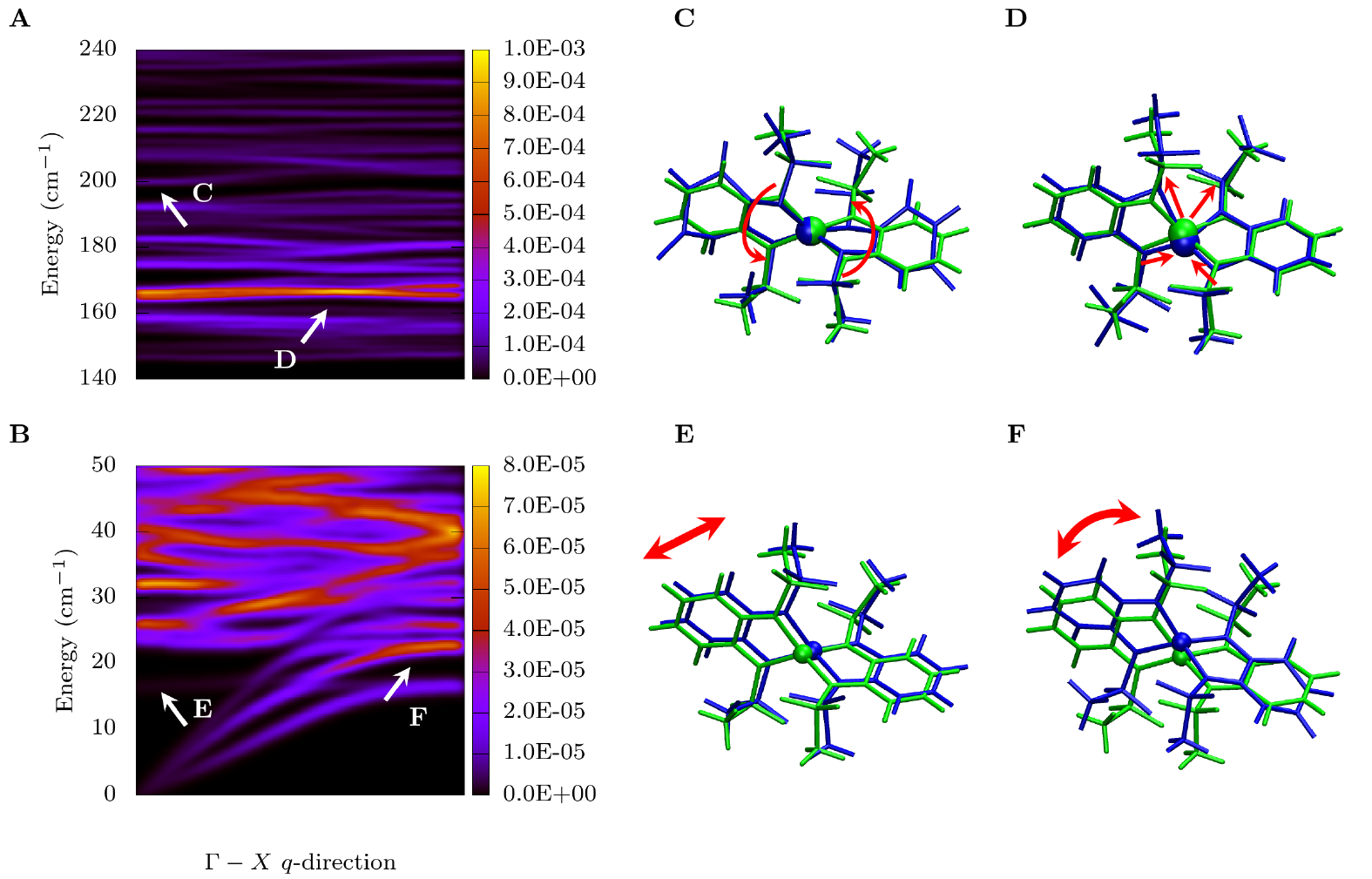}
 \end{center}
 \caption{\textbf{Spin-phonon coupling in CoL$_{2}$.} Panel A reports the spin-phonon coupling intensity for the high-energy window where vibrations are resonant with the excited KD energy. Panel B reports the low energy window proper of acoustic and the first optical modes. Panels C-F depict the molecular vibrations with energy indicated by the white arrows in panels A and B. Reprinted from ref. \cite{lunghi2020multiple}, with the permission of AIP Publishing.}
 \label{CoModes}
\end{figure}

Finally, we present the case of the high-spin Co$^{2+}$ complex [CoL$_{2}$]$^{2−}$ (H$_{2}$L=1,2-bis(methanesulfonamido)benzene)\cite{rechkemmer2016four}. [CoL$_{2}$]$^{2−}$ exhibit a tetrahedral coordination and a large uni-axial magnetic anisotropy with $D\sim 115$ cm$^{-1}$. [CoL$_{2}$]$^{2−}$ represents the only case other than VO(acac)$_{2}$ where phonon dispersions and their role in spin-phonon coupling have been studied with \textit{ab initio} methods. Fig. \ref{CoModes} reports the results of Ref. \cite{lunghi2020multiple}, where the spin-phonon coupling intensity was reported as function of frequency and $\mathbf{q}$-vector along an arbitrary direction.
Also in this case, optical modes are found to be significantly more coupled than acoustic phonons, except at border-zone, where they have similar energies to the optical ones. The left panel of Fig. \ref{CoModes} reports the atomic displacements associated with some of the phonons and confirm that local distortions of the coordination shell strongly affect the magnetic degrees of freedom. Here we note a typical scenario, where low energy optical modes are an admixture of rigid rotations and delocalized intra-molecular translations, while high-energy optical modes are strongly localized on specific intra-molecular degrees of freedom. When the latter involve the first-coordination shell of the metal, they posses a large coupling with the spin.\\

Finally, it is worth mentioning that the experimental estimation of spin-phonon coupling is also possible. Far infra-red or Raman spectroscopies recorded in the presence of external magnetic fields have been used to extract the spin-phonon coupling coefficients for the modes in resonance with spin transitions\cite{moseley2018spin,stavretis2019spectroscopic,moseley2020inter,blockmon2021spectroscopic,marbey2021}. Once combined with ab initio simulations, these experimental techniques provide one of the most direct insights on spin-phonon coupling. 

\newpage
\subsection*{The coordination bond and ligand field contributions}

The studies discussed so far highlight important features of spin-phonon coupling in crystals of magnetic molecules but do not address the critical point of how the chemistry of a compound influences this interaction and the nature of the lattice vibrations.\\

\begin{figure}[h!]
 \begin{center}
  \includegraphics[scale=1.5]{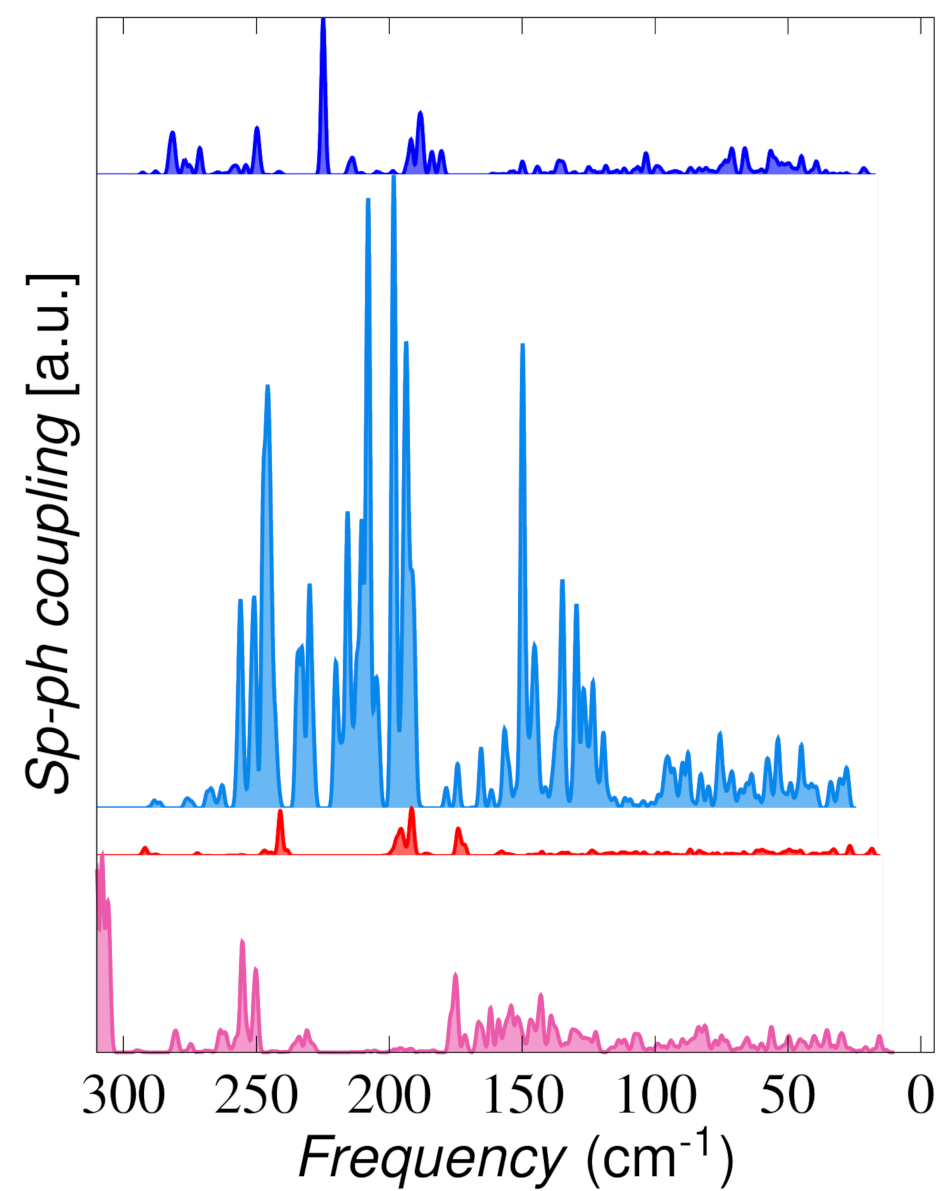}\includegraphics[scale=1.5]{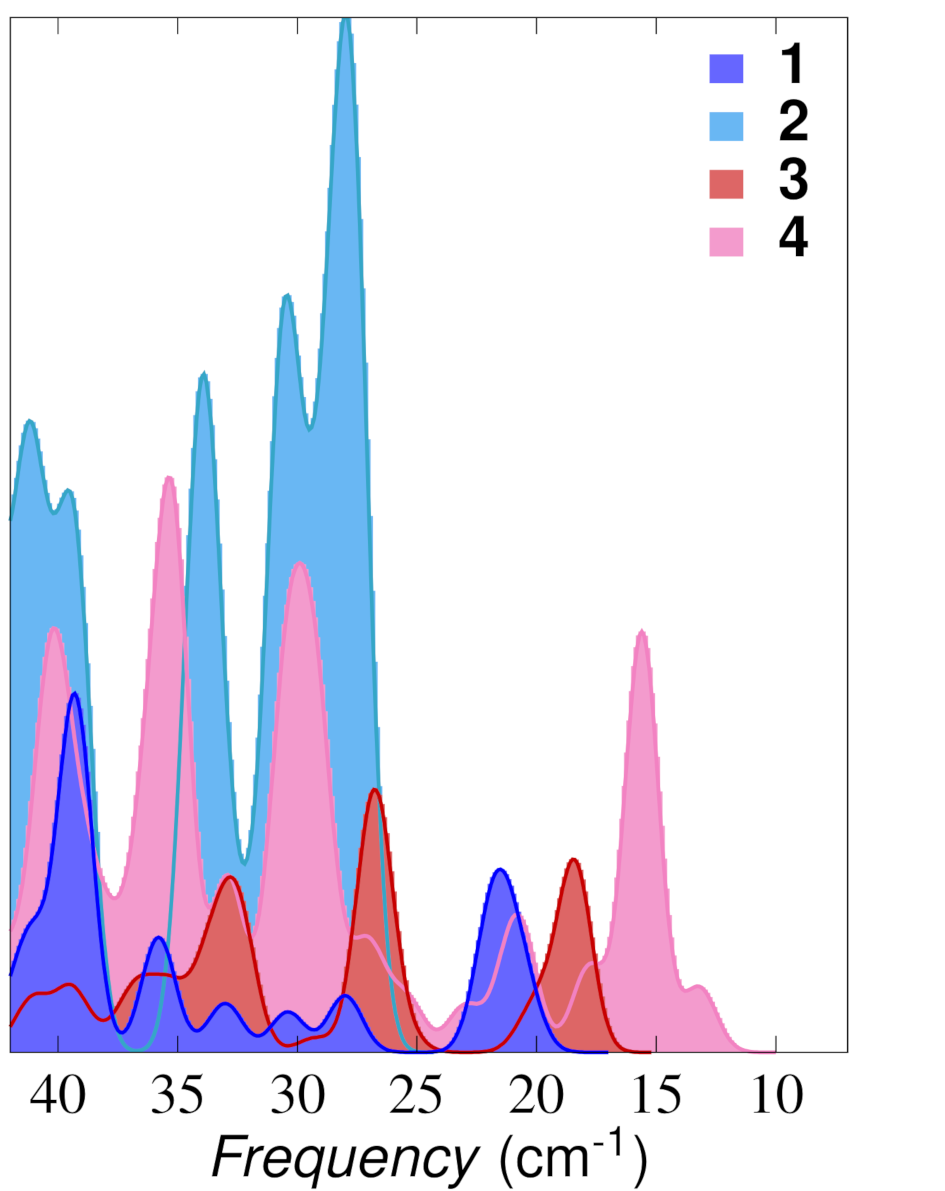}
 \end{center}
 \caption{\textbf{Spin-phonon coupling in V$^{4+}$ complexes.} The spin-phonon coupling density due to the g-tensor anisotropy is computed as function of vibrational modes' frequency for [VO(cat)$_{2}$]$^{2-}$ (1), [V(cat)$_{3}$]$^{2-}$ (2), [VO(dmit)$_{2}$]$^{2-}$ (3), and [V(dmit)$_{3}$]$^{2-}$ (4). Reprinted with permission from ref. \cite{albino2019first}. Copyright 2019 American Chemical Society.}
 \label{albino}
\end{figure}

Let us start once again from $S=1/2$ systems. In the work of Albino et al.\cite{albino2019first}, four complexes were studied: two penta-coordinated vanadyl (VO$^{2+}$) and two hexa-coordinated V$^{4+}$ molecules, where the coordination is
obtained by catecholate and dithiolate ligands, namely [VO(cat)$_{2}$]$^{2-}$, [V(cat)$_{3}$]$^{2-}$, [VO(dmit)$_{2}$]$^{2-}$, and [V(dmit)$_{3}$]$^{2-}$, where cat=catecholate and dmit=1,3-dithiole-2-thione-4,5-dithiolate. The analysis of spin-phonon coupling strength distribution as function of the frequency is reported in Fig. \ref{albino} for all four compounds. The notable result is that the molecules with cat ligands present higher frequencies of vibrations with respect to the corresponding dmit ligands, but at the same time overall larger spin-phonon coupling coefficients. The first observation can be traced back to the lower mass and harder nature of the Oxygen donor atom of Cat with respect to the Sulphur of dmit. To understand the effect ligand's chemical nature on spin-phonon coupling coefficients it was instead proposed a correlation with the static $g$-shifts, which depend on the position of the electronic excited states, in turns correlated to ligand's donor power and covalency of the coordination bond. These observations also advances the hypothesis that the $g$-tensor could be used as a proxy for spin-phonon coupling, pointing to isotropic ones associated with low spin-phonon coupling and \textit{vice versa}. \\

These results for the spin-phonon coupling associated to the modulation of the $g$-tensor have also been formalized in the context of ligand field theory\cite{mirzoyan2020dynamic} and multi-reference perturbation theory\cite{lunghi2019ligand}. Following the formalism of dynamical ligand field proposed by Mirzoyan et al.\cite{mirzoyan2021deconvolving}, we express the $g$-tensor in terms of second-order perturbation theory over $d$ orbitals 
\begin{equation}
 g_{ij}=g_{e}-\sum_{v}\frac{\zeta}{S\Delta_{0v}}\langle \psi_{0} | \hat{l}_{i} | \psi_{v} \rangle\langle \psi_{v} | \hat{l}_{j} | \psi_{0} \rangle\:,
\label{gxx}
\end{equation}
where $\hat{l}_{i}$ is a component of the electronic angular momentum of the metal ion, $\Delta_{0v}$ is the energy separation between the ground state and the $v$-excited state characterized by the wave functions $\psi_{0}$ and $\psi_{v}$, respectively. Finally, $\zeta$ is the free-ion spin-orbit coupling constant and $S$ is the value of the molecular ground-state spin.
Considering once again the case of [V(cat)$_{3}$]$^{2-}$, where the V$^{4+}$ ion experience a distorted-octahedral crystal field, the ground-state and the first excited-state wave functions $\psi_{0}$ and $\psi_{1}$ can be written as
\begin{align}
& | \psi_{0} \rangle = \alpha |d_{z^{2}} \rangle + \alpha' |\phi_{0} \rangle \\
& | \psi_{1} \rangle = \beta |d_{yz} \rangle + \beta' |\phi_{1} \rangle
\label{ligwave}
\end{align}
where $\alpha$ and $\beta$ describe the contribution of V$^{4+}$'s $d$-like orbitals that participate to $g_{xx}$, while $\alpha'$ and $\beta'$ represent the contribution of the ligands' orbitals. Assuming that only the portion of the orbitals localized on the V$^{4+}$ centre can contribute to the integrals of Eq. \ref{gxx} and substituting Eq. \ref{ligwave} into Eq. \ref{gxx}, one obtains that 
\begin{equation}
\delta g_{xx}=g_{xx}-g_{e}=-6\zeta\frac{\alpha^{2}\beta^{2}}{\Delta_{01}}=-\zeta_{\mathrm{eff}}\Delta_{01}^{-1}\:, 
\label{gshlf}
\end{equation}
It is then possible to derive an analytical expression for spin-phonon coupling by taking the derivative of Eq. \ref{gshlf} with respect to the atomic positions $X_{is}$
\begin{align}
 \Big(\frac{\partial g_{xx}}{\partial X_{is}}\Big)_{0} & = 6\zeta \frac{\alpha^{2}\beta^{2}}{\Delta_{01}^{2}}\Big(\frac{\partial \Delta_{01}}{\partial X_{is}}\Big)_{0}- 6\zeta \Delta_{01}^{-1}\Big(\frac{\partial (\alpha^{2}\beta^{2})}{\partial X_{is}}\Big)_{0}  \\
 & = -\frac{\delta g_{xx}}{\Delta_{01}}\Big(\frac{\partial\Delta_{01}}{\partial X_{is}}\Big)_{0}- 6\zeta \Delta_{01}^{-1}\Big(\frac{\partial (\alpha^{2}\beta^{2})}{\partial X_{is}}\Big)_{0}\:.
\label{gsphlf}
\end{align}
Eq. \ref{gsphlf} shows that there are two possible contributions to spin-phonon coupling due to the modulation of the $g$-tensor. The first one is due to the modulation of the first excited state's energy and the second one is due to the modulation of the metal-ligand hybridization due to the coordination bond. Both DFT and CASSCF results confirm that the modulation of covalency is not the dominant contribution to Eq. \ref{gsphlf}\cite{mirzoyan2020dynamic,lunghi2019ligand}, thus confirming the important role of $\Delta_{01}$. These observations also lead to the conclusion that spin-phonon coupling in $S=1/2$ systems is proportional to the static $g$-tensor, which can be used as a simple rule-of-thumb to rank potential slow relaxing molecules. This hypothesis has also been recently tested on a series of iso-structural $S=1/2$ complexes with V$^{4+}$, Nb$^{4+}$, and Ta$^{4+}$, where it was possible to correlate relaxation time to the experimental $g$-shift\cite{chakarawet2021effect}. \\

These studies on the contributions to spin-phonon coupling also make it possible to rationalize a body of recent literature that proposes a variety of apparently orthogonal strategies to improve spin-lattice relaxation time in magnetic molecules. For instance, in light of Eq. \ref{gsphlf}, several experimental works reporting respectively $g$-tensor isotropy\cite{ariciu2019engineering}, large molecular covalency\cite{fataftah2019metal,briganti2019covalency} and molecular rigidity\cite{bader2014room,atzori2016room} as primary ingredients for long spin-lattice relaxation times in molecular qubits, can be reconciled by recognizing that they all point to different contributions to spin-phonon coupling and that they all must be considered at the same time in analyzing spin-lattice relaxation times. This analysis holds for $S=1/2$ in high-field, were the $g$-tensor is likely to drive relaxation. The study of $S>1/2$ transition metals or lanthanides needs to follow a different strategy. However, the analysis of $g$-tensor contributions to spin-phonon coupling brings forward an important general point: strong spin-phonon coupling arises from the modulation of those molecular degrees of freedom that more intensely contribute to the magnitude of the static spin Hamiltonian parameters. Interestingly, this analysis can also be extended to the discussion of molecular symmetry\cite{santanni2020probing,Kazmierczak2021}.\\

Similarly to the case of $g$-tensor, magnetic anisotropy $\mathbf{D}$ in transition metals can be described within a ligand field formalism and second-order perturbation theory. Taking analytical derivatives of $\mathbf{D}$ one obtains an expression equivalent to Eq. \ref{gshlf}, pointing once again to the energy of excited electronic states as the key variable influencing spin-phonon coupling\cite{higdon2020spin}. In the case of transition metals with very large anisotropy, electronic excited states are generally so close to the ground state that the perturbation theory used to describe $D$ in the ligand field approach breaks down. This electronic configuration leads to a pseudo-degenerate ground state which maximise the effect of spin-orbit coupling beyond perturbative regime, similarly to what happens in lanthanides\cite{atanasov2011detailed}. The molecular vibrations that contribute to the removal of such electronic degeneracy will then be those with large spin-phonon coupling. In the case of [(tpaPh)Fe]$^{−}$ it was in fact observed that the vibrations that maximally couple to spin are those Jahn-Teller-like distortions that remove the C$_{3v}$ symmetry axis of the molecule by bending the in-plane $\widehat{NFeN}$ angles, thus quenching the residual angular momentum\cite{lunghi2017intra}.\\

\begin{figure}[h!]
 \begin{center}
  \includegraphics[scale=0.7]{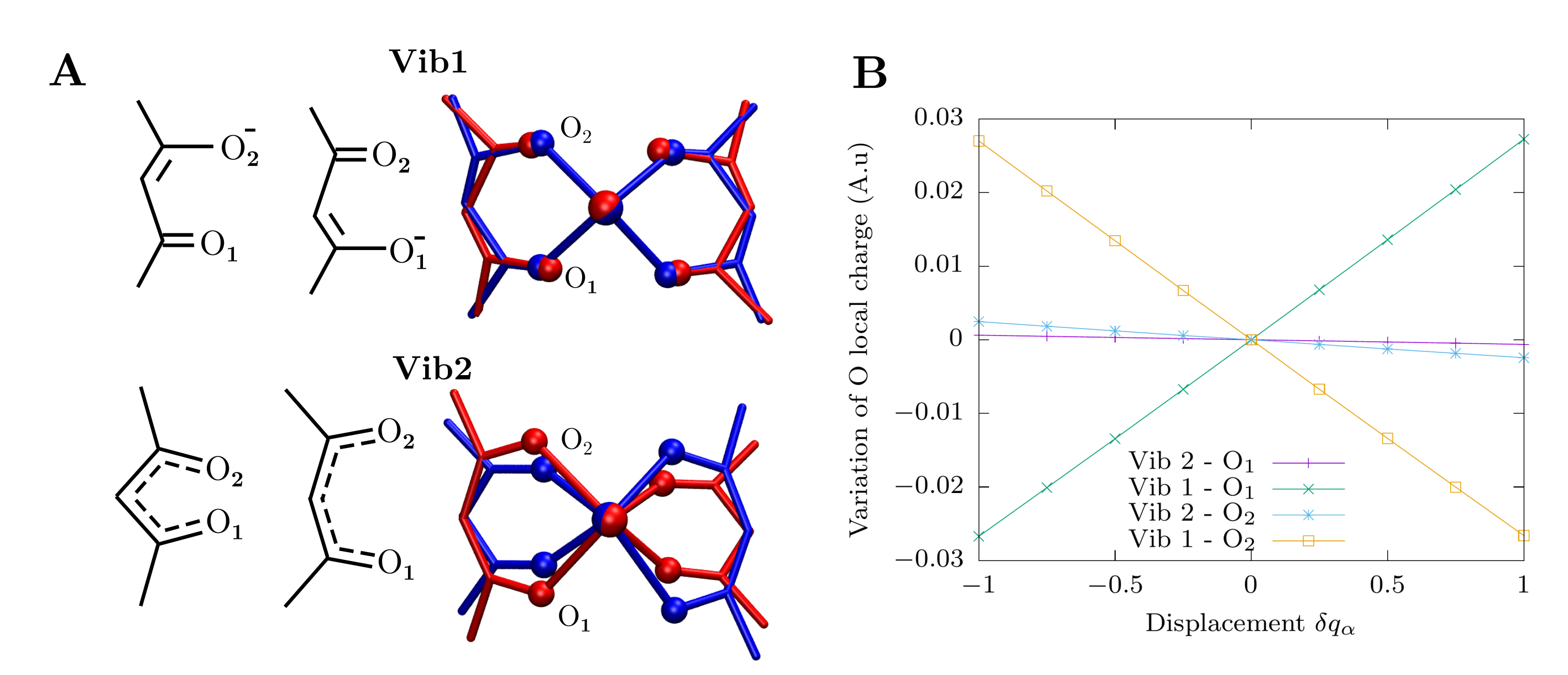}
 \end{center}
 \caption{\textbf{Spin-phonon coupling in Dy(acac).} Panel A schematically shows two different types of vibrations similarly coupled to spin. Panel B reports the variation of local charge density on the Oxygen atoms during the molecular motion. Reprinted with permission from ref.\cite{briganti2021}. Copyright 2021 American Chemical Society.}
 \label{Dyacac}
\end{figure}

The spin properties of single-ion Lanthanide complexes are largely dominated by electrostatic contributions, as discussed by Rinehart et al\cite{rinehart2011exploiting}. Depending on the element of the Ln series, the electronic density associated to different $M_{J}$ states can either be prolate, oblate or spherical. Crystal field interactions would then stabilize certain shapes of electronic density depending on the spatial orientation of the ligands. For instance, the states with $M_{J}=\pm 15/2$ and $M_{J}=\pm 1/2$ of Dy$^{3+}$ ions are associate with an oblate and prolate electronic density, respectively. A very axial distribution of ligands around the Dy ion would then stabilize in energy the $M_{J}=\pm 15/2$ states with respect to $M_{J}=\pm 1/2$, and \textit{vice versa} for equatorial crystal fields\cite{ungur2011magnetic, chilton2015first,chilton2015design,ungur2016strategies}. As a consequence, it is natural to expect that vibrations able to break the axial symmetry of the Dy ion's crystal field will be able to strongly couple to its magnetic moment. It has also been show that a second coupling mechanism can arise\cite{briganti2021}. This is due to the fact that the effective charge of the first coordination sphere, determining the crystal field felt by the ion, is also modulated during molecular motion. This adds a second channel of spin-phonon coupling and in Ref. \cite{briganti2021} it was shown that even vibrations that little affect the shape of the first coordination shell are able to strongly couple to spin if the second-coordination shell motion is able to modulate the ligands' local charges. Fig. \ref{Dyacac} provides a schematic representation of these two different spin-phonon coupling mechanisms for the molecule Dy(acac)$_{3}$(H$_{2}$O)$_{2}$\cite{jiang2010mononuclear}.
\textbf{Vib1}, schematically represented in the left panel of Fig. \ref{Dyacac}, is a molecular vibration where the first coordination shell of Dy is little distorted, but at the same time the charges of the ligands show large fluctuations due to the vibrations involving the second coordination shell. \textit{Vice versa}, \textbf{Vib2} represents the typical case where local charges remain largely constant during vibration but the first coordination shell gets significantly distorted. Both types of vibrations were found to be strongly coupled to spin.

\newpage
\section{The mechanism of spin-phonon relaxation in magnetic molecules} \label{abinitio-relaxation}

In this last section of the chapter we will finally address the mechanism of spin-phonon relaxation and the prediction of its rate. \\

\subsection*{Direct and Raman Relaxation in $S=1/2$ systems}

\begin{figure}[h!]
 \begin{center}
  \includegraphics[scale=1]{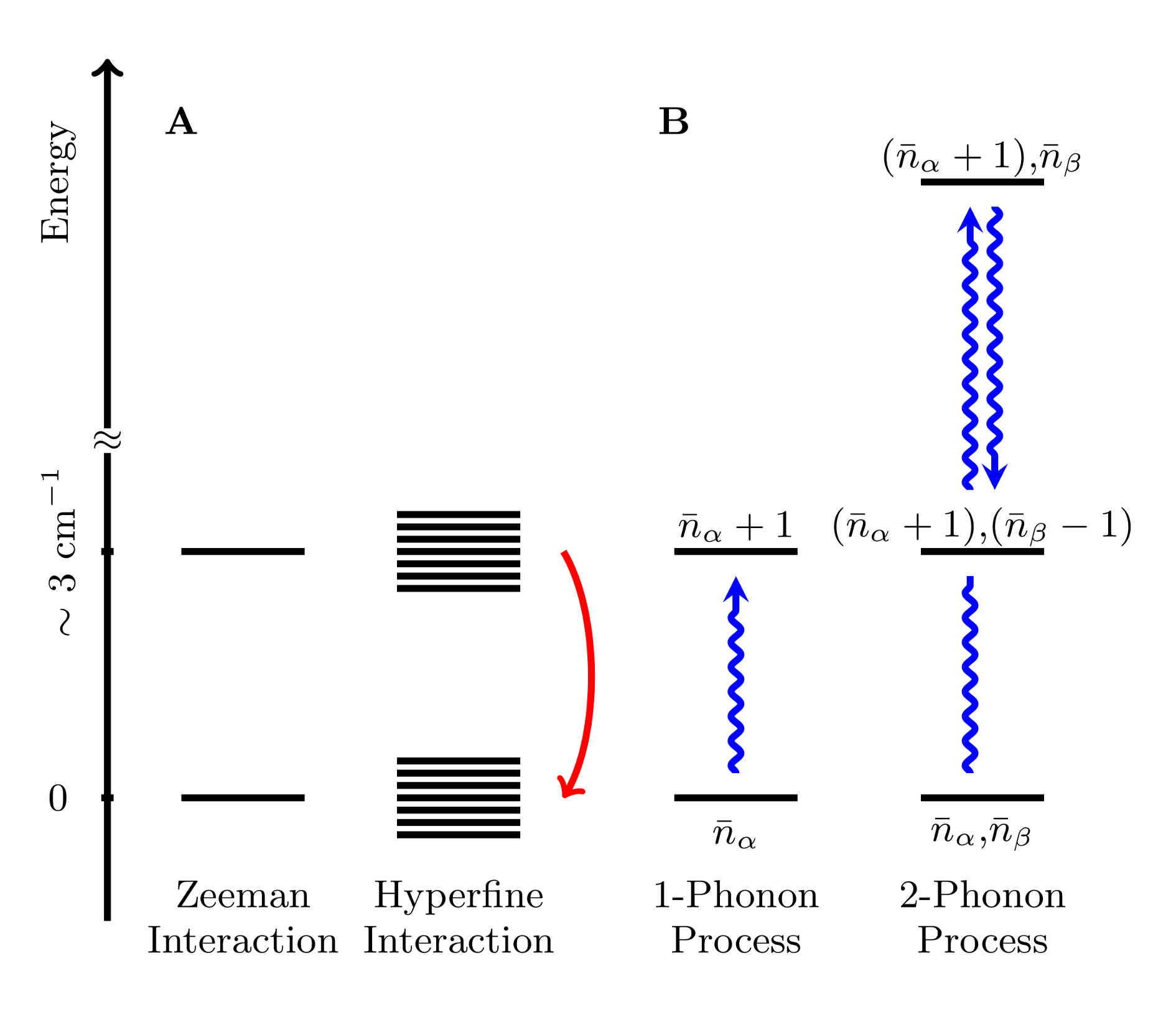}
 \end{center}
 \caption{\textbf{Relaxation mechanism in VO(acac)$_{2}$.} The spin levels are reported in the section A of the scheme. The part B of the scheme instead report the vibrational levels involved in direct and Raman relaxation. Reprinted with permission from ref. \cite{lunghi2020limit}. Copyright 2020 American Chemical Society.}
 \label{VOlevels}
\end{figure}

\begin{figure}[h!]
 \begin{center}
  \includegraphics[scale=1]{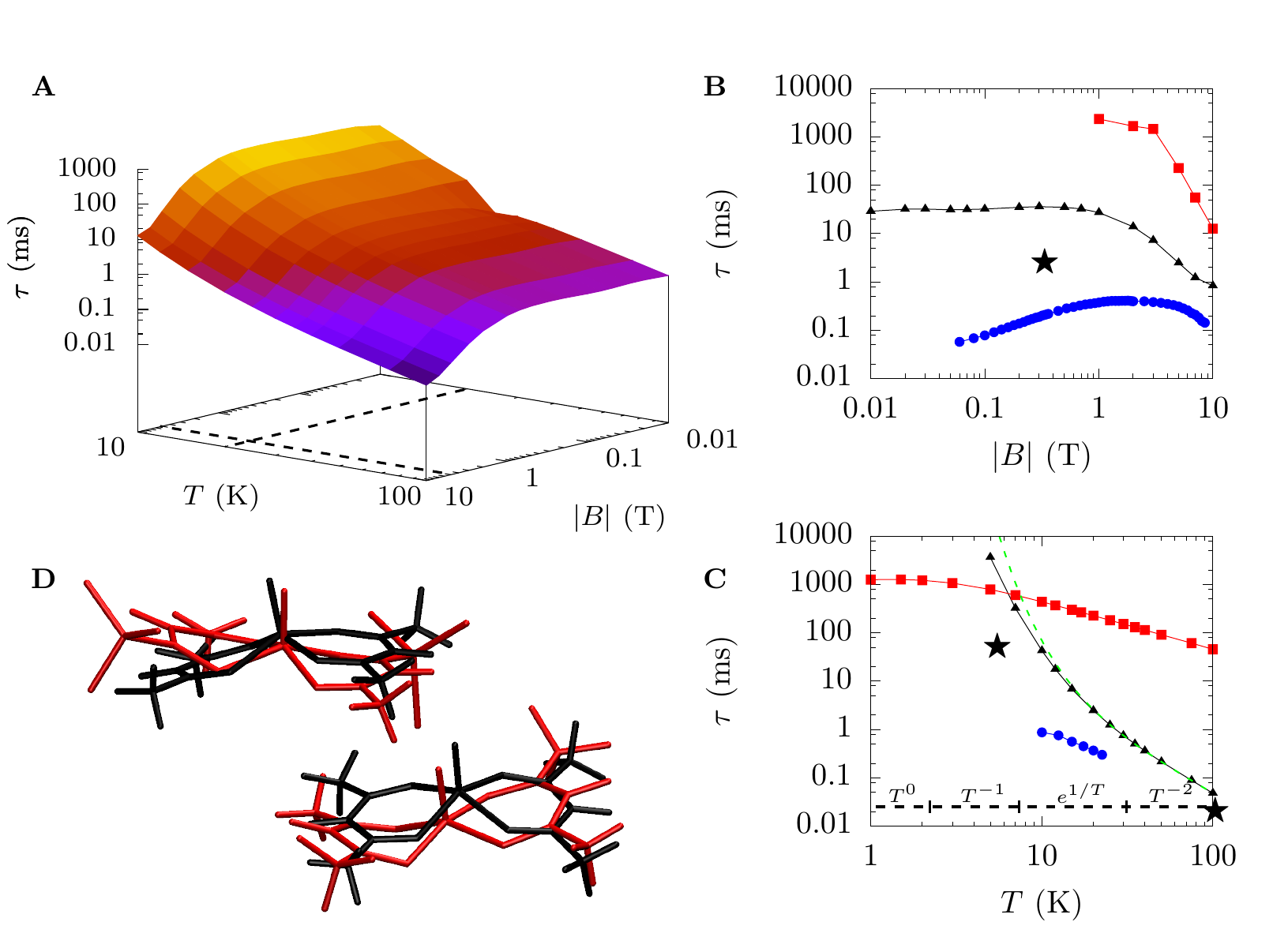}
 \end{center}
 \caption{\textbf{Spin-phonon relaxation in VO(acac)$_{2}$.} Panel A reports the simulated Raman relaxation time as function of field and temperature. Panel B and C reports both Raman (black) and direct (red) simulated relaxation time together with experimental values from AC measurements (blue) and inversion recovery on diluted sample of similar molecules (black star). Panel B refers to 20 K and Panel C to 5 T. Panel D reports the molecular displacements associated to the first two optical modes at the $\Gamma$-point. Reprinted with permission from ref. \cite{lunghi2020limit}. Copyright 2020 American Chemical Society.}
 \label{VOacac_relax}
\end{figure}

The molecule VO(acac)$_{2}$ will serve once again as benchmark system to present the main features of spin-phonon relaxation in $S=1/2$ systems. In the case of a two-level system, one- and two-phonon transition rates are determined by second-order perturbation theory, and first- and second-order coupling strength, respectively. In Refs \cite{lunghi2019phonons} and \cite{lunghi2020limit} the numerical strategies of section \ref{abinitio-methods} were used to compute the Redfield matrices of Eqs. \ref{Red21} and \ref{Red22} and determine the time evolution of the magnetization moment. The main results of that studies are reported in Figs. \ref{VOlevels} and \ref{VOacac_relax}\\

Let us begin our analysis from direct spin relaxation. This relaxation mechanism involves transitions between spin states due to resonant one-phonon absorption or emission, as depicted in Fig. \ref{VOlevels}. \textit{Ab initio} spin dynamics simulations were carried out as function of magnetic field intensity and showed that two possible relaxation regimes are possible: the first one is active at low field, where the modulation of hyperfine coupling drives relaxation, and the second one is active at high field, where the modulation of the $g$-tensor is instead the main source of relaxation. For VO(acac)$_{2}$ the transition field between the two regimes was estimate to be around $B^{*}\sim$ 1 T. Importantly, the two regimes have a different field dependence, with the former going as $\tau\sim B^{-2}$ and the latter as $\tau\sim B^{-4}$. Regarding the temperature dependence of relaxation, it was observed that $\tau\sim T^{-1}$ at high $T$, and $\tau\sim T^{0}$ for $T\rightarrow0$. \\

Interestingly, \textit{ab initio} spin dynamics simulations reveal a picture of direct relaxation in agreement with the conclusions of section \ref{oldmodels}. As anticipated in section \ref{abinitio-sph}, this is because at very low vibrational frequencies typical of Zeeman/hyperfine splittings, the acoustic modes of a molecular crystal qualitatively behave as prescribed by the Debye model. Therefore, for Zeeman splittings much lower than the energy of the optical phonons, the qualitative $\tau$ vs $B/T$ behaviour expected from classical models based on the Debye model holds. It is important to note, however, that the energies of VO(acac)$_{2}$'s optical modes are on the high-end side of this class of compounds, with several examples known where the first optical vibrations fall as low as 10 cm$^{-1}$\cite{albino2019first}, as opposite to 40-50 cm$^{-1}$ in VO(acac)$_{2}$\cite{lunghi2019phonons}. In these compounds severe deviations from the conclusions of the Debye model should be expected for fields as low as a few Tesla. \\

Let us now turn to the Raman relaxation, which it was found to be driven by simultaneous absorption and emission of two phonons, as represented schematically in Fig. \ref{VOlevels}. Also in the case of Raman relaxation, it was showed that two field regimes are potentially active, with hyperfine driving relaxation as $\tau\sim B^{0}$ and $g$-tensor as $\tau\sim B^{-2}$\cite{lunghi2020limit}. Once again, this is in agreement with the conclusions reached in section \ref{oldmodels}. The temperature dependence of Raman relaxation was found to behave as $T\sim T^{-2}$ for $T>20$ K and then rapidly increase for lower $T$. In section \ref{oldmodels} we derived a similar behaviour of $\tau$ vs $T$ coming from the acoustic phonons as accounted in the Debye model. However, as we have noted in the previous section, optical phonons are much more strongly coupled with spin, and since they fall at surprisingly low energies, it is natural to look at them as source of spin-phonon relaxation instead of acoustic ones. The results for VO(acac)$_2$ were then interpreted by noticing that the Fourier transform of the two-phonon correlation function $G^{\mathrm{2-ph}}$, reported in Fig. \ref{MLVOacac}C for the process of simultaneous absorption and emission of two phonons at $T=20$ K, shows a sharp maximum at some frequency close to the first $\Gamma$-point optical frequency. Since only the phonons in a small energy windows seem to contribute to relaxation, we can treat them as an effective localized mode. According to the structure of Eq. \ref{Red22}, a pair of quasi-degenerate phonons contributes to $\tau$ as
\begin{equation}
 \tau^{-1}\sim \frac{e^{\beta\omega}}{(e^{\beta\omega}-1)^{2}}
 \label{ramanT2}
\end{equation}
which gives the correct simulated behaviour in the limit of k$_{B}T>\omega$. A relaxation process that follows a single contribution such as Eq. \ref{ramanT2} is often associated with the local-mode relaxation mechanism\cite{eaton2002relaxation}. The latter was originally proposed to origin in the presence of local low-energy optical modes in a magnetically doped crystal due to the mismatch in vibrational frequency between the magnetic impurities and the diamagnetic host\cite{klemens1962localized}. However, in the case of magnetic molecules there is no reason to invoke such a mechanism as plenty of low-lying optical modes are always present, even in defect-free crystals\cite{garlatti2020unveiling}.\\

The $T^{-2}$ limit closely matches the experimental observation of relaxation following a low-exponent power law. However, deviations from $T^{-2}$ are often observed in favour of power-law dependencies $T^{-n}$ with $n>2$\cite{atzori2016room}. Three possible explanations are likely to apply to this scenario. When several strongly-coupled low-energy optical modes are present, multiple contributions in the form or Eq. \ref{ramanT2} overlap, resulting in a more complex $T$ dependency. On the other hand, when relaxation time is fast and rapidly exits the window of measurable values $\tau$, it is hard to demonstrate that the high-$T$ limit has been fully established. In this scenario, the profile of $\tau\: \text{vs}\: T$ might still be affected by the pseudo-exponential regime and therefore results might erroneously be fitted with a power-law with exponent larger than $n=2$. Finally, at this stage it is impossible to exclude the contribution of spin-phonon coupling strength beyond the quadratic order and that should be considered in future works. Despite the limitations of Eq. \ref{ramanT2}, we suggest that it should be preferred to power laws when fitting experimental data. The latter, unless supported by a physical picture as the one presented in section \ref{oldmodels}, do not provide insights on the spin relaxation mechanism. On the other hand, Eq. \ref{ramanT2} contains the frequency of the phonons contributing to relaxation and therefore is able to provide a clearer interpretation of relaxation experiments. Such approach has been successfully applied to a series of spin 1/2 systems\cite{santanni2020probing,de2021exploring,Kazmierczak2021,pfleger2021terminal}. \\

It is important to remark that the agreement between experiments and simulations become acceptable only in the regimes of high-$T$/high-$B$. This is due to the fact that in the opposite regime, dipolar spin-spin coupling, not included in the simulations, becomes the dominant source of relaxation. Indeed, when diluted samples are measured with EPR inversion recovery, a much better agreement is observed\cite{lunghi2021towards}.\\  

\textit{Ab initio} spin dynamics simulations also make it possible to determine the phonons responsible for relaxation. In the case of direct relaxation, only phonons in resonance with the transition are able to contribute. As depicted in Figs. \ref{DOS1VOacac_dec}, at fields lower than 10 T only acoustic phonons are present, and their intra-molecular component is the main drive for relaxation. In the case of Raman relaxation instead, all the phonons in the spectrum are potentially able to contribute. However, only low-energy optical phonons significantly contribute to spin dynamics as they are at the same time thermally populated and strongly-coupled. Indeed, the population of high-energy phonons diminish exponentially with their frequency, while the acoustic phonons, despite being thermally populated, have a low-density of states. This can also be appreciated from Fig. \ref{DOS1VOacac_dec}, where the density of states rapidly drops below 40-50 cm$^{-1}$. Fig. \ref{VOacac_relax}D shows typical molecular distortions associated with the first few optical modes at the $\Gamma$-point. \\

\subsection*{Orbach and Raman Relaxation in $S>1/2$ systems with uni-axial anisotropy}

Our second case-study is a high-spin Co$^{2+}$ complex ($S=3/2$) with large uni-axial magnetic anisotropy\cite{rechkemmer2016four}, namely with $D=-115$ cm$^{-1}$ and vanishing rhombicity $|E/D|$. In the absence of an external field, the direct transitions within Kramers' doublets $M_{S}=\pm n/2$ are prohibited by time-reversal symmetry and one-phonon relaxation must occur through the Orbach mechanism. Interestingly, the same selection-rule applies to two-phonon transitions due to second-order perturbation theory and quadratic spin-phonon coupling, as the matrix elements of Eqs. \ref{Red21} and \ref{Red22} have the same dependency on spin operators and $\hat{H}_{s}$ eigenfunctions. The source of two-phonon relaxation is therefore to be looked for in Eq. \ref{Red41}, which uses fourth-order perturbation theory. 
\begin{figure}[h!]
 \begin{center}
  \includegraphics[scale=1]{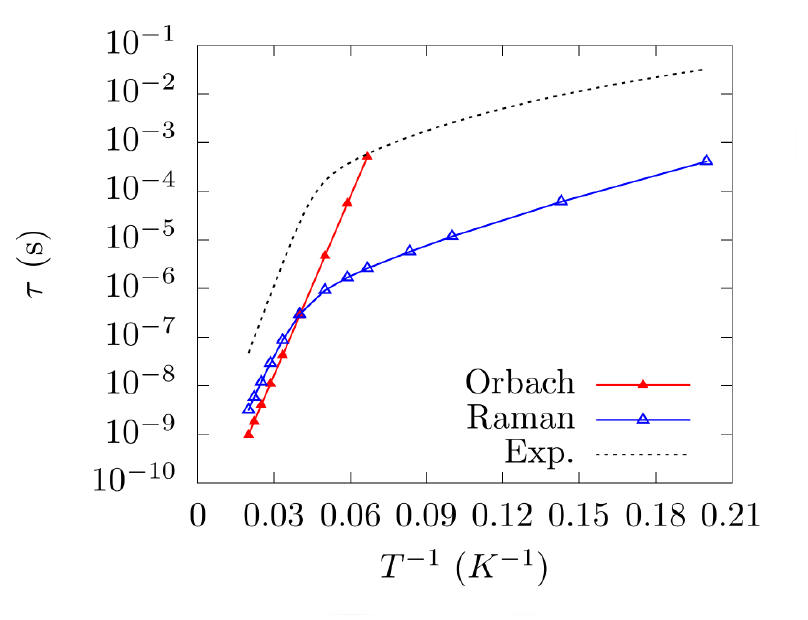}\includegraphics[scale=1]{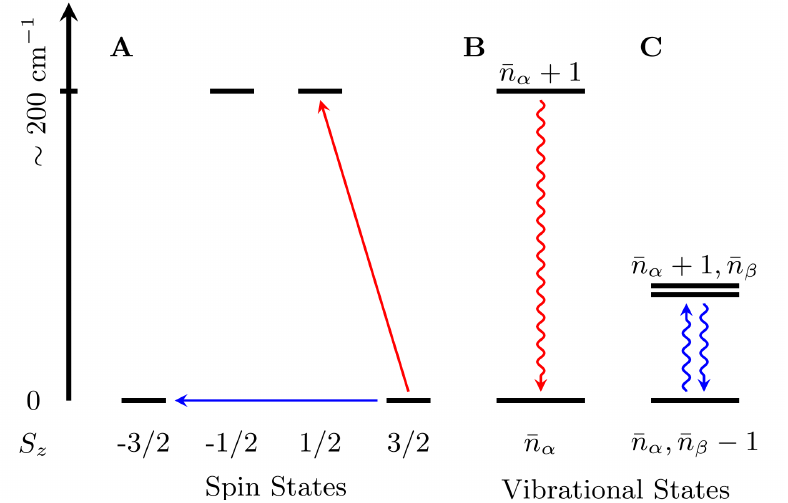}
 \end{center}
 \caption{\textbf{Orbach and Raman Relaxation in a Co$^{2+}$ complex.}
Left Panel: dashed lines corresponds to the measured relaxation time in zero external field. The red line and triangles is the computed Orbach relaxation and the blue line and triangles is the computed Raman relaxation. Right Panel: (\textbf{A}) The spin states of Co(pdms)$_{2}$ are reported as function of their energy and the nominal $S_{z}$ value. The red and blue arrows represent possible relaxation pathways. (\textbf{B}) An electronic excitation from the ground state to an excited state (red arrow in panel \textbf{A}) can be accompanied by the absorption of a phonon with energy in resonance with the spin transition (Orbach process). (\textbf{C}) The direct transition $S_{z}$=3/2 $\rightarrow$ $S_{z}$=-3/2 (blue arrow in panel \textbf{A}) can be induced by the simultaneous absorption and emission of two phonons with the same energy (Raman process). Reprinted from ref. \cite{lunghi2020multiple}, with the permission of AIP Publishing.}
 \label{Copdms}
\end{figure}
The main results of Ref. \cite{lunghi2020multiple} are presented in Fig. \ref{Copdms}. At high temperature the Orbach mechanisms becomes the most favourable relaxation pathway and it is dominated by one-phonon absorption of phonons resonant with the $M_{S}=\pm 3/2 \rightarrow M_{S}=\pm 1/2$ transition, as depicted in the right panel of Fig. \ref{Copdms}. At low temperature, the Raman mechanism instead drives relaxation and show a pseudo-exponential trend with respect to $T$. From the analysis of the transition rates computed with Eq. \ref{Red41}, it was possible to interpret the Raman relaxation as a direct spin flip $M_{S}=\pm 3/2 \rightarrow M_{S}=\mp 3/2$ due to the simultaneous absorption and emission of two degenerate phonons. This process is represented in the right panel of Fig. \ref{Copdms}. Overall, the temperature behaviour of Orbach and Raman spin-relaxation were found to follow the expression 
\begin{equation}
 \tau^{-1} = \tau_{0}^{-1} e^{-\beta U_{\mathrm{eff}}} + \tilde{V}_{\mathrm{2-sph}}\frac{e^{\beta\tilde{\omega}}}{(e^{\beta\tilde{\omega}}-1)^{2}}\:,
 \label{orbach_raman}
\end{equation}
In Eq. \ref{orbach_raman}, $U_{\mathrm{eff}}$ is the energy of the excited Kramers doublet and $\tilde{\omega}$ is a value of energy compatible with the lowest-lying optical modes, computed around $\sim 20$ cm$^{-1}$. Similarly to the case of $S=1/2$, Raman relaxation receives contributions from all the vibrational spectrum. Among all the phonons, those contemporaneously populated and coupled to spin will drive relaxation. However, in the case of $S=3/2$, the matrix element of Eq. \ref{Red41} will also favour phonons in resonance with the spin excited states. In highly anisotropic compounds such as this, the excited Kramers doublet lies at a much higher energy than the first optical modes, suggesting that phonons well above the first optical transitions should be involved in Raman relaxation. However, it should be noted that the resonant condition with the excited states is only imposed at the power of two (see Eq. \ref{Red41}), while the population of phonons decreases exponentially with their energy, as dictated by the Bose-Einstein population. At low temperature the latter condition overcomes the former, leading to the conclusion that low-energy modes are the main drive for Raman relaxation in both $S=1/2$ and $S>1/2$. As discussed for $S=1/2$, acoustic modes do not generally contribute significantly to Raman relaxation in molecules as they have a very low density of states. However, at very low temperature ($T<5$ K), when even optical modes are strongly un-populated, then acoustic phonons might become relevant. Similarly to what discussed for relaxation in $S=1/2$, the use of Eq. \ref{orbach_raman} should be preferred to power laws, as commonly done.\\

As it can be appreciated from Fig. \ref{orbach_raman}, the accuracy reached in Ref. \cite{lunghi2020multiple} is not fully satisfactory and the error between experiments and predictions is around one and two orders of magnitude. As discussed in Ref. \cite{lunghi2021towards}, this is almost entirely due to the misuse of the secular approximation, which does not totally decouple coherence and population terms of the density matrix for Kramers systems in zero external field. Once the secular approximation is correctly carried out, the agreement between experiments and theory becomes virtually exact\cite{lunghi2021towards}.\\

\begin{figure}[h!]
 \begin{center}
  \includegraphics[scale=1]{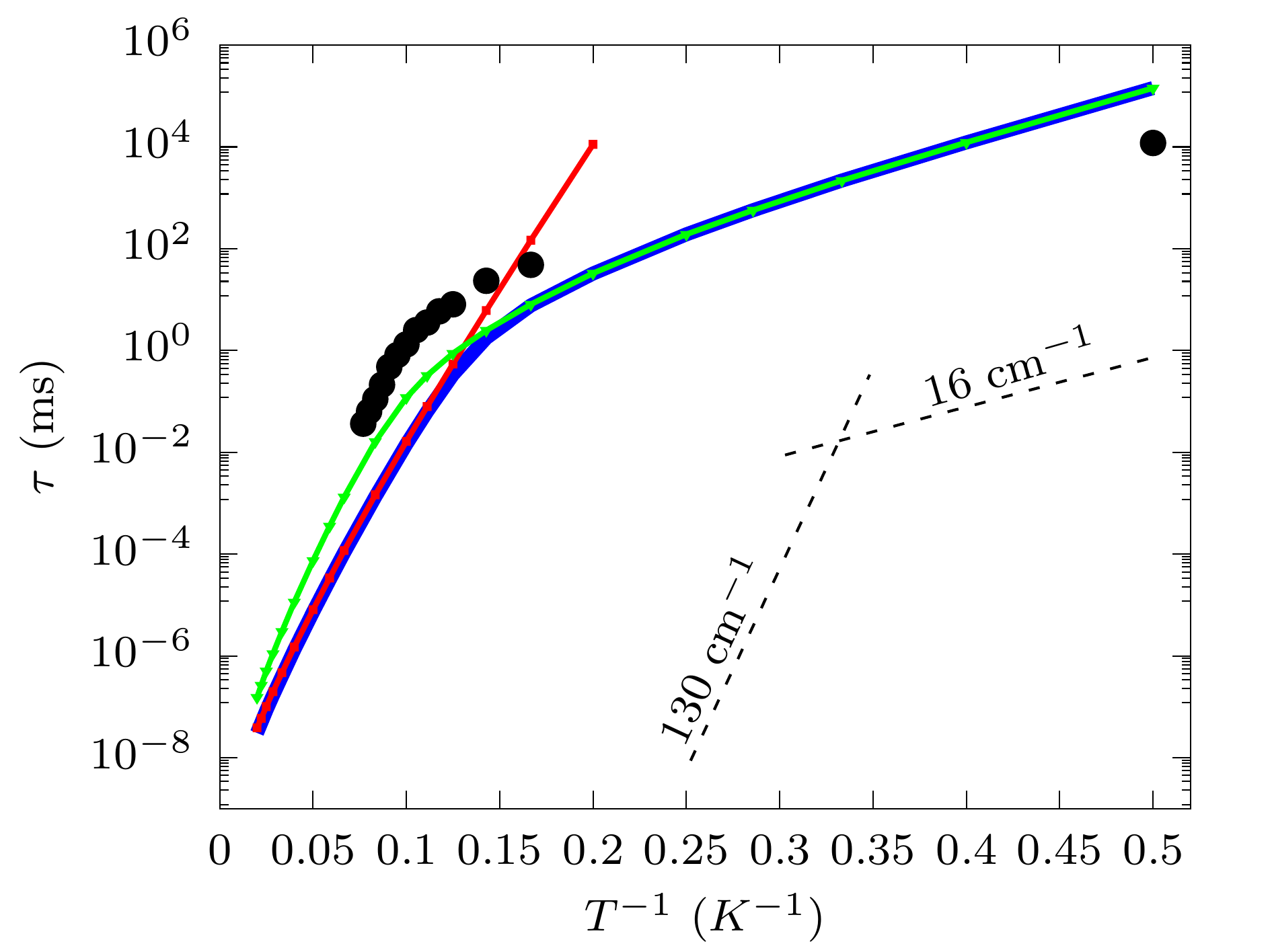}
 \end{center}
 \caption{\textbf{Orbach and Raman Relaxation in Dy$^{3+}$ complexes.} Experimental relaxation time for Dy(acac) (black dots) together to the simulated Orbach (red line and symbols) and Raman (green line and symbols) ones. Adapted with permission from ref. \cite{briganti2021}. Copyright 2021 American Chemical Society.}
 \label{Dyrelax}
\end{figure}

Dy$^{3+}$ complexes with axial symmetry have received a large attention from the molecular magnetism community in virtue of their slow spin relaxation\cite{luzon2012lanthanides}. Several molecules have then been studied by means of \textit{ab initio} methods to elucidate the role of spin-phonon coupling and predict spin relaxation rates. Ref. \cite{goodwin2017molecular} represents the first attempt at computing the relaxation times in Ln complexes. The system studied was [DyCp$^{\textrm{ttt}}_2$]$^{+}$ (Cp$^{\textrm{ttt}}$=[C$_{5}$H$_{5} ^{\;\;\textrm{t}}$Bu$_{3}$-1,2,4]), DyCp in short, \textit{i.e.} the first single-molecule magnet with a blocking temperature approaching Nitrogen boiling point. DyCp exhibits a Dy$^{3+}$ ion with a strongly anisotropic ground state $J=15/2$ due to an almost perfectly axial coordination geometry. As mentioned in section \ref{abinitio-sph}, this situation is ideal to create very large energy gaps between the KDs with $M_{J}=\pm 15/2$ and $M_{J}=\pm 1/2$, which is estimated to be around 1500 cm$^{-1}$ in DyCp. The simulations presented in in Ref. \cite{goodwin2017molecular} were able to capture the experimental results in the high-$T$ regime ($T>60$ K) with only one order of magnitude of error. Simulations shown that the modulation of the effective crystal field operator was responsible for transitions involving up to the 6th excited Kramers doublet, supporting the use of the first term of Eq. \ref{orbach_raman} to fit the Orbach relaxation regime and determining an effective anisotropy barrier $U_{\mathrm{eff}}\sim 1225$ cm$^{-1}$. 
The prediction of Orbach rates with \textit{ab initio} spin dynamics has been further validated through the study of a series of Dy$^{3+}$ molecules with long relaxation time, revealing interesting correlations between the molecular motion of high-energy modes in resonance with spin Kramers doublet and spin relaxation\cite{evans2019monophospholyl,yu2020enhancing,reta2021ab}. In particular, these studies showed that relaxation may occurs though complex relaxation pathways, often involving several of the 16 states of the Dy$^{3+}$ ground-state multiplet. Moreover, it was shown that, although the performance of slow-relaxing single-molecule magnets is primarily due to the axiality and strength of the crystal field splitting, among molecules showing similar values of $U_{\mathrm{eff}}$, the energy alignment and the careful design of the vibrational density of states can play a fundamental role in bringing down relaxation rates even further\cite{reta2021ab}. These series of studies were performed including only gas-phase vibrations as well as considering only one-phonon processes. As a consequence they were not able to predict spin relaxation at low temperature, where a second pseudo-exponential relaxation regime appears, similarly to what observed in Co$^{2+}$ (see Fig. \ref{Copdms}).\\

The compound Dy[(acac)$_{3}$(H$_{2}$O)$_{2}$], Dyacac in short, was also studied with \textit{ab initio} spin dynamics, this time including all the phonons of the crystal's unit-cell and both one- and two-phonon transitions\cite{briganti2021}. Similarly to what discussed for Co$^{2+}$, relaxation occurs through Raman mechanism at low temperature while Orbach is the dominating mechanism at high temperature. Once again the relaxation time depends on temperature as reported in Eq. \ref{orbach_raman} and depicted in Fig. \ref{Dyrelax}. Differently from DyCp, the Orbach relaxation in Dyacac was observed to be mediated by the first two excited Kramers doublets, due to the low symmetry of the coordination of the Dy ion and the consequently large admixing between different $M_J$. Regarding the Raman mechanism it was instead observed that only the first few optical modes at the $\Gamma$-point contribute to relaxation in virtue of a larger thermal population at low temperature. \\

Simulations for DyCp were repeated using a similar approach to the one employed for Dyacac\cite{lunghi2021towards}, confirming that a two-phonon Raman mechanism is able to explain spin relaxation at low-$T$. A similar interpretation to the dynamics in Dyacac also applies to DyCp. Despite the much higher energy of the spin transitions in DyCp respect to Dyacac ( the first spin excited state falls at $\sim$ 450 cm$^{-1}$ in DyCp and  $\sim$ 100 cm$^{-1}$ in Dyacac), simulations confirm that low-energy vibrations up to a few THz are the only contribution to Raman spin-phonon relaxation. On the other hand, only high-energy optical phonons in resonance with spin transitions are responsible for Orbach relaxation. Importantly, the results on DyCp's spin dynamics in zero-field showed a remarkable dependence on the inclusion of coherence terms in the spin reduced density matrix, with deviations of up to 9 orders of magnitude from the experimental results when neglected. \textit{Vice versa}, as observed for Co$^{2+}$ system discussed above, once the secular approximation is correctly implemented, simulations become in perfect agreement with experiments over the entire range of temperature\cite{lunghi2021towards}, therefore providing a conclusive proof that spin-phonon relaxation in magnetic molecules of Kramers ions presenting easy-axis magnetic anisotropy is fully understood.

\clearpage

\section*{Outlook}

In this chapter we have illustrated how the theory of open quantum systems can be adapted to the problem of spin-phonon interaction and provide a quantitative ground for the study of spin dynamics. Despite the handful of systems studied so far, \textit{ab initio} simulations have already provided deep insights into the nature of spin-phonon coupling and spin relaxation in crystals of magnetic molecules. Here we will attempt to summarize the most important points. \\

Most of our understanding of spin relaxation in solid-state comes from the application of the Debye model and perturbation theory. However, differently from the solid-state doped materials initially studied, molecular crystals posses many low-lying optical modes. Moreover, as one of the main goals pursued in the field of molecular magnetism is the enhancement of magnetic anisotropy, the energy splitting among spin states have been increasing by orders of magnitudes along the years\cite{zabala2021single,duan2021data}. This situation made it so that the assumptions of section \ref{oldmodels} have lost their general validity. \textit{Ab initio} simulations showed that for molecular materials, acoustic modes are not particularly relevant and that optical ones are instead responsible for relaxation. Moreover, it was found that optical modes appear at surprisingly low energies, so that even at very low temperature they remain sensibly populated. Thanks to \textit{ab initio} spin dynamics, it has now been possible to demonstrate the nature of spin relaxation mechanism in Kramers systems. Direct and Orbach mechanisms are due to the absorption and emission of resonant phonons\cite{goodwin2017molecular,lunghi2019phonons}. The origin of Raman relaxation is attributed to different mechanisms for $S=1/2$ and $S>1/2$ systems. In the former case, quadratic spin-phonon coupling is responsible for two-phonon transitions\cite{lunghi2020limit}, while in the latter case, fourth-order perturbation theory is necessary to explain spin-relaxation in zero-external field\cite{chiesa2020understanding,gu2020origins,lunghi2020multiple,briganti2021}. In both cases, Raman is mediated by low-energy optical phonons exhibiting local molecular rotations admixed to delocalized intra-molecular distortions.\\

These findings have major consequences for molecular magnetism practices. In addition to well established approaches to increase the axiality of the crystal field\cite{rinehart2011exploiting} and exchange coupling multiple ions\cite{Gould2022ultrahard}, \textit{ab initio} spin dynamics results point to the design of rigid molecular structures as key to reduce spin-phonon relaxation rates\cite{lunghi2017intra}. Indeed, by increasing molecular rigidity, optical modes are shifted up in energy with a two-fold effect: i) they become less admixed with acoustic modes\cite{lunghi2019phonons}, so reducing the rate of direct relaxation, and ii) they become less populated so that Raman relaxation rate would also slow down\cite{lunghi2020limit,lunghi2020multiple,briganti2021}. A second approach suggested by \textit{ab initio} spin dynamics involves the careful disalignement of spin and vibrational spectra in order to reduce the effect of resonant Orbach process\cite{goodwin2017molecular,ullah2019silico,yu2020enhancing}.\\

Finally, although \textit{ab initio} spin dynamics is not yet a fully quantitative computational approach, we have shown that it is rapidly maturing into one. Since its first proposals in 2017\cite{escalera2017determining,lunghi2017role,lunghi2017intra,goodwin2017molecular}, a level of accuracy down to one order of magnitude has now been proved for different systems across different relaxation regimes\cite{lunghi2021towards}. Although further benchmarking is needed, this level of accuracy already hints at the possibility to rank the spin lifetime of different compounds in advance of experimental characterization. This is an important milestone for the field of molecular magnetism, as it opens the gates to a long-sought fully \textit{in silico} screening of new compounds.\\

Let us now attempt to provide an overview of future developments as well as a brief discussion of the potential impact of \textit{ab initio} spin dynamics for other adjacent fields. \\

\textit{Ab initio} theory of spin relaxation now includes all the terms up to two-phonon processes. This is essentially the same level of theory employed in the past to derive the canonical picture of spin-phonon relaxation and we can therefore claim that the theoretical formulation of such theories in terms of \textit{ab initio} models is available. However, this is far from claiming that no further work will be necessary to obtain a conclusive and universal picture of spin relaxation in solid-state. On the contrary, this theoretical advancements form the stepping stone for an in-depth analysis of relaxation phenomenology, for instance going beyond the classical Born-Markov approximation and the inclusion of other decoherence channels such as magnetic dipolar interactions. These two avenues in fact represent some of the most urgent development of the proposed theoretical framework and have the potential to account for phonon-bottleneck\cite{tesi2016giant,ullah2021insights} and magnetization quantum tunnelling effects\cite{ding2018field}, respectively. These contributions to relaxation are indeed ubiquitous in experiments and must find their place into a quantitative theory of spin dynamics. Moreover, it is important to remark that so far \textit{ab initio} spin dynamics has been applied to only a handful of systems and there is large scope for investigating the large and varied phenomenology of molecular magnetism. Examples includes the study of exchange-coupled poly-nuclear magnetic molecules\cite{zabala2021single}, surface-adsorbed systems\cite{mannini2009magnetic} and molecules in glassy solutions\cite{sato2007impact}.\\

In order to reach a fully quantitative ground, \textit{ab initio} spin dynamics now requires a careful testing of the electronic structure methods underlying it. It is well known that the inclusion of electronic correlation in the simulations is key to achieve accurate prediction of spin Hamiltonian parameters\cite{atanasov2015first,ungur2017ab}, and the same effect  can be expected for spin-phonon coupling coefficients. On the other hand, the simulation of phonon modes requires significant advancements. The introduction of sophisticated vdW corrections to DFT\cite{tkatchenko2012accurate} and the inclusion of anharmonic lattice effects\cite{ruggiero2020invited} stand out as two important avenues for future development. The development of efficient simulation strategies is another necessary step towards the widespread use of \textit{ab initio} spin dynamics. On the one hand, the \textit{ab initio} methods underlying spin relaxation theory are particularly heavy in computational terms. Magnetic molecules indeed often crystallize in large unit cells and the simulations of phonons beyond the $\Gamma$-point is a daunting task. Similarly, the calculation of spin-phonon coupling coefficients requires the quantum chemistry simulations for many similarly distorted molecular geometries, rapidly leading to immense computational overheads. In section \ref{abinitio-sph} we have hinted to machine learning as a game-changer in this area. Machine learning, learning the underlying features of a distribution from examples drawn out it, makes it possible to interpolate complex functions such as the potential energy surface of a molecular crystal\cite{lunghi2019unified,lunghi2020multiple} or the relation between spin Hamiltonian and molecular structure\cite{lunghi2020surfing}. After the training stage, ML models can be used to make predictions at very little computational cost, therefore offering a significant speed-up of these simulations. Further adaptation of machine learning schemes to this tasks is needed, but encouraging proof-of-concept applications have recently been presented\cite{lunghi2020surfing,lunghi2020limit,lunghi2020multiple,lunghi2020insights,nandy2021computational,zaverkin2021thermally}. Another interesting strategy involve the use of parametrized Hamiltonians to compute the spin-phonon coupling coefficients. When an accurate parametrization is possible, extensive speed-ups can be achieved\cite{escalera2020design}. \\

Although this chapter has been focusing on crystals of magnetic molecules, many more systems are within the reach of the methods we have illustrated. Indeed, the only strong assumptions of \textit{ab initio} spin dynamics are that the system must be well described with a spin Hamiltonian formalism and that it must exist an accurate \textit{ab initio} method to predict lattice and spin properties. The computational scheme we have discussed should readily apply to any kind of system exhibiting localized magnetism. Indeed, solid-state defects or impurities in solid-state semiconductors are commonly described with the same theoretical tools used in the field of molecular magnetism, and DFT-based schemes for the simulation of their properties already exists\cite{ivady2018first}. Many of these systems are currently under scrutiny for quantum sensing technologies and there is a large interest in their relaxation properties\cite{wolfowicz2021quantum}. Another type of system that we believe is easily within the reach of the presented methods is nuclear spin. Indeed, electronic structure methods are routinely used to predict nuclear spin Hamiltonian's parameters\cite{helgaker1999ab}, and when such nuclear spins are part of a molecular system, there is virtually no difference from the treatment of electron spin dynamics.\\

In conclusion, we have provided a comprehensive overview of the state-of-the-art in \textit{ab initio} spin dynamics as well as a practical guide to its application to crystals of magnetic molecules. We have shown that this novel computational method is rapidly maturing into a quantitative tool able to provide unique insights into spin relaxation, thus holding great promise for the field of molecular magnetism and beyond. As such, we hope that this work will serve as a starting point for the readers interested in adopting this strategy as well as a review of the strives so far for expert readers. \\

\noindent
\textbf{Acknowledgements}\\

We acknowledge funding from the European Research Council (ERC) under the European Union’s Horizon 2020 research and innovation programme (grant agreement No. [948493])\\

\newpage

\bibliographystyle{naturemag}
\bibliography{lib,lib_manual}

\end{document}